\documentclass[useAMS,usenatbib]{mn2e}
\usepackage{epsf}
\usepackage{graphicx,verbatim,amsfonts,amsmath,natbib,amssymb}
\usepackage{bm}
\usepackage{times}

\newcommand{\hmsolar}{h^{-1}{\rm\,M_{\sun}}}

\newcommand{\hmpc}{h^{-1}{\rm\,Mpc}}

\newcommand{\ihmpcC}{h^3 {\rm\,Mpc}^{-3}}
\newcommand{\kmsmpc}{{\rm\ km\ s^{-1}\ Mpc^{-1}}}

\newcommand{\etal}{et al.}
\newcommand{\beq}{\begin{equation}}
\newcommand{\eeq}{\end{equation}}
\newcommand{\beqa}{\begin{eqnarray}}
\newcommand{\eeqa}{\end{eqnarray}}

\newcommand{\galex}{{\sl GALEX}\,}

\title[$UV$-- Optical Clustering]{The $UV$--Optical Color Dependence of Galaxy Clustering in the Local Universe.}
\author[Y.-S. Loh et al.]{Yeong-Shang Loh$^{1}$\thanks{E-mail: yeongloh@astro.ucla.edu},
 R. Michael Rich$^{1}$, S\'{e}bastien Heinis,$^{9}$ Ryan Scranton$^{13}$,
\newauthor Ryan P. Mallery$^{1}$, Samir Salim$^{14}$, D. Christopher Martin,$^{2}$ Ted Wyder$^{2}$, St\'{e}phane Arnouts$^{9}$,
\newauthor Tom A. Barlow$^{2}$, Karl Forster$^{2}$, Peter G. Friedman$^{2}$, Patrick Morrissey$^{2}$, Susan G. Neff$^{5}$,
\newauthor David Schiminovich$^{6}$, Mark Seibert$^{2}$, Luciana Bianchi$^{7}$, Jose Donas$^{8}$,
\newauthor Timothy M. Heckman$^{9}$, Young-Wook Lee$^{10}$, Barry F. Madore$^{11}$, Bruno Milliard$^{8}$,
\newauthor Alex S. Szalay$^{9}$, Barry Y. Welsh$^{12}$ and Suk Young Yi$^{10}$\\
$^{1}$Department of Physics and Astronomy, University of California, Los Angeles, CA 90095-1562\\
$^{2}$California Institute of Technology,MC 405-47, 1200 East California Boulevard, Pasadena, CA 91125\\
$^{3}$Max-Planck Institut f\"ur Astrophysik, D-85748 Garching, Germany\\
$^{4}$Institut d'Astrophysique de Paris, CNRS, 98 bis boulevard Arago, F-75014 Paris, France\\
$^{5}$Laboratory for Astronomy and Solar Physics, NASA Goddard
Space Flight Center, Greenbelt, MD 20771\\
$^{6}$Department of Astronomy, Columbia University, New York, NY 10027\\
$^{7}$Center for Astrophysical Sciences, The Johns Hopkins
University, 3400 N. Charles St., Baltimore, MD 21218\\
$^{8}$Laboratoire d'Astrophysique de Marseille, BP 8, Traverse
du Siphon, 13376 Marseille Cedex 12, France\\
$^{9}$Department of Physics and Astronomy, The Johns Hopkins
University, Homewood Campus, Baltimore, MD 21218\\
$^{10}$Center for Space Astrophysics, Yonsei University, Seoul
120-749, Korea\\
$^{11}$Space Sciences Laboratory, University of California at
Berkeley, 601 Campbell Hall, Berkeley, CA 94720\\
$^{11}$Observatories of the Carnegie Institution of Washington,
813 Santa Barbara St., Pasadena, CA 91101\\
$^{12}$Space Sciences Laboratory, University of California at
Berkeley, 601 Campbell Hall, Berkeley, CA 94720\\
$^{13}$Department of Physics and Astronomy, University of Pittsburgh, 
3941 O’Hara St., Pittsburgh, PA 15260\\
$^{14}$National Optical Astronomy Observatory, 950 North Cherry Avenue, Tucson, AZ 85719}
\begin{document}

\pagerange{\pageref{firstpage}--\pageref{lastpage}} \pubyear{0000}

\maketitle

\label{firstpage}

\begin{abstract}
We measure the UV-optical color dependence of galaxy clustering in the local universe. Using the clean separation of the red and blue 
sequences made possible by the $NUV - r$ color-magnitude diagram, we segregate the galaxies into red, blue and intermediate ``green'' 
classes. We explore the clustering as a function of this segregation by removing the dependence on luminosity and by excluding
edge-on galaxies as a means of a non-model dependent veto of highly extincted galaxies.   We find that $\xi(r_p, \pi)$ for 
both red and green galaxies shows strong redshift space distortion on small scales -- the ``finger-of-God''  effect, with green galaxies 
having a lower amplitude than is seen for the red sequence, and the blue sequence showing almost no distortion.  On large 
scales, $\xi(r_p,\pi)$ for all three samples show the effect of large-scale streaming from coherent infall. On scales 
$1 \hmpc < r_p < 10 \hmpc$, the projected auto-correlation function $w_p(r_p)$ for red and green galaxies fits a power-law with slope 
$\gamma \sim 1.93$ and amplitude $r_0 \sim 7.5$ and $5.3$, compared with $\gamma \sim 1.75$ and $r_0 \sim 3.9 \hmpc$ for blue sequence galaxies. 
Compared to the clustering of a fiducial $L^*$ galaxy, the red, green, and blue have a relative bias of $1.5$, $1.1$, and $0.9$ respectively. 
The $w_p(r_p)$ for blue galaxies display an increase in convexity at $\sim 1 \hmpc$, with an excess of large scale clustering. Our results suggest that 
the majority of blue galaxies are likely central galaxies in less massive halos, while red and green galaxies have larger satellite fractions, 
and preferentially reside in virialized structures. If blue sequence galaxies migrate to the red sequence via processes like mergers or quenching that
take them through the green valley, such a transformation may be accompanied 
by a change in environment in addition to any change in luminosity and color.
\end{abstract}


\section{Introduction}\label{sec:intro}

With the advent of the Sloan Digital Sky Survey (SDSS;\citealt{Yor00}) and its value added galaxy catalogs, it has been possible to study 
the subject of galaxy bimodality and its relationship to fundamental properties, such as stellar mass and star-formation history 
\citep[e.g.,][]{Kau04,Sch07,Sal07}. The broad division of galaxies into star forming disks and quiescent early type galaxies is the 
fundamental principle of Hubble's tuning fork system of classification and is well established. In a plot of optical $g-r$ color vs $M_r$, 
red galaxies define a clear sequence, while the locus of blue galaxies is broadened into the so-called "blue cloud". The red sequence has 
been shown to maintain its integrity with look-back time \citep{Bow92} and has grown in mass since redshift $\sim 1$. 
Studies by \cite{Bel04}, \cite{Bla06}, \cite{Fab07}, and \cite{Bro08} argue that the stellar mass contained within the red population has increased 
by roughly a factor of two in half the Hubble time.

A significant breakthrough in expressing this blue/red dichotomy occurred when photometry from the {\sl Galaxy Evolution Explorer} (\galex), notably the 
near-ultra violet ($NUV$) band, was matched with SDSS photometry \citep{Mar07,Wyd07,Sch07,Sal07}. When the diagram is plotted using $NUV-r$ as the color, 
two clear sequences emerge: the familiar red sequence and a new blue sequence in place of the ''blue cloud'' of optical studies. Between the blue and red 
sequences there are galaxies present in a so-called ``green valley''. Many of these are spectroscopically classified Type II active galactic
 nuclei \citep{Ric05,Mar07,Sal07}.

\cite{Fab07}, \cite{Mar07}, and \cite{Sch07} propose several paths by which galaxies might transition from the blue to the red sequence. 
The presence of AGN in the green valley suggests that AGN activity is associated with a quenching of star formation \citep{Sil98,Hop06,Hop07}. 
Other paths from the blue to the red sequence might, hypothetically, involve gas-rich mergers of blue galaxies \citep{Too72}, or virial shock heating of cold 
gas streams \citep{Dek06}. The red sequence might consolidate in luminosity 
via dissipationless mergers of red galaxies, or low luminosity blue galaxies might acquire bulges through mergers with starbursts, retaining 
sufficient mass to land the evolved galaxy on the red sequence. However, the green valley might also be populated by casual visitors --- red galaxies 
that acquire gas and form stars or feed a central AGN. Following this brief burst of star formation, these galaxies might ultimately return to the red 
sequence from which they started.

In \cite{Sal07}, a plot of mass against specific star formation rate reveals a clear division between lower mass, star forming, blue sequence galaxies, 
and more massive AGN, which are not detected in large numbers until stellar mass $M\sim 3\times 10^{10}M_\odot$. The process responsible for populating 
the green valley and for potentially contributing to evolution from blue to red must bear some relationship to environment and to the dark matter 
halos in which the galaxies reside. In this study, we investigate the clustering environment of the blue and red sequences, and for the green valley.

The current paradigm of structure formation assumes that galaxies are assembled in dark-matter halos. The dependence of the clustering on galaxy 
properties may provide clues to the baryonic processes that are important to galaxy formation and evolution. The dependence of galaxy clustering 
on galaxy type has been known since the earliest studies of extragalactic astronomy \citep{Hub36, Zwi68}. In the modern era of large-scale galaxy 
surveys, \cite{Dav76} showed that the angular auto-correlation of ellipticals has a steeper power-law slope that those of spirals. Recent redshift 
surveys using the Two-Degree Field Galaxy Survey (2dF) and SDSS confirms these earlier results and the apparent bimodal nature of 
galaxy clustering \citep{Mad03,Bud03,Zeh05,Li06,Wan07}.

Studies using SDSS have further revealed that galaxy color is the property most predictive of local environment. \cite{Bla05b} found that at fixed 
luminosity and color, density does not correlate with surface brightness nor the Sersic index, and argue that morphological properties of galaxies 
are less closely related to environment than their star-formation history, and are traced by broadband optical colors. (See \cite{Par07} for an 
alternative analysis and point of view.) \cite{Li06} found that the dependence of clustering on optical $g-r$ color and $D_{4000}$ is much stronger 
than structural parameters like concentration and surface brightness, and extend to $5 \hmpc$, beyond what is expected from the localized
halo paradigm of structure formation. They concluded that at fixed stellar mass, the clustering properties of the surrounding dark matter 
haloes are correlated with the color of the selected galaxies. They further argued that different physical processes may be required to 
explain environmental trends in star formation, distinct from those established by galaxy structure.

In this paper, we will consider the color dependence of the two-point physical correlation function of galaxies using samples constructed from \galex, 
augmented with redshift and optical data from SDSS. In particular, we use the natural separation from the $NUV-r$ color to assign galaxies into three 
subsamples of red, green and blue galaxies. Our study complements recent work by \cite{Hei08} who investigate the physical clustering of galaxies as a 
function of star-formation history in the local universe, as well as earlier studies by  \cite{Mil07}, \cite{Hei07} and \cite{Bas08} who investigate 
the angular-correlation function of rest frame $UV$-selected galaxies and their evolution. We measure the auto-correlation function each of the different 
subsamples of galaxies, as well as the cross-correlation function between the subsamples. In section \ref{sec:data}, we describe in detail the data used 
in this analysis. In section \ref{sec:methods} we describe the method for estimating correlation functions. We present our results in 
section \ref{sec:results}, and discuss their implications for the nature of green valley galaxies and the formation of red sequence galaxies. 
We summarize our findings in section \ref{sec:summary}.

\section{Data}\label{sec:data}

\subsection{\galex and SDSS Data}

The ultraviolet imaging portion of the data-set is from the Galaxy Evolution Explorer Satellite (\galex) that was launched in 2003 April 
\citep{Mar05,Mor05,Mor07}. \galex obtains wide field imaging in both the far-UV ($FUV$; centered at $1540$\AA) and the near-UV ($NUV$, $2300$\AA) 
over a $1.2^\circ$ diameter field of view, with $5^{\prime\prime}$ images. Here we use data from the Medium Imaging Survey (MIS); these 
images are $\approx 1500$ sec in duration reaching $NUV \approx 23$ mag, covering an orbital shadow crossing. 
The MIS pointing that defines our sample targets the North Galactic Cap, which overlaps the SDSS spectroscopic footprint; this part of the program was 
designed from studies cross-matching 
SDSS and \galex data. The data-set used for our current analysis is from the Galaxy Release 3 (GR3) which is available from the Multi-mission 
Archive at STScI (MAST). The \galex pipeline uses SExtrator \citep{Ber96} to detect sources and measure fluxes. We use the ``MAG\_AUTO'' output 
from SExtrator as our default flux measurement; it is essentially a \cite{Kro80} magnitude with an elliptical aperture.

Because our analysis requires each galaxy to have a spectroscopic redshift, we start our cross-matching with a galaxy from the SDSS Main spectroscopic survey. For each 
SDSS galaxy, we search for the closest \galex detection within $4\arcsec$ radius from the location of the SDSS spectroscopic fiber. Only \galex 
sources within $0.55^\circ$ of the tile center field-of-view (FOV) are retained, since astrometry degrades toward the periphery of the FOV (a problem 
which will be resolved in later releases) and the incidence of artifacts increases as well. After the matching of \galex and SDSS sources, we further 
trim the sample to create a statistically complete data-set following the procedure of \cite{Wyd07}. We will refer the reader to their table 1 for 
the full details. Here, we list a few essential parameters and the minor modifications we employed: $14.0 < r < 17.6$, $0.03 < z < 0.25$, 
$\sigma_r < 0.2$, $z_{\rm conf} > 0.67$, \galex exposure time $t > 750$s and $16.0 < NUV < 23.0$.\footnote{Our faint-end limit of $23.0$ is 
about $0.5$ mag fainter than those employed by \cite{Wyd07}.} All magnitudes are AB magnitudes and corrected for Galactic foreground extinction.

\subsection{SDSS Large-Scale Structure Sample} 

Considerable effort has been invested by the SDSS team to prepare the redshift data for large-scale structure studies. This secondary data-set known 
as the New York University Value Added Catalog (NYU-VAGC) is documented in \cite{Bla05a} and available from the NYU website. Proprietary versions of 
this catalogue have been used by many groups within the SDSS collaboration for various investigations of clustering and luminosity function of galaxies. 
We match the \galex-SDSS catalog constructed above with the SDSS large-scale structure (LSS) DR5 sample. The version used for our analysis includes 
all of the detailed radial and angular selection functions for the various statistical subsamples used in previous 
analyses \citep[e.g.][]{Zeh05}. 
This sample is ideal because we can compare our result with that of  \cite{Zeh05} which was based solely 
on selection from optical criteria.

\subsection{\galex--SDSS Overlapping Footprint} 

In order to statistically define our combined \galex--SDSS survey, we need to have the understanding of the angular sampling function of the two surveys, 
which varies across different regions of the sky. \galex's Medium Imaging Survey (MIS) survey consists of overlapping circular tiles (radius $0.55^\circ$), while the SDSS 
spectroscopic survey is a combination of circular spectroscopic plates, but with fiber placement based on a rectangular imaging survey that runs along 
great circles. To combine the footprints of both surveys, we use the GESTALT footprint server. 
GESTALT \footnote{http://nvogre.phyast.pitt.edu:8080/gestalt\_tutorial/ } uses a hierarchical pixelization system, enabling one to encode 
observations of arbitrary geometry while tagging information about completeness and the masking of artifacts. We first encode the \galex MIS survey 
using GESTALT. We then obtain the detailed observational footprint of the SDSS Large-scale structure sample from the NYU-VAGC website. The SDSS footprint 
is expressed as a set of disjoint polygons using the software MANGLE \citep{Ham02} which takes into account the complex angular mask and geometry of the 
SDSS survey. We convert these polygons into the pixelization scheme of GESTALT and consider the intersection of the two surveys. There are approximately 
490 square deg in the \galex--SDSS overlapping footprints after masking for holes, bright stars and satellite trails, and excluding defects.

In order to measure the correlation function, a random sample needs to be constructed to normalize the galaxy pair counts. 
Angular sampling completeness as a function of position in the sky encoded in the footprint server is used to generate 
random samples of density roughly 50 times the galaxy density. We adopt the method proposed by \cite{Li06} where we assign 
each galaxy in our sample to a random position on the sky but keep all other attributes the same (e.g. redshift, color, magnitude). 
This random sample by construction has the same redshift distribution as the original sample, and thus does not smooth out the 
redshift structure like those generated via the luminosity function. As noted in \cite{Li06}, this approach works well in surveys 
with a wide-angular sky coverage (e.g. much larger than the typical large-scale structure), and with small variation in survey depth. 
We note here that our random sample would inherit the redshift correlation function of the color-magnitude distribution of the parent galaxy 
sample, only spatial distribution has been randomized, hence any excess in clustering must be due to positional differences. 

In the SDSS spectroscopic survey, no two galaxies with separation $\theta$ less than $55\arcsec$ can both be assigned spectroscopic 
fibers for observation on any given observing plate. Hence, a large fraction of galaxy pairs with $\theta < 55\arcsec$ are missing. 
We correct for this ``fiber-collision'' problem by using the observed angular correlation function, a method first suggested \cite{Li06}, 
and described in detail in the companion paper by \cite{Hei08}. In brief, the observed projected two-point angular correlation 
function for both the photometric sample $w_{\rm ph}(\theta)$ and the spectroscopic sample $w_{\rm sp}(\theta)$ are measured and used 
to construct the pair weighting ratio:
\beq
F(\theta) = \frac{w_{\rm ph}(\theta) + 1}{w_{\rm sp}(\theta) + 1}
\eeq
as a function of separation. Empirically, \cite{Hei08} finds $F(\theta) \sim 3$ for $\theta < 55\arcsec$ and zero otherwise for the all 
\galex--SDSS cross-matched galaxies. We apply this correction to each of the red, green and blue subsamples equally.    

\section{Methodology} \label{sec:methods}
\subsection{The Two-point Correlation Function} 
In brief, the two-point auto-correlation function $\xi(r)$ measures the excess probability of finding a galaxy pair with 
separation $r$ from a random galaxy distribution, 
\beq dP = n[1 + \xi(r)]dV \eeq 
where $n$ is the mean number density of galaxy sample. For the last forty years, the correlation function has served as the primary method for cosmologists to 
quantify the clustering properties of galaxies from large scale surveys \citep{Tot69,Pee80}. If the underlying density distribution is Gaussian, 
then the correlation 
function fully describes all statistical properties of a given distribution. Since early cosmological models are often based on primordial Gaussian dark 
matter density fields, the use of the correlation function leads to a straightforward comparison between empirical studies on the statistical distribution of 
galaxies with such models. 
Recently, with the widespread adoption of a standard $\Lambda$-dominated cosmology \citep{Ost95,Rie98,Per99,Spe07}, 
the correlation function is 
used instead to probe the growth of structure in the universe and the range of formation scenarios for varying kinds of galaxies. \footnote{The use of
correlation function to probe the growth of structure predate the advent of $\Lambda$-CDM, especially in the study of faint blue galaxies \citep[e.g. ][]{Efs01}. }
  The correlation length -- the 
amplitude from a power-law correlation function -- provides information on the mass of dark-matter halos in which the various galaxies reside, 
linking observation with theoretical description of structure formation \citep{Bow06,Cro06}

For a galaxy survey with a well-defined angular selection function, $\xi$ can be estimated using an optimal estimator like the \cite{Lan93} estimator: 
\beq 
\widehat{\xi}_{LS} = \frac{DD-2DR+RR}{RR} 
\eeq
where $DD$, $DR$ and $RR$ are normalized counts of galaxy pairs in the data-data, data-random and random-random 
catalogs.\footnote{Normalized counts of galaxy pairs are weighted by the selection function of the galaxies involved.} 
The \cite{Lan93} estimator is preferred because it is relatively insensitive to the size of the random catalog and to edge corrections \citep{Ker00}.

Because we observe galaxies in redshift and projected space and not in physical space, in practice, the correlation function is first measured on a 
two-dimensional grid of separations: $\pi$ along the line of sight (redshift space) and $r_p$ for angular separation on the sky. In addition to providing 
information about the underlying mass distribution through the amplitude of the clustering signals, the two-dimensional correlation function $\xi(\pi,r_p)$, 
contains additional information about the dynamics of the galaxies \citep{Pee80}. At small projected separations $r_p$, random motions within virialized
 over-density (e.g. clusters of galaxies) causes an elongation along the line-of-sight ($\pi$ direction) known as the ``finger-of-God'' effect. On 
large scales, coherent streaming of galaxies into potential wells causes an apparent compression of structure along the line-of-sight \citep{Sar77,Kai87,Ham92}. 
Various studies have used $\xi(\pi,r_p)$ to extract cosmological and dynamical information \citep[e.g.,][]{Tin07}.

\subsubsection{Real-Space Correlation}

Because redshift space distortion only affects the line-of-sight component of $\xi(\pi,r_p)$, we can recover the true space correlation 
function $\xi(r)$ by following the standard procedure of computing the projected correlation function:
\beq 
w_p(r_p) = 2 \int_0^{\infty} d\pi \xi(r_p, \pi). 
\eeq
In practice, we integrate along the line-of-sight direction out to $\pi = 30 \hmpc$. This is large enough to include almost all correlated pairs 
but also stable enough to suppress noise from distant uncorrelated pairs. The projected correlation function can in turn be related to the real 
space correlation function $\xi(r)$,
\beq
w_p(r_p) = 2 \int_0^{\infty} rdr \xi(r)(r^2 - r_p)^{-1/2}.
\eeq
If the real-space correlation function follows a power-law $\xi(r) = (r/r_0)^\gamma$, we can infer its parameters: the correlation length $r_0$ and 
the power-law slope $\gamma$ from the best-fit power-law to $w_p(r_p)$ using the following deprojection: 
\beq
w_p(r_p) = r_p \left(\frac{r_p}{r_0}\right)^{-\gamma}\Gamma\left(\frac{1}{2}\right) \Gamma\left(\frac{\gamma-1}{2}\right)\,
          \Bigr/ \,\Gamma\left(\frac{\gamma}{2}\right).
\label{eq:powerlaw}
\eeq

\subsubsection{Cross-correlation}\label{sec:cross_meth}
Related to the auto-correlation function is the cross-correlation function between two classes of galaxies. The cross-correlation 
function $\xi_{1,2}(r)$, measures the clustering of one type of galaxy around another. $\xi_{1,2}(r)$ is essentially the probability of 
finding a galaxies of type 1 around a galaxy of type 2 as a function of separation $r$. For our analysis, we use the cross-correlation 
version of the classical \cite{Dav83} estimator:
\beq 
\widehat{\xi}_{1,2} = \frac{D_1D_2}{D_1R_2} - 1 
\eeq 
where $D_1D_2$ are the normalized counts of cross-pairs; while $D_1R_2$ are cross-pairs of type 1 with random galaxies having the same 
redshift distributions as galaxies of type 2.

Recently, the cross-correlation function has been used extensively in clustering studies of galaxy properties \citep{Zeh05,Wan07,Coi08,Pad08,Che09} 
since the auto-correlation function alone tells us little about how galaxies of various types relate to one another. Two populations of galaxies may 
both have comparable correlation strength, and yet be physically unrelated if they are spatially segregated. Specifically, the cross-correlation function 
may be used in conjunction with the auto-correlation functions of the galaxies to glean information as to how they mix statistically. For example, 
if the two populations are well mixed, i.e. they are consistent with being drawn stochastically from their respective auto-correlation function equally, 
their projected cross-correlation function $w_p^{1,2}(r_p)$ would 
follow that of the geometric mean of their auto-correlation functions
\beq
w_p^{\overline{1,2}}(r_p) = \sqrt{w_p^{1,1}(r_p)w_p^{2,2}(r_p)}.
\eeq 
On the other hand, if the populations were segregation between the populations and do not distribute evenly in all space, e.g. they sit on different halos, 
the amplitude of $w_p^{1,2}(r_p)$ would be lower than that of $w_p^{\overline{1,2}}(r_p)$ known in the literature as stochastic anti-bias \citep[e.g.][]{Bla00,Swa08}

\subsubsection{Bootstrap Errors}
We estimate the errors for the correlation measurements using a modified bootstrap \citep{Efr81} method for spatial statistics known as marked-point 
bootstrap \citep{Loh08}. 
We first estimate the contribution of each galaxy to the correlation function -- the marks: 
\beq
\xi_i(r) = \sum_{j=1,j\not=i}^{n} \phi\{|x_i-x_j| \in (r-dr,r+dr)\}
\eeq
such that
\beq
\xi(r) = \sum_{i=1}^{n} \xi_i
\eeq
where $n$ is the number of galaxies and $\phi$ is the chosen correlation estimator (e.g. $\widehat{\xi}_{LS}$). Hence, the mark $\xi_i$ associated 
with galaxy $x_i$ is the number of excess pairs at a distant $r$ from $x_i$. These $\xi_i$ are (roughly) independent, and identically distributed, and 
can be used to replicate $B$ bootstrap samples by the sampling with replacement. An estimate of the correlation function $\widehat{\xi_j^*}$ can be 
computed from each of the $j$-th bootstrap samples. We use $B = 999$ bootstrap samples for our analysis. 
The distribution of $\widehat{\xi^*}(r_i)$ can be use to estimate the error bars of the correlation function at each separation $r_i$. 
For an equivalent of one sigma errors, we ranked $\widehat{\xi^*}(r_i)$ and take the $159^{nd}$ and $840^{th}$ for the lower and upper error bounds.
Because the $\widehat{\xi}$ errors for the $r_i$ are correlated, when we estimate parameters for the two parameter power-law model (equation \ref{eq:powerlaw}),  
we refitted each bootstrap sample to obtain a two dimension distribution of the parameters. 

We chose this method over the conventional jackknife methods because of the fragmented nature of the \galex--SDSS footprint. Our bootstrap errors 
are consistent with the jackknife errors estimated by excluding one \galex tile at a time. We note here that jackknife errors are known to overestimate 
the variance on small-scale, and often bias the overall correlation estimates \citep{Nor09}. Since bootstrap is a form of internally estimated errors,
and internally estimated errors are known not to reproduce externally estimated errors accurately \citep{Nor09}, our clustering results should 
be treated with caution.

\begin{figure*}
\includegraphics[width=0.33\textwidth,height=0.33\textwidth]{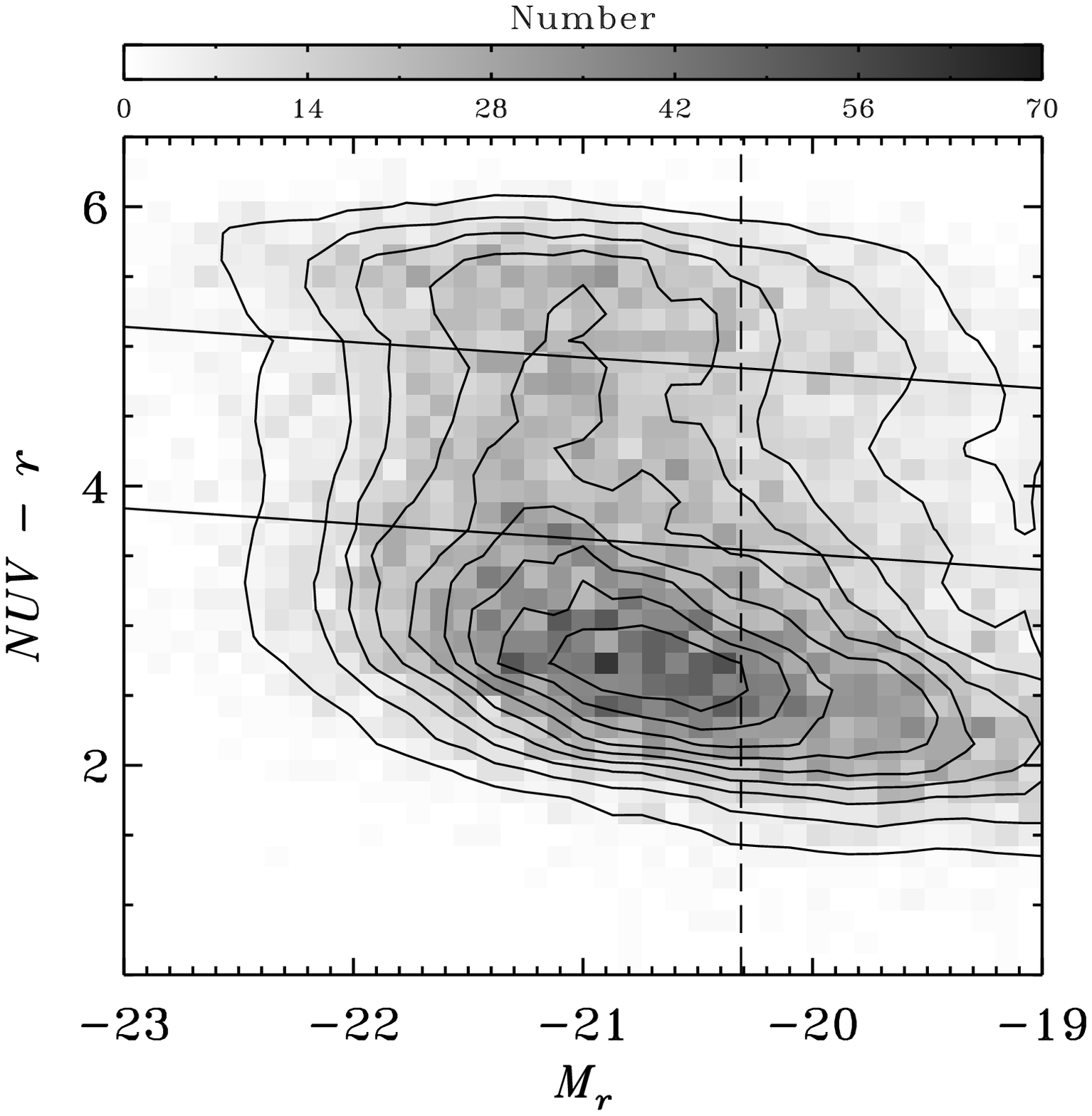}
\includegraphics[width=0.33\textwidth,height=0.33\textwidth]{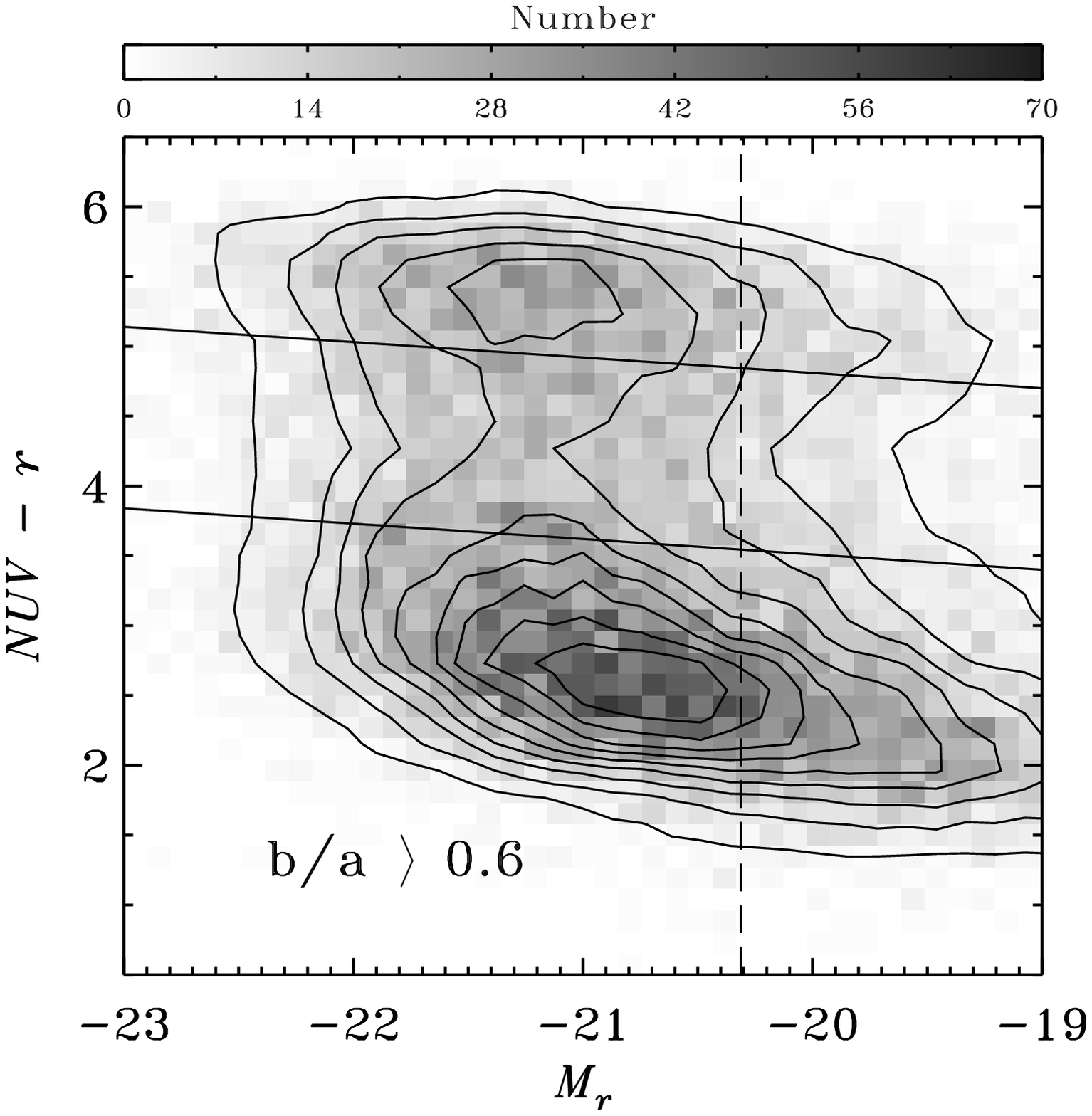}
\includegraphics[width=0.33\textwidth,height=0.33\textwidth]{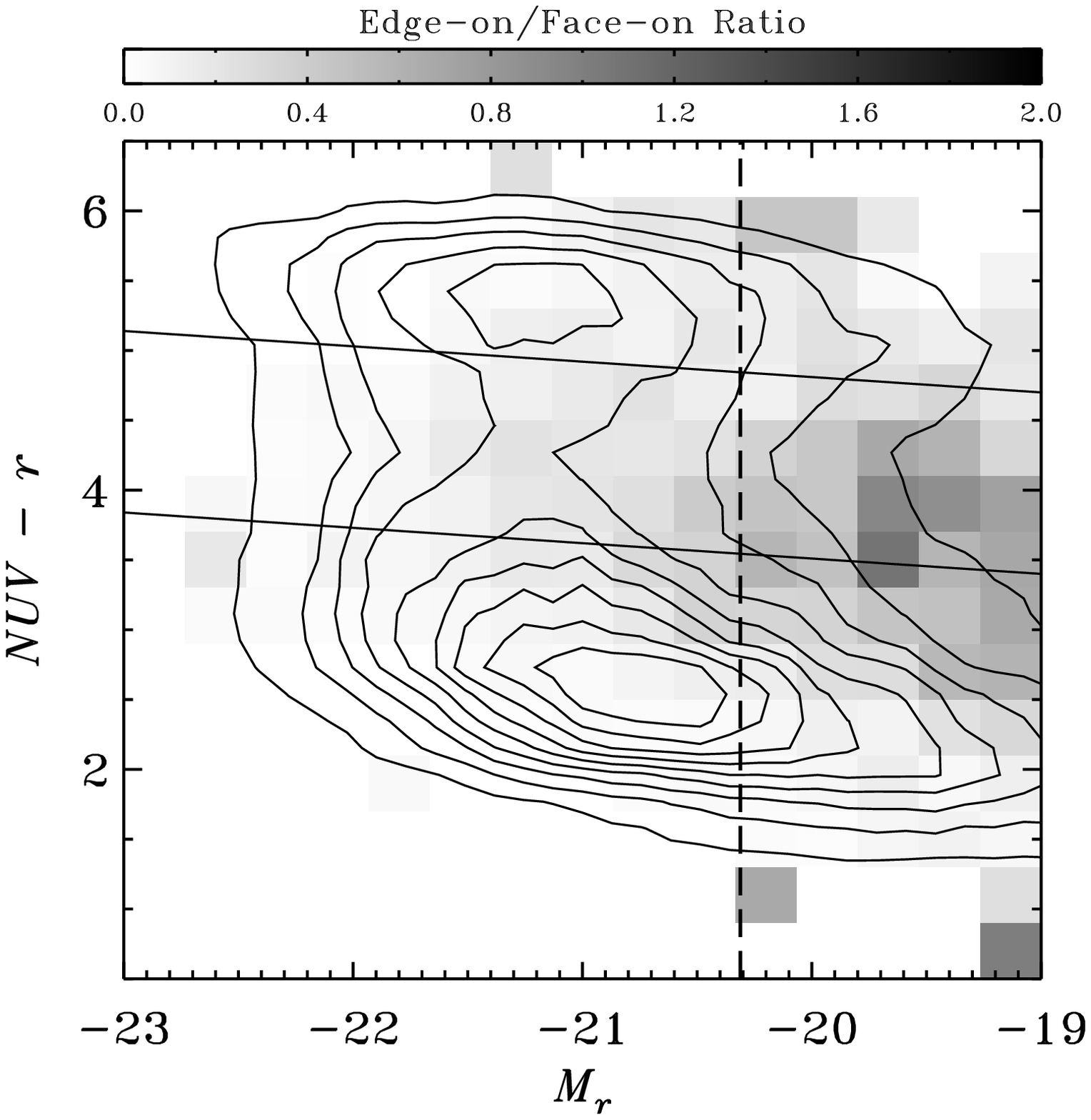}
\caption{$NUV-r$ vs $M_r$ color-magnitude diagram (CMD) of galaxies in the local Universe.
The left panel shows the full distribution. The middle panel shows the distribution drawn 
from face-on galaxies with $b/a > 0.6$, used to correct for the effect of dust 
on the CMD. The right panel shows the distribution of edge-on/face-on ratio overlay onto the CMD contours.  
The green valley is dominated by region of the CMD with high relative fraction of edge-on galaxies. 
By limiting ourselve to galaxies with $M_r < -20.3$ (dashed vertical lines) for the volume-limited sample, 
we also eliminate region of the CMD with the highest fraction of edge-on galaxies.
\label{fig:f1}}
\end{figure*}

\section{Dependence of Clustering on $UV$--Optical Color}\label{sec:results} 
Figure \ref{fig:f1} shows the $NUV - r$ color-magnitude diagram (CMD) in the local universe. All absolute magnitudes are $k$-corrected using the 
K\_CORRECT program by \cite{Bla07} and are quoted in units with Hubble constant $H_{\rm 0} = 100 h \kmsmpc$ , i.e. $M_r \equiv M_r - 5 \log (h = 1)$. 
Compared with the optical color-magnitude diagram (e.g. $g-r$ vs $M_r$), the $NUV-r$ color axis has a higher dispersive power. The scatter plot shows 
two clear sequences: red and blue, and an intermediate ``valley'' of ``green'' population. Conventional optical diagrams only display a single red 
sequence with an extended blue cloud. Hence, diagnostic diagrams like Figure \ref{fig:f1} separate galaxies into three natural groupings: (1) a red 
sequence of bulge-dominated galaxies with old stellar systems, (2) a blue sequence of star-forming systems consisting of mainly late-type galaxies, and 
(3) galaxies in a ``green valley'' that in principle might exhibit transitional properties \citep{Mar07,Sch07}. Detailed luminosity functions and 
physical properties analysis of these galaxies derived from this $UV$--optical CMD has been reported elsewhere \citep{Wyd07,Mar07,Sal07,Sch07}. Here, 
we investigate the clustering properties of these three subpopulations of galaxies, separated by the two tilted horizontal lines, using the two-point 
auto-correlation function, as well as their relation to each other using the cross-correlation function. Our results are not sensitive to $0.25$ mag 
shift in the oblique equation used to classify the galaxies.

\subsection{Dust}\label{sec:dust}
Dust content within each galaxy can modify their broadband colors, hence the resultant color-magnitude diagram, as galaxies move from one group to 
another \citep{Mar07,Sch07}.  This is a concern, as it might affect our estimate of the correlation function by either increasing the scatter between 
groups, or bias the result in a systematic but unknown manner. Indeed, a fraction of the green valley galaxies are dusty star forming galaxies whose 
intrinsic color would have placed them on the blue sequence in the absence of dust \citep{Wyd07}. However, the available procedures for dust correction 
are highly uncertain. They give inconsistent results depending on whether one uses a primarily photometric approach \citep{Sal07,Joh07}, or one based on 
spectroscopic indices \citep{Kau04}. The former, for example, has the side effect of reducing the red sequence number counts 
substantially \citep{Sch07,Hei08}, while the latter approach requires spectroscopy of modest S/N, which is not available for part of our sample. 
A comprehensive analysis of the clustering as a function of star-formation history with a dust-corrected color-magnitude diagram using the method 
of \cite{Joh07} is done in \cite{Hei08}. 

Here, we adopt a geometric approach. Because dust lanes in galaxies usually appear when viewed edge-on, we can 
reduce the effect of dust on galaxy color by restricting our analysis to galaxies that are primarily face-on.  The middle panel of Figure \ref{fig:f1} 
shows the ``dust-corrected'' color-magnitude diagram obtained by including only galaxies with isophotal (minor-over-major) axis ratio $b/a > 0.6$. To 
the extent that each of the subpopulations have the same intrinsic distribution of $b/a$, removing edge-on galaxies will give a correct mixture of 
galaxies that mimic the intrinsic color-magnitude distribution. Comparing the dust-corrected CMD with the CMD on the left panel (hereafter the full 
distribution), the density of green galaxies around $M_r \sim -21$ is reduced substantially, with the contours showing a more pronounced ``valley'' 
separating the sequence of galaxies. 

The righthand panel of Figure \ref{fig:f1} shows the median ratio of edge-on ($b/a < 0.4$) to face-on ($b/a > 0.8$) as 
a function of color and magnitude, with the same density contours from the middle panel. The region with the highest edge-on/face-on ratio falls in
the green valley and towards the faint end of the luminosity distribution \citep{Cho07,Mar07,Sch07}. For the galaxies with the magnitude range 
with $-23.5 < M_r < -19.0$ used to construct our flux-limited sample, the median $b/a$ for red and blue sequence galaxies are $0.73$ and $0.72$ 
respectively, while for green valley galaxies it is $0.66$. \cite{Cho07} find that edge-on galaxies are also statistically fainter due to the internal 
extinction, and it affects morphologically late-type galaxies (classified by eye) more than early-type galaxies. Hence by limiting our analysis to
face-on galaxies, we would reduce biases associated with differential dimming in addition to the biases from color shifts. Following \cite{Cho07}, 
we restrict our analysis to galaxies with $b/a > 0.6$ as a proxy for a galaxy distribution derived from a dust-corrected CMD. 

\begin{figure*}  
\includegraphics[width=0.44\textwidth,height=0.44\textwidth]{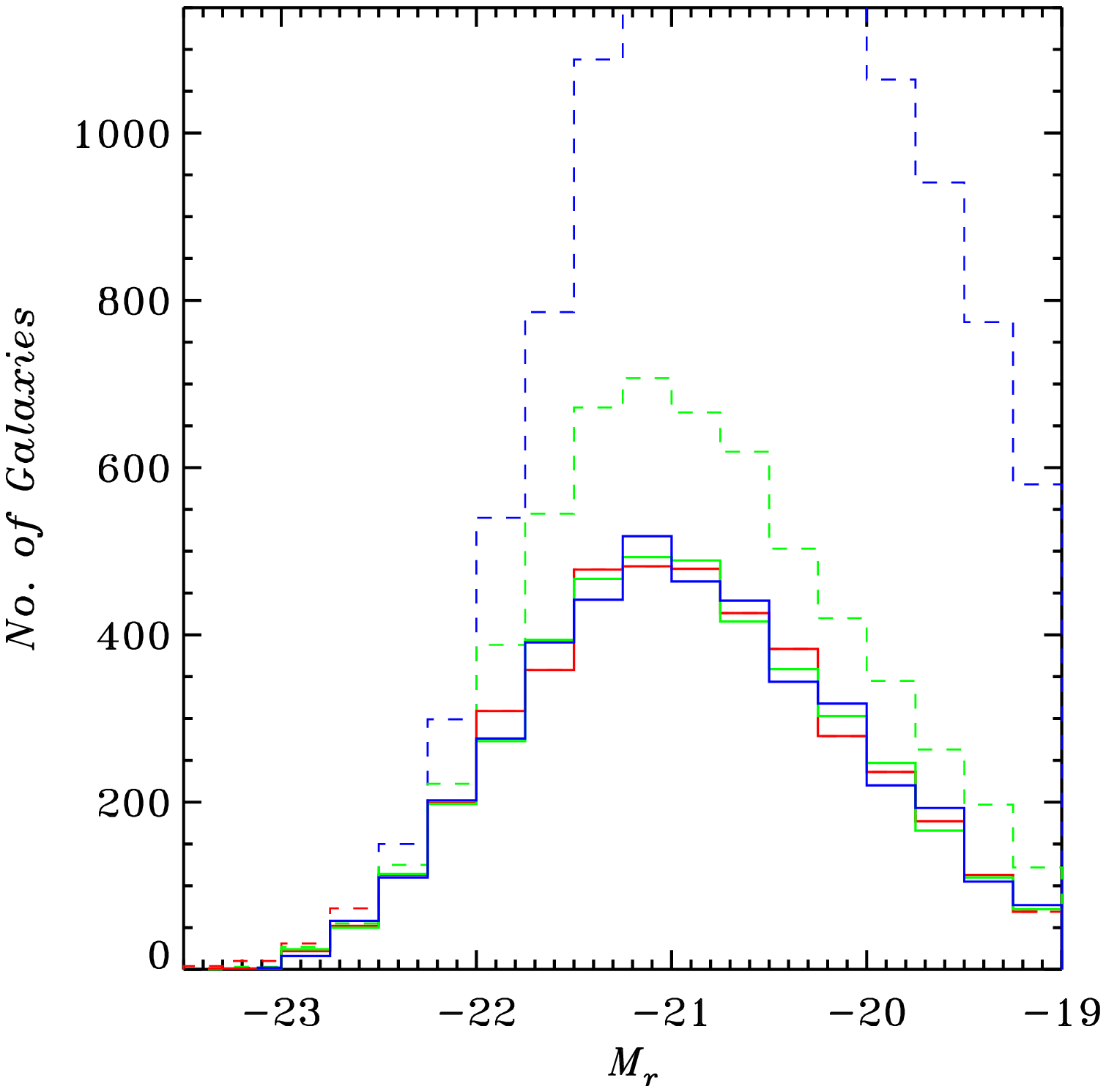}
\includegraphics[width=0.44\textwidth,height=0.44\textwidth]{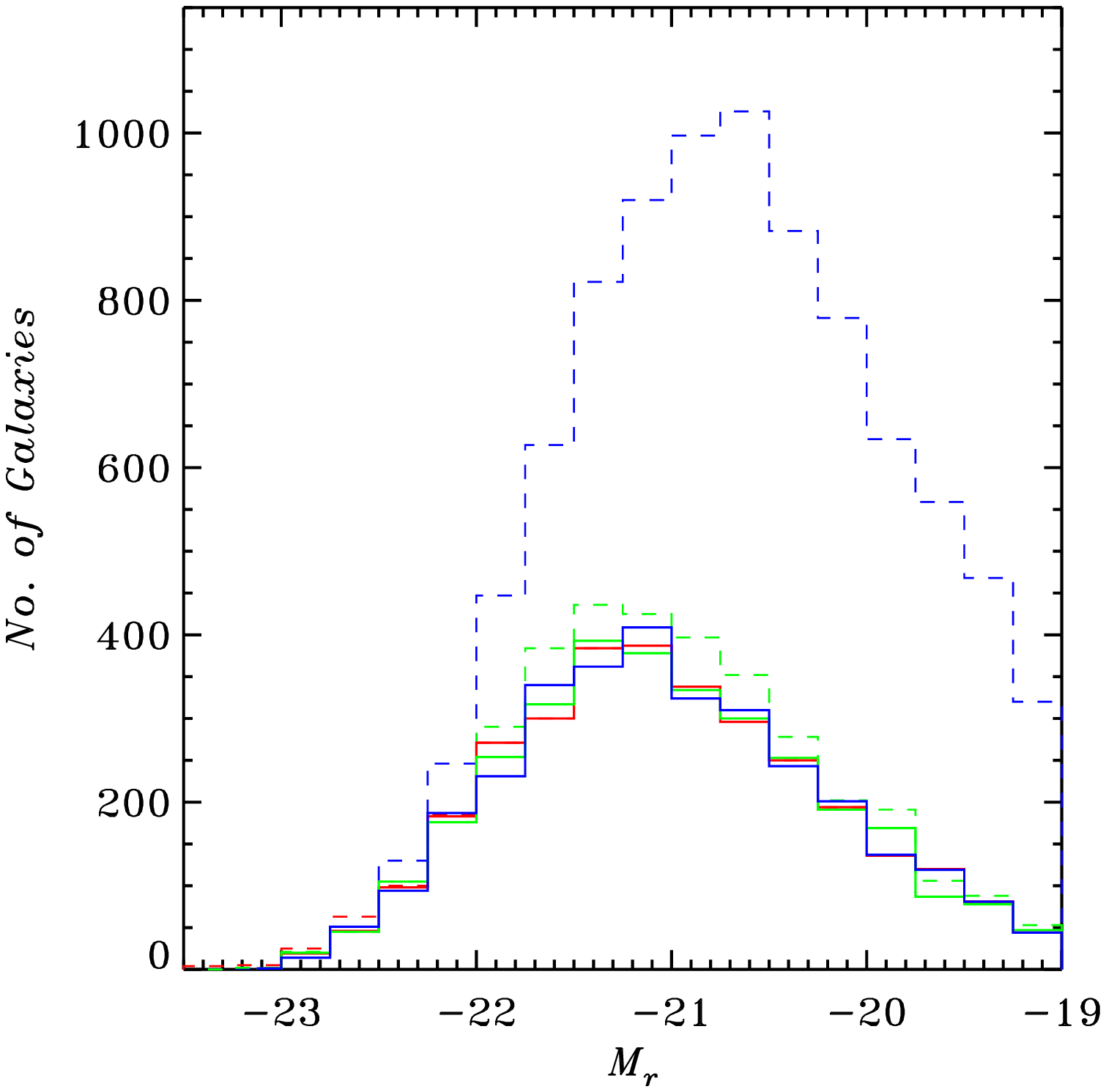}
\includegraphics[width=0.44\textwidth,height=0.44\textwidth]{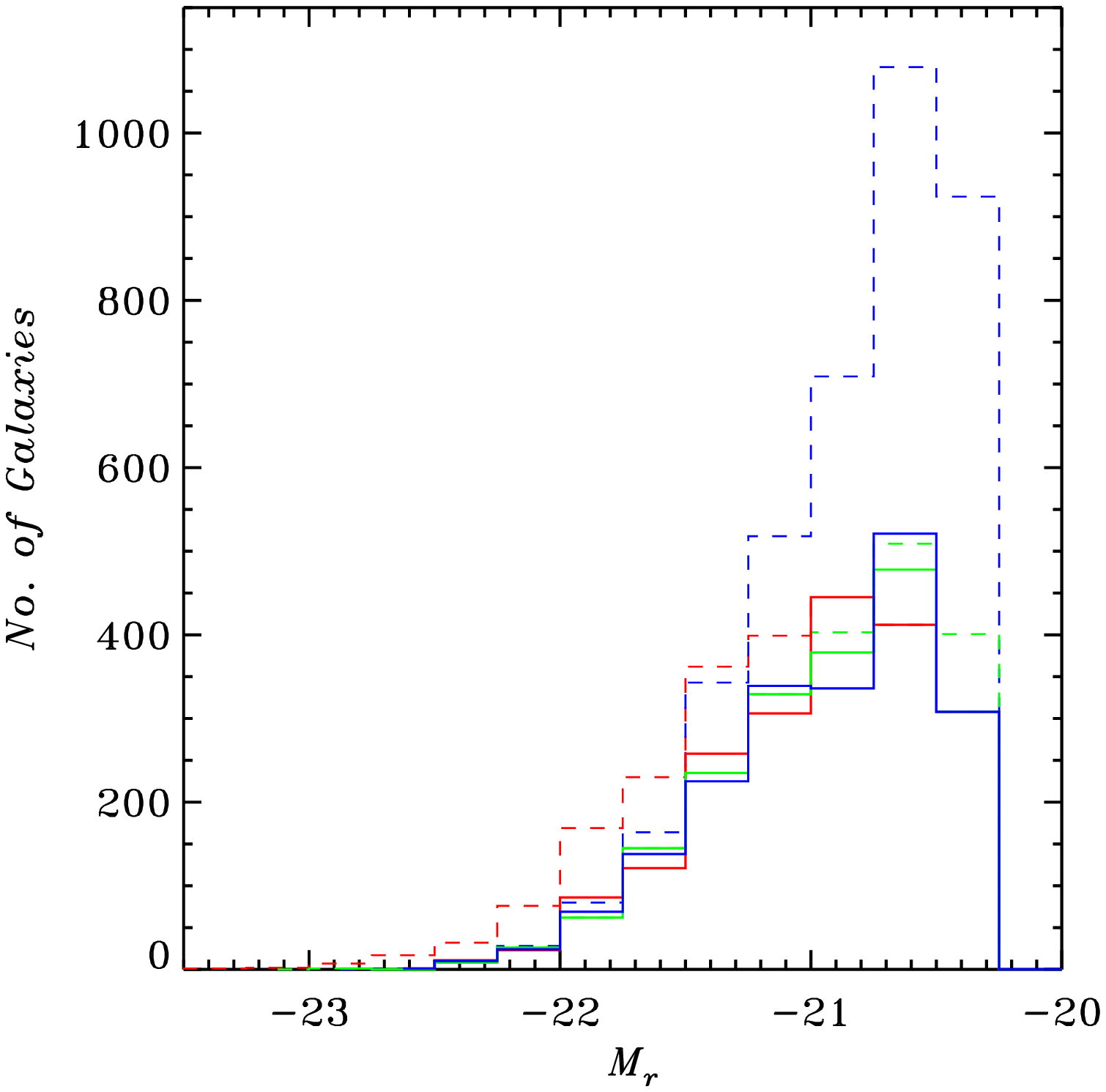}
\includegraphics[width=0.44\textwidth,height=0.44\textwidth]{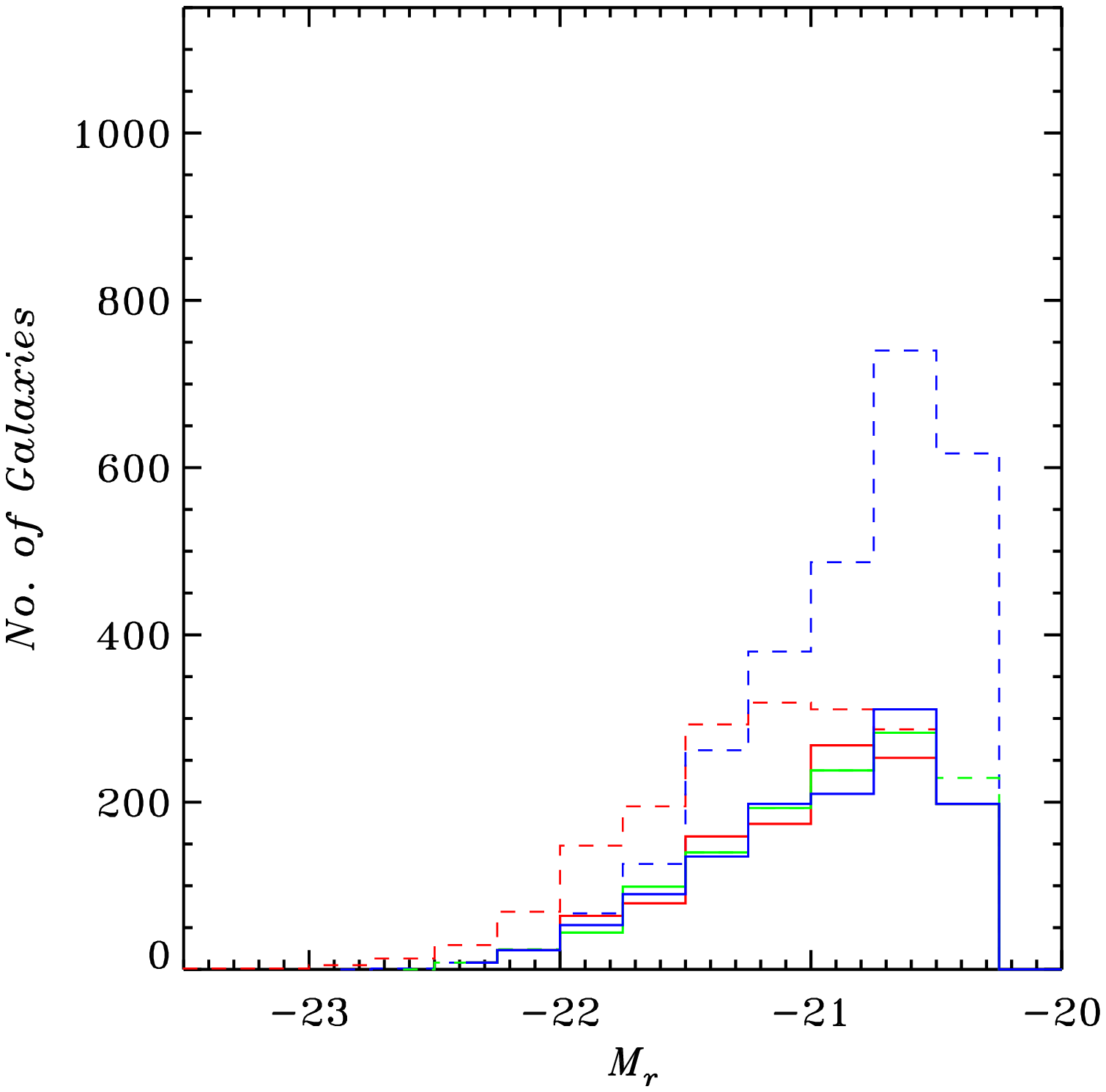}
\caption{The luminosity distribution of our flux-limited (top row) and volume-limited (bottom row) samples. 
The two panels on the left are for the full distribution, the right is for dust-corrected CMD, using only galaxies with $b/a > 0.6$. 
The dashed histograms show the original luminosity distribution for each of the subsamples
of red, green and blue galaxies. We resampled the luminosity to match the number counts of the minimal of 
the three distribution at each magnitude bin, shown as solid histograms. 
Even for the volume-limited samples, the variations in the original luminosity distribution among the three subsamples are substantial. 
Blue galaxies are heavily weighted towards the less luminous end compared to red or green galaxies. 
\label{fig:f2}}
\end{figure*}

\subsection{Luminosity bias}\label{sec:lum_bias}
The clustering of galaxies is luminosity dependent. \cite{Nor02}, \cite{Zeh05} and \cite{Li06} show that the amplitude of the projected correlation 
function $w_p$ increases monotonically as a function of luminosity (or stellar mass) on scales ranging from $0.2 \hmpc$ to $10 \hmpc$. In order to 
study the color 
dependence on clustering, or to compare the clustering intrinsic to the membership of discrete subpopulations, it is vital to remove this known luminosity 
dependence. The top two panels of Figure \ref{fig:f2} shows the luminosity distribution of the red, green and blue population of galaxies with 
$M_r < -19$ from our flux-limited sample. The series of dashed histograms (in red, green and blue) show the original luminosity distributions. Blue 
galaxies are more numerous and less luminous, on average, compare to red and green galaxies, reflecting a steeper blue luminosity function at the 
faint-end \citep{Wyd07}. To remove this luminosity dependence, we resample the luminosity histograms to match the number counts of the smallest of 
the three distributions at each magnitude bin. The result of this resampling is shown by the solid histograms. The left panel shows histograms from the 
full distribution, while the right shows those from a dust-corrected distribution. There are 4177 (full) and 3148 (dust-corrected) galaxies in each of 
the resampled luminosity distribution of red, green and blue galaxies, with a common median luminosity of $M_r \sim -21.0$, about half a magnitude 
brighter than $M^*$, the typical luminosity \citep{Bla03}. Note that if the fundamental attribute that drive clustering is stellar mass, removing the
luminosity dependence like what we have done here would still leave residual clustering due to the difference in mass-to-light ratio of the respective color
selected sample. 

\subsection{Volume-limited sample}\label{sec:vol_lim}
While the resampled red, green and blue galaxies are matched in luminosity, they are not matched in volume and have different redshift 
distributions. This makes the interpretation of the physical correlation function problematic. To this end, we select the largest possible volume 
within our catalog, using a redshift cut of $0.03 < z < 0.12$ and an additional luminosity cut at $M_r < -20.3$ to construct a volume-limited 
sample.\footnote{This is not strictly volume-limited since $NUV$ is not complete at these redshifts. An additional weighting is applied using the
luminosity function of \cite{Wyd07}.} The luminosity distributions of this sample are shown on the bottom row of Figure \ref{fig:f2}. As 
before, the left panel is for the full CMD, while the right panel are for the dust-corrected (restricted) CMD. Similar to the flux-limited case, 
the (original) blue dashed histograms are substantially different the red and the green, and are weighted more heavily towards the faint-end.  
We resample the histograms to match in luminosity. Note that in the case, the procedure outlined in section \ref{sec:lum_bias} essentially amounts to looking
for an optimal function $W(L)$ such that $W(L)\Phi(L)$ gives the minimum luminosity function of the three subsamples.
The new red, green and blue volume-limited subsamples each consists of 1971 (full) and 1226 
(dust-corrected) galaxies with a common median luminosity of $M_r \sim 20.9$, almost identical to the flux-limited case.

\subsection{Red, Green, Blue Auto-correlation}\label{sec:auto} 
\begin{figure*}  
\includegraphics[width=0.33\textwidth,height=0.33\textwidth]{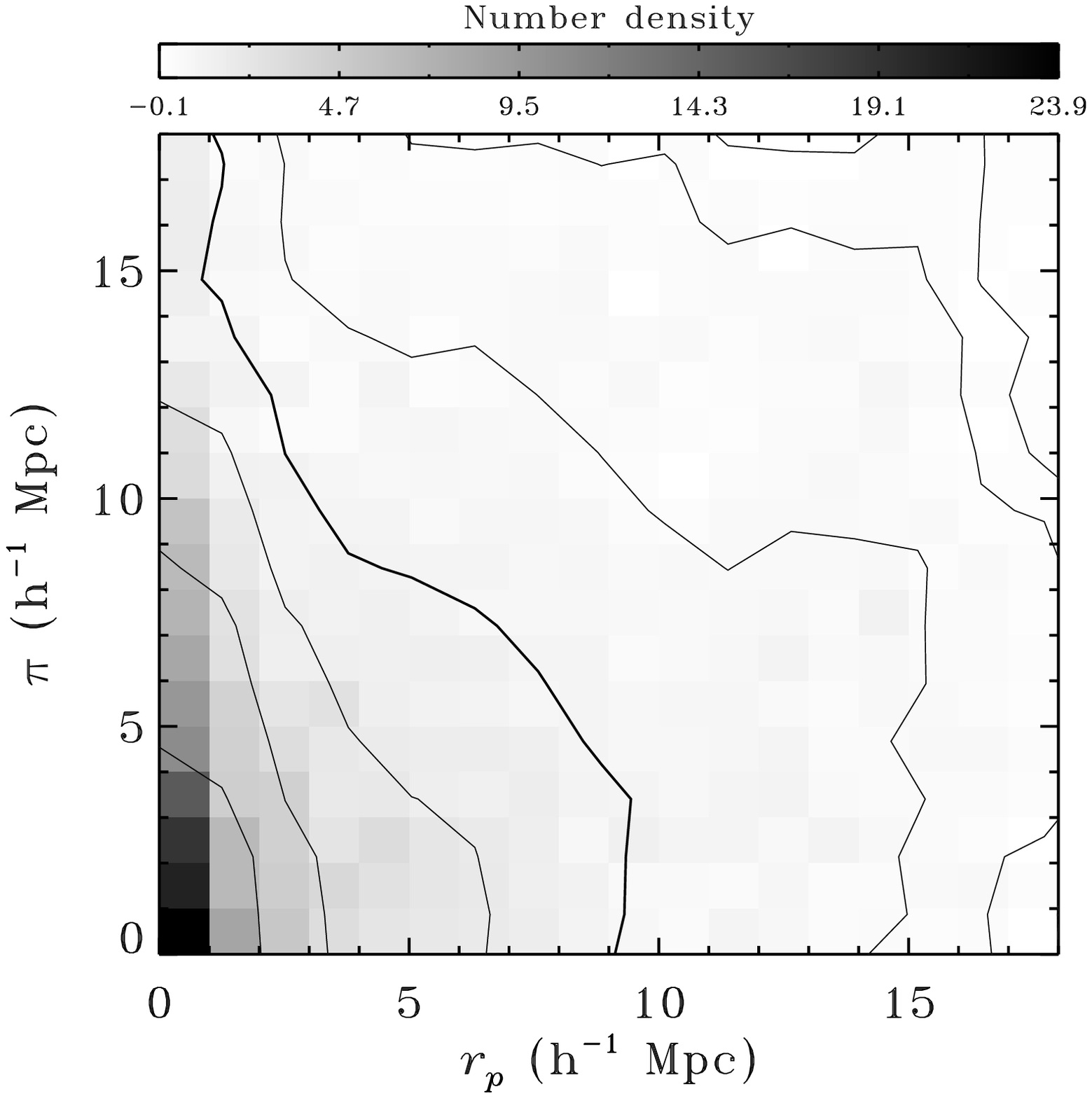}
\includegraphics[width=0.33\textwidth,height=0.33\textwidth]{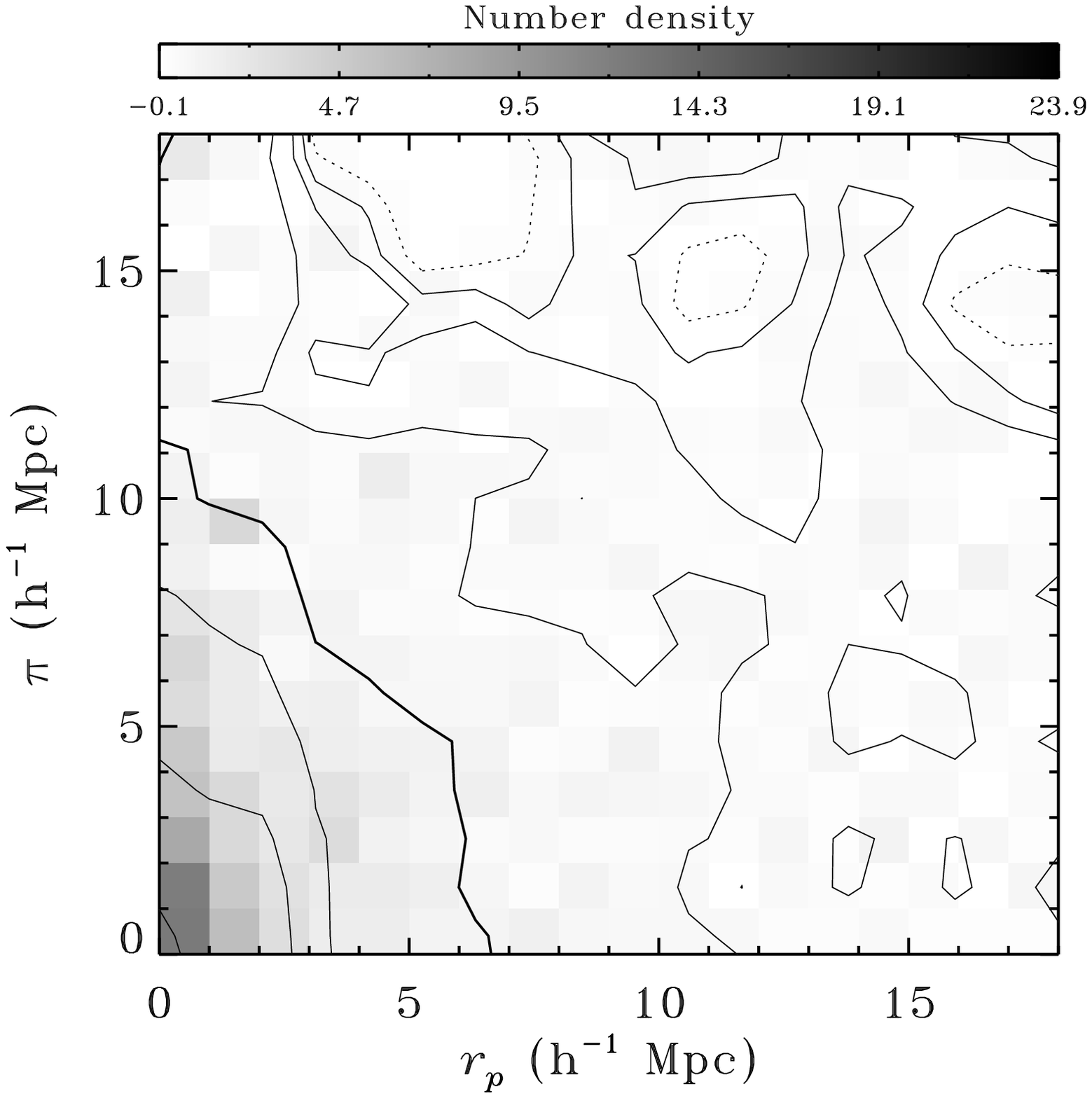}
\includegraphics[width=0.33\textwidth,height=0.33\textwidth]{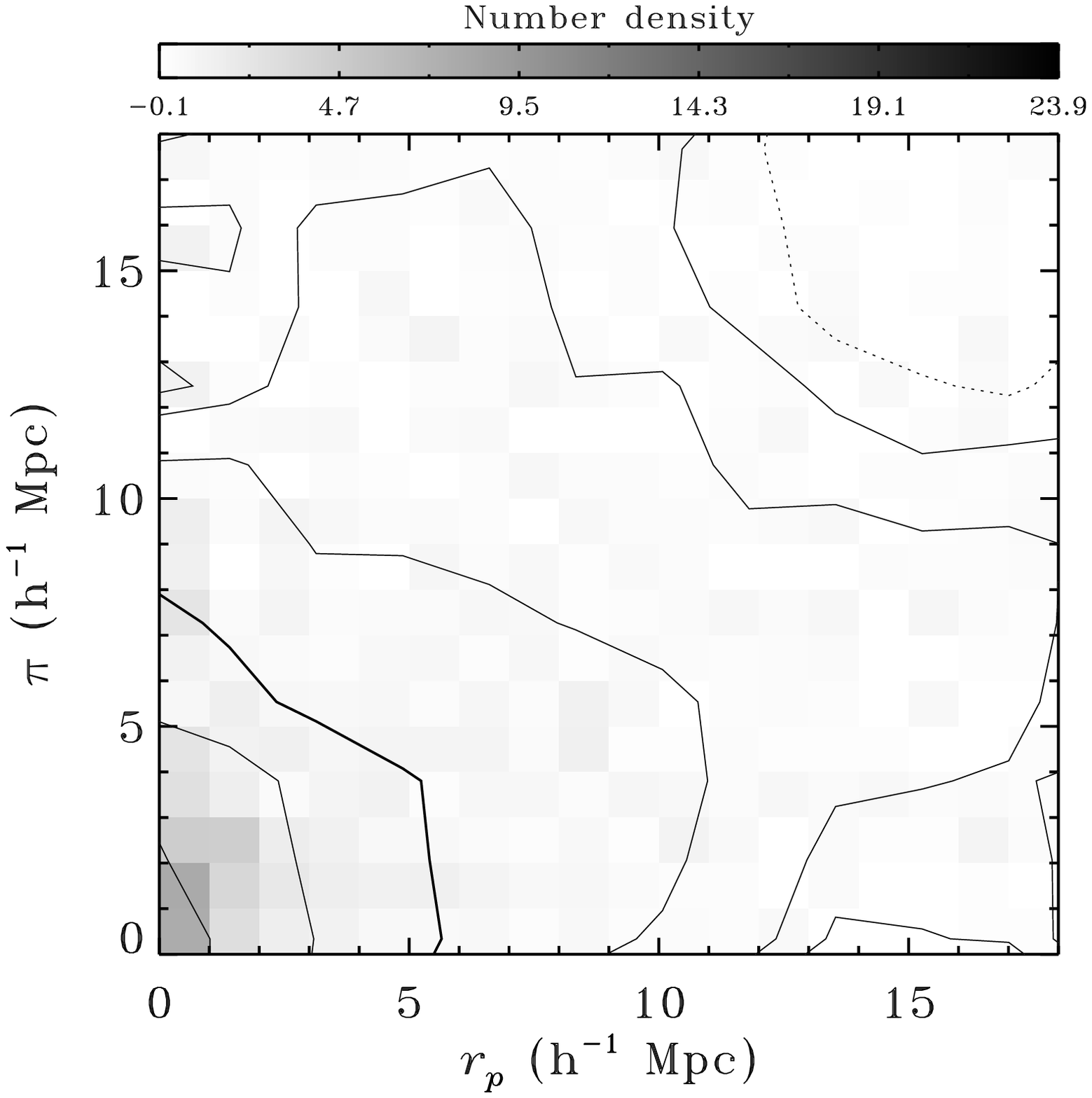}

\includegraphics[width=0.33\textwidth,height=0.33\textwidth]{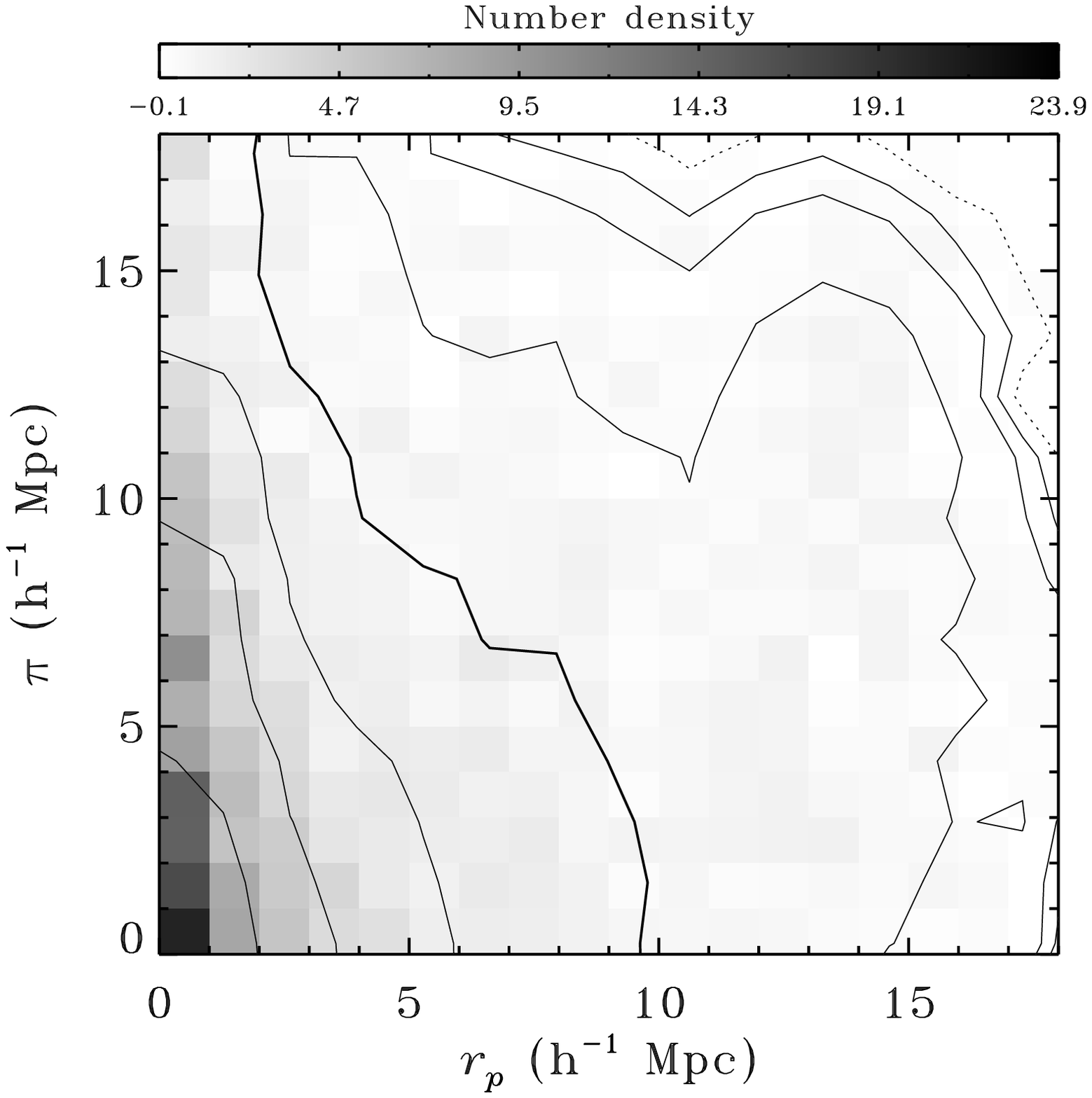}
\includegraphics[width=0.33\textwidth,height=0.33\textwidth]{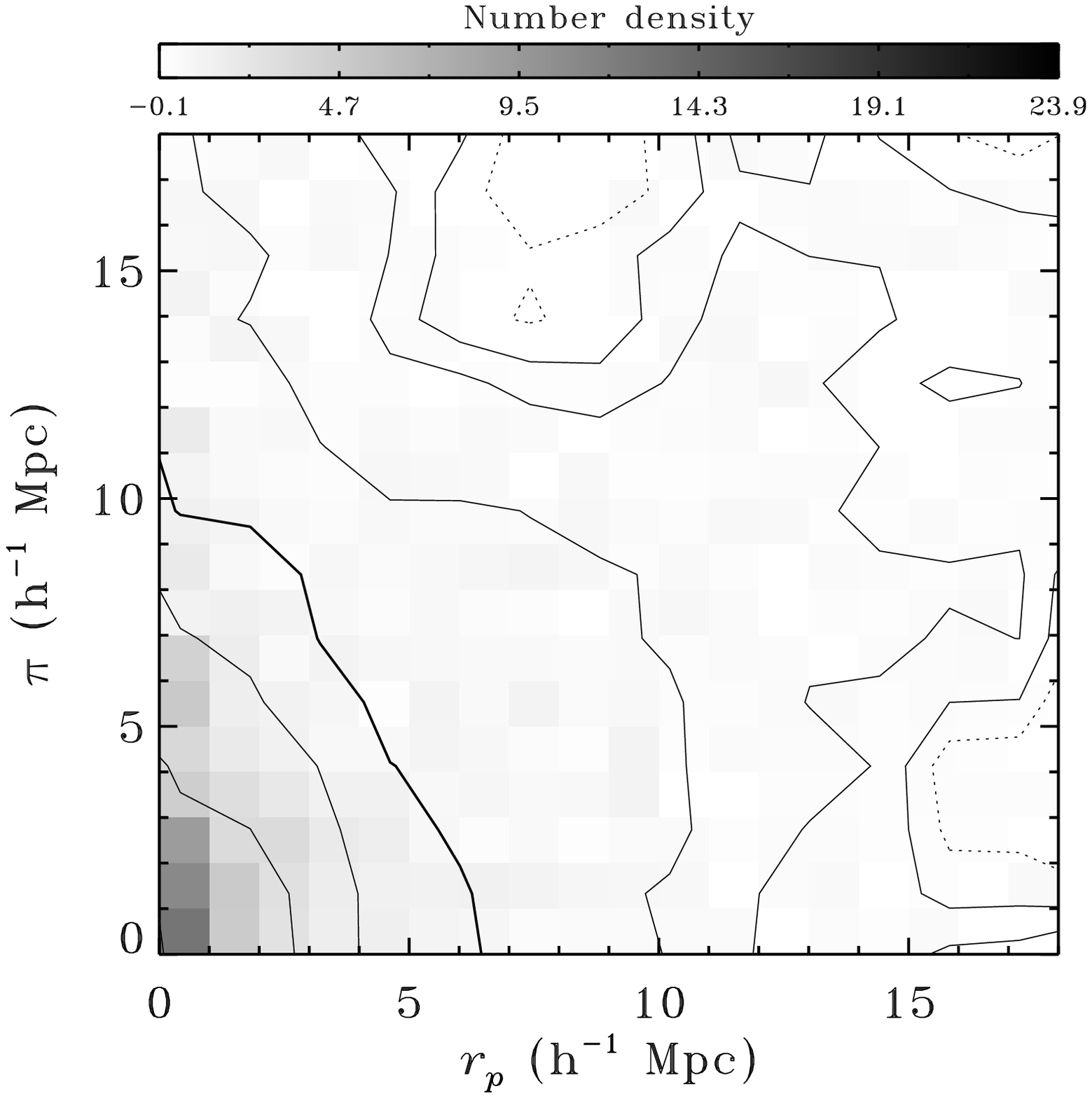}
\includegraphics[width=0.33\textwidth,height=0.33\textwidth]{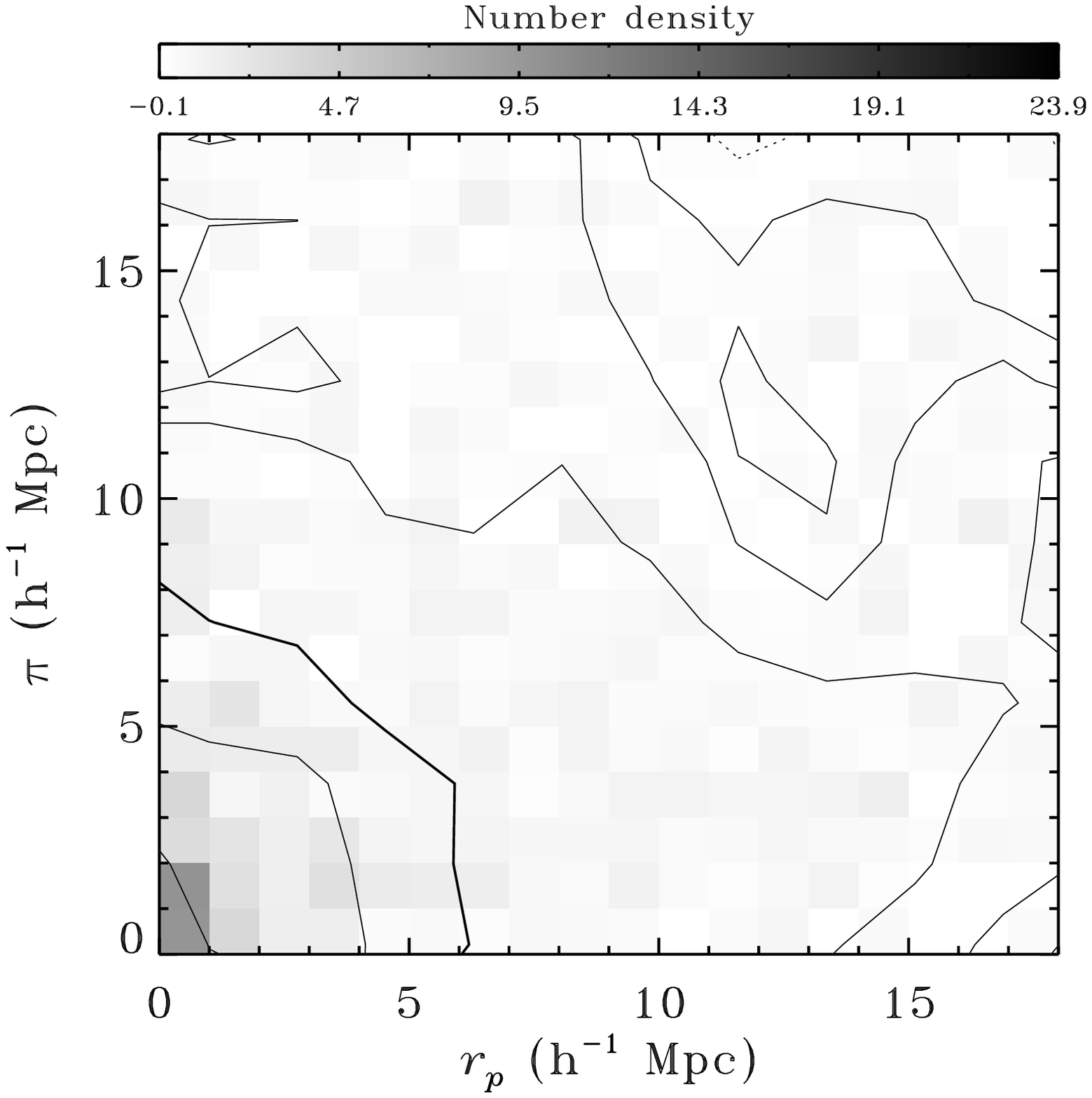}
\caption{Contours and normalized counts of the two-dimensional correlation function $\xi(\pi,r_p)$ for
red (left), green (middle) and blue (left) populations for the flux-limited sample. The panels on the top 
row are for the full distribution, bottom row are restricted to 
face-on galaxies ($b/a > 0.6$). The contours obtained after 2 x 2 $\hmpc$ boxcar smoothing.
The levels are 0.0 (dotted lines), 0.25, 0.5, 1.0 (heavy lines), 2.0, 4.0, 8.0 and 16.0.
Red galaxies have the strongest finger of God effect (extension in the $\pi$ direction).  
The finger-of-God effect is seen in the green sample as well, while it is not present in the blue.   
Galaxies in the green valley appear to have clustering characteristics of the red sequence.
\label{fig:f3}}
\end{figure*}
\begin{figure*}  
\includegraphics[width=0.33\textwidth,height=0.33\textwidth]{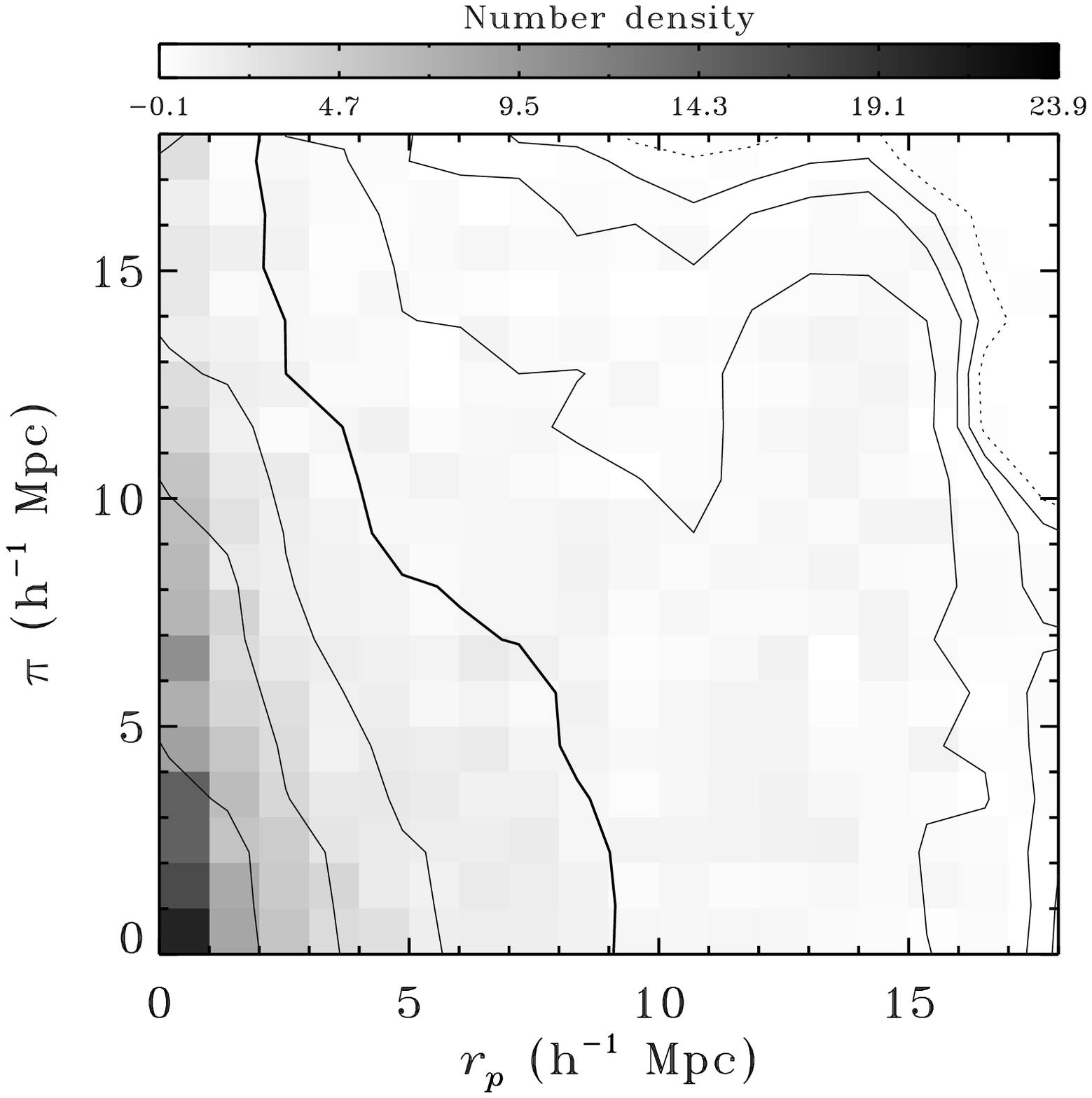}
\includegraphics[width=0.33\textwidth,height=0.33\textwidth]{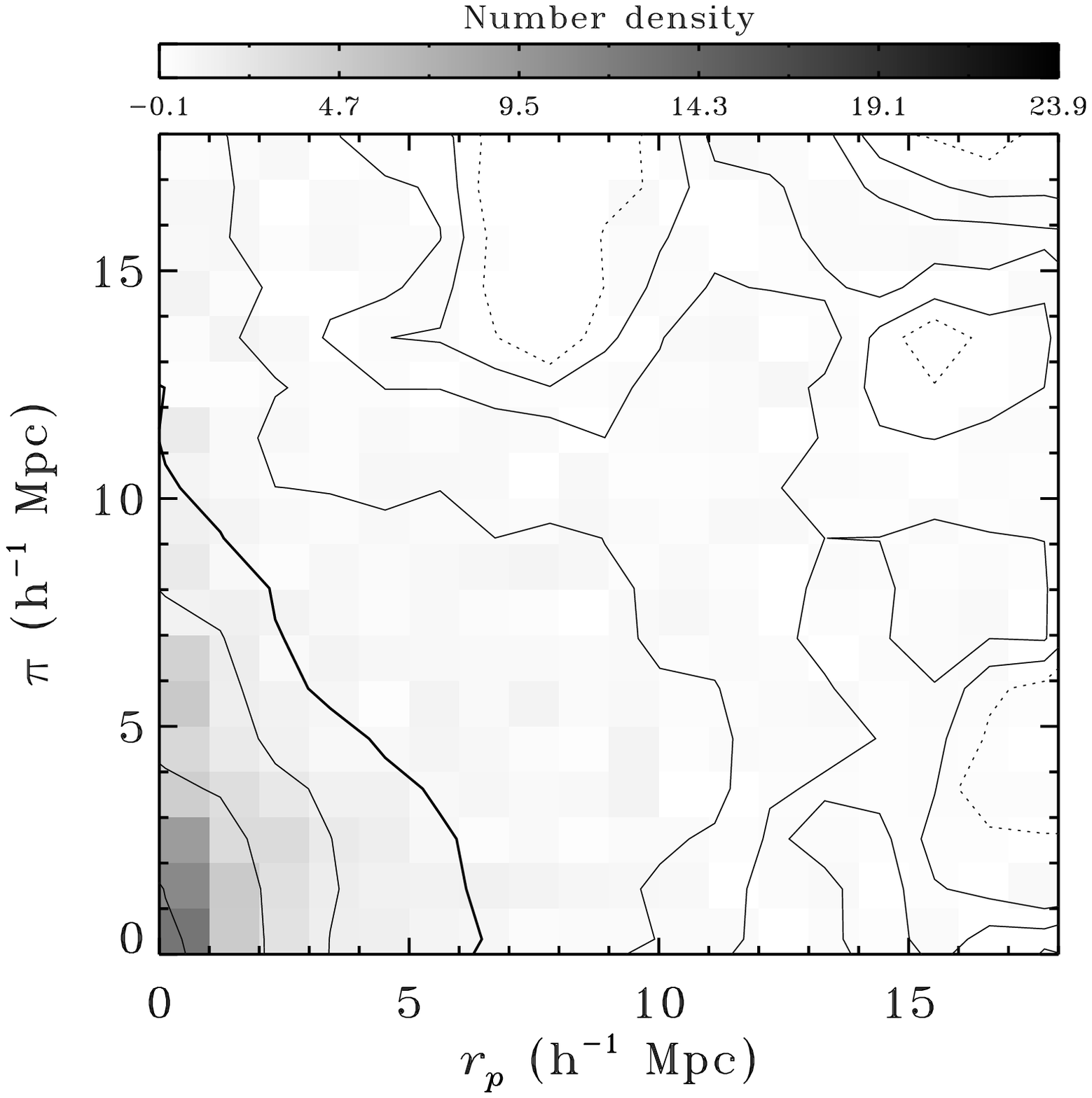}
\includegraphics[width=0.33\textwidth,height=0.33\textwidth]{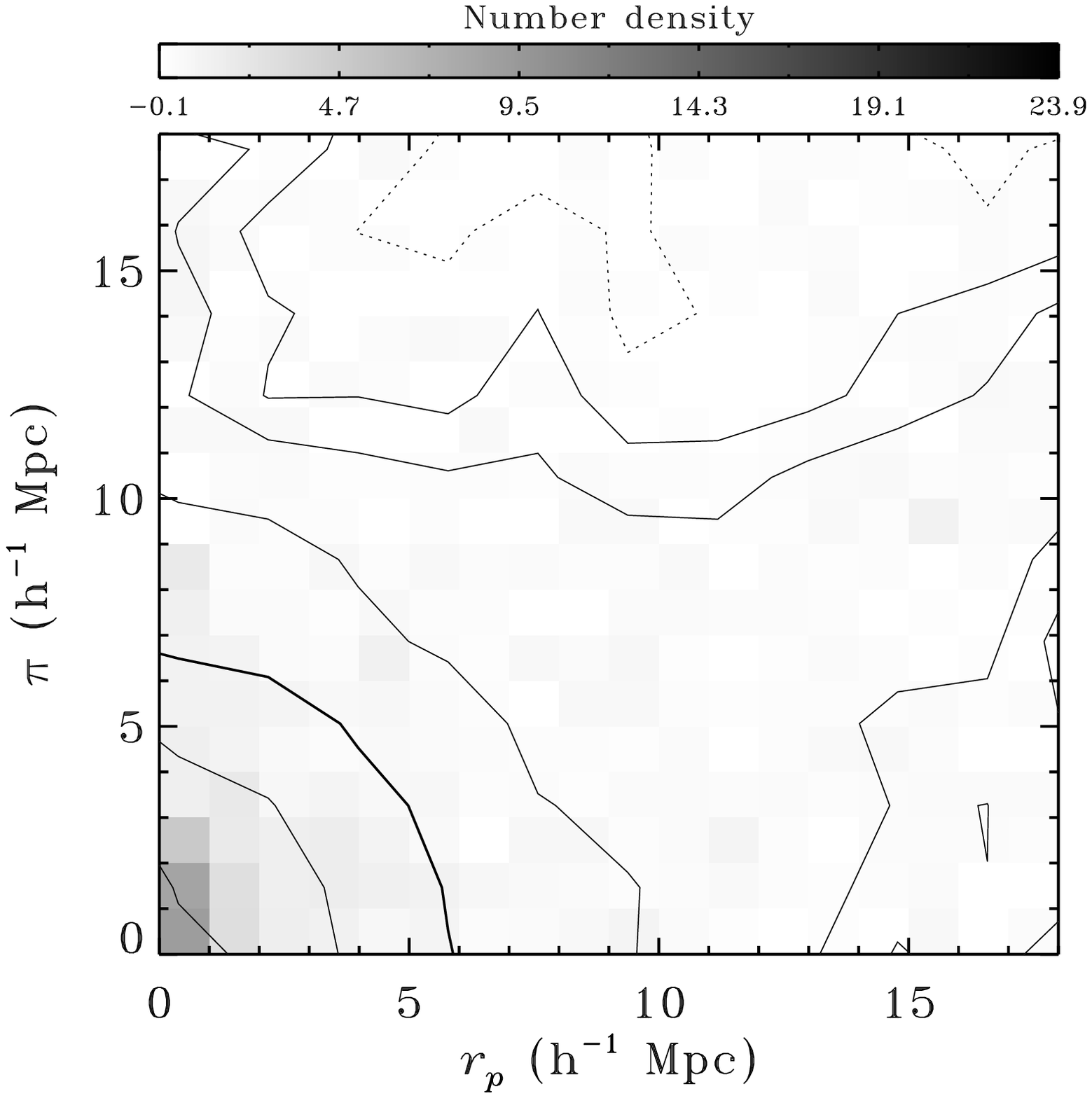}

\includegraphics[width=0.33\textwidth,height=0.33\textwidth]{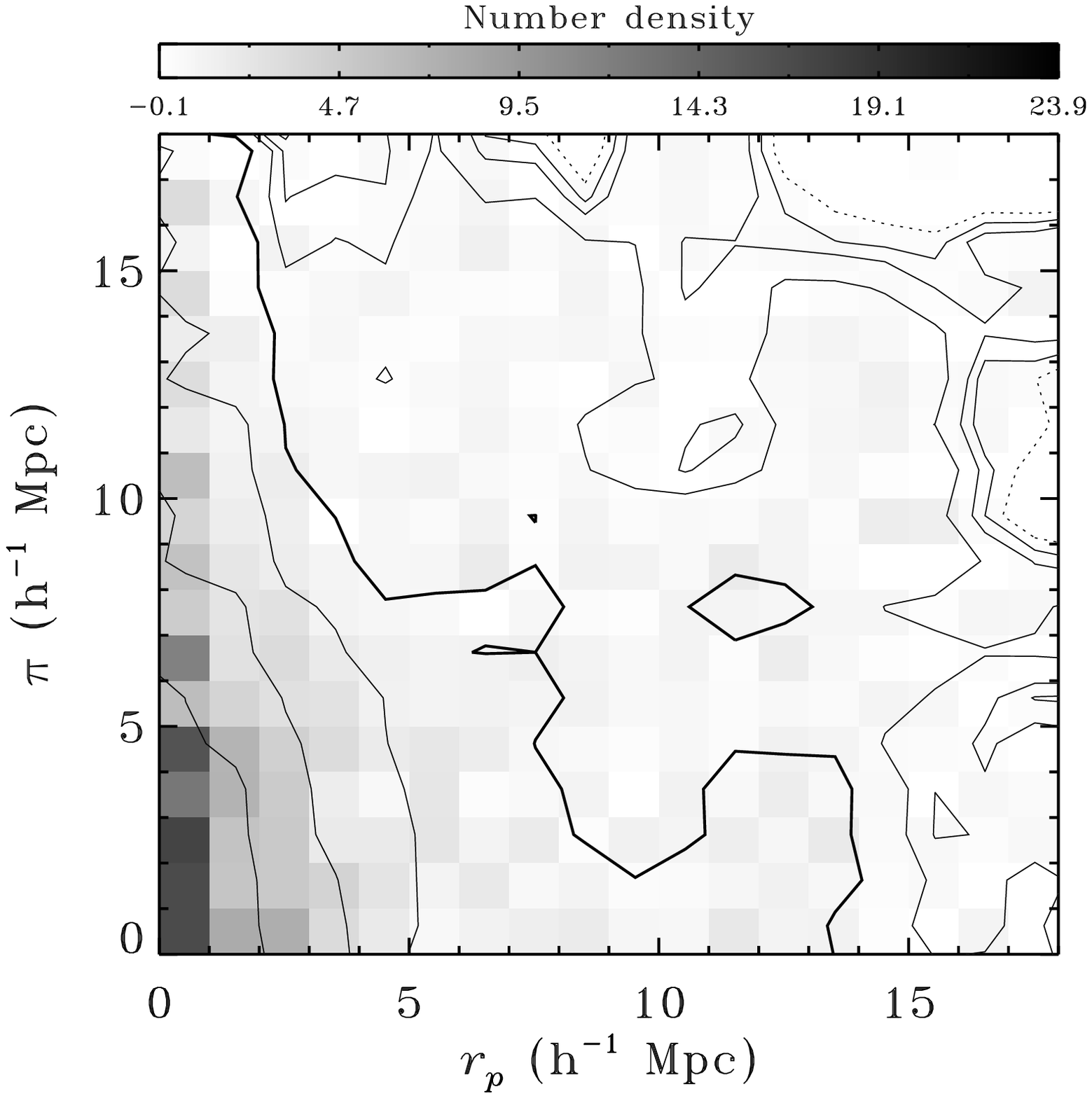}
\includegraphics[width=0.33\textwidth,height=0.33\textwidth]{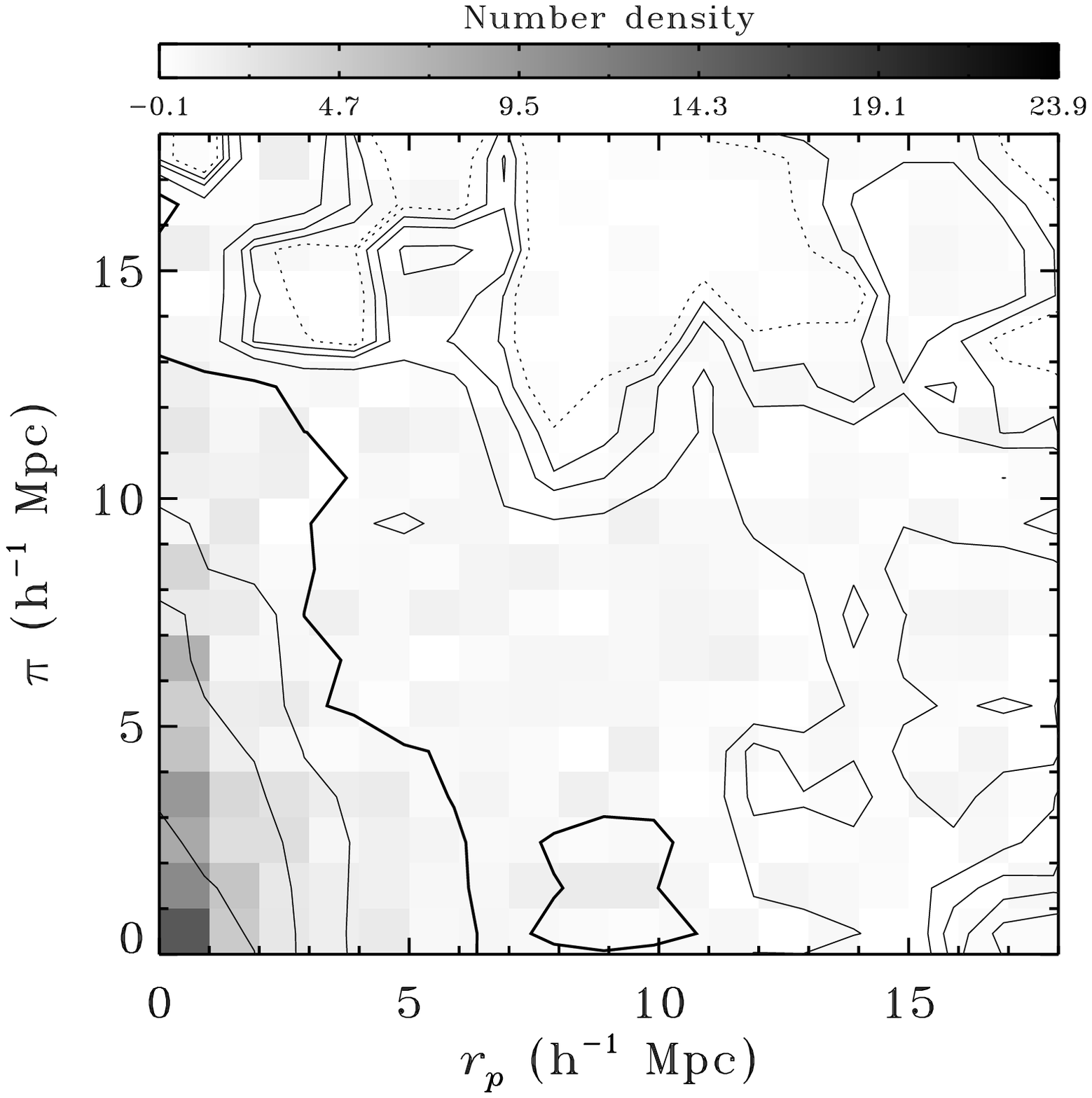}
\includegraphics[width=0.33\textwidth,height=0.33\textwidth]{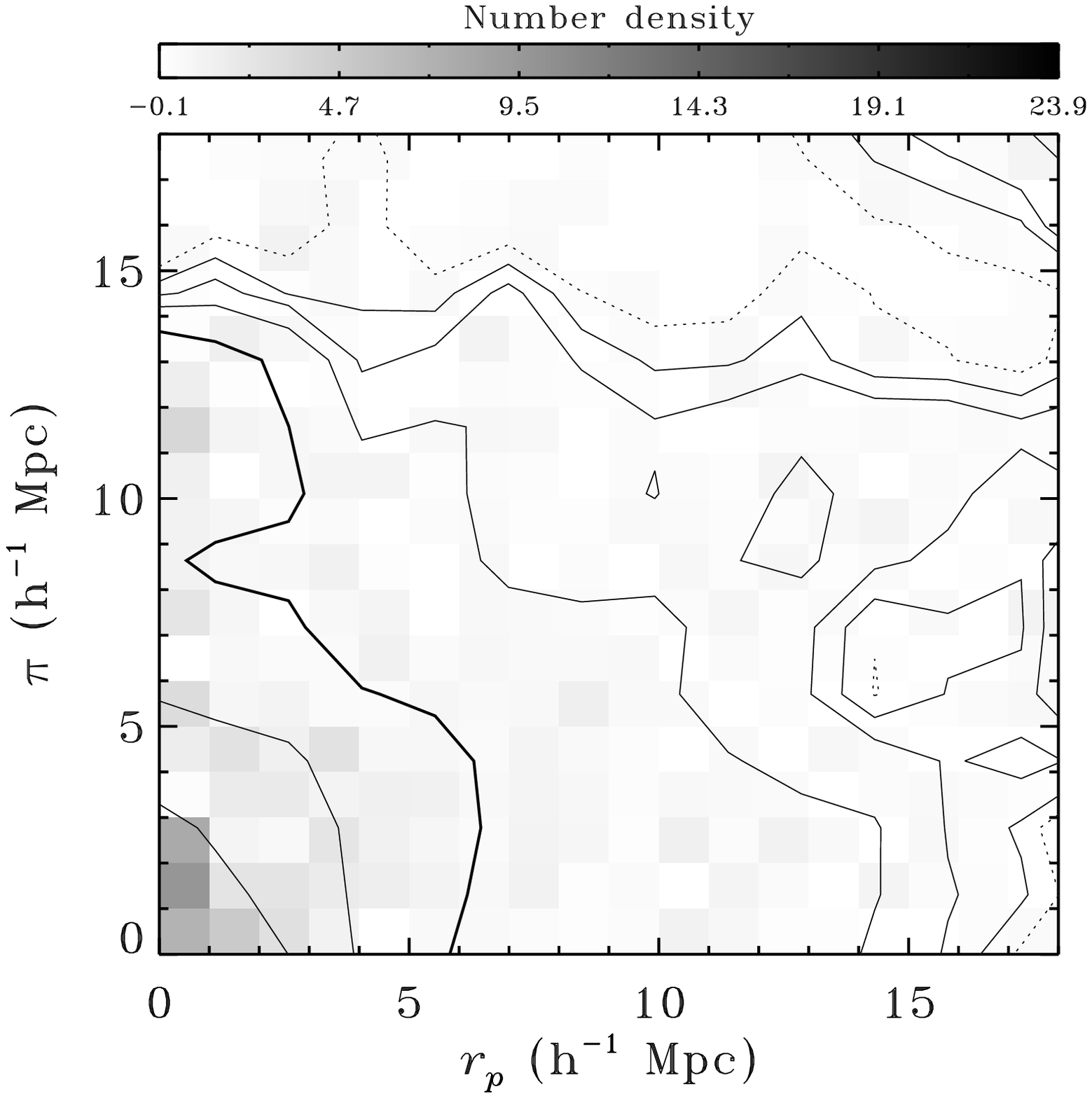}
\caption{Contours and normalized counts of the two-dimensional correlation function $\xi(\pi,r_p)$ for
red (left), green (middle) and blue (left) populations for the volume-limited sample. The panels on the top 
row are for the full distribution, bottom panels are restricted to 
face-on galaxies ($b/a > 0.6$). The contours obtained after 2 x 2 $\hmpc$ boxcar smoothing.
The levels are 0.0 (dotted lines), 0.25, 0.5, 1.0 (heavy lines), 2.0, 4.0, 8.0 and 16.0.
Red galaxies have the strongest finger of God effect (extension in the $\pi$ direction).  
Similar to Figure \ref{fig:f3}, but with lower signal-to-noise, 
the finger-of-God effect is seen in the green sample as well, while it is not present in the blue.   
\label{fig:f4}}
\end{figure*}
Figure \ref{fig:f3} and \ref{fig:f4} shows the two-dimensional auto-correlation function $\xi(\pi,r_p)$ as a 
function of line-of-sight ($\pi$) and projected ($r_p$) separation for the flux-limited and volume-limited 
samples. The panels on the first row in each figure are from the full distribution, while the second row are 
from the dust-corrected distribution. The panels are for red sequence (left), green valley (middle) and blue 
sequence (right) galaxies. We have binned $\pi$ and $r_p$ linearly at $1 \hmpc$. For all panels, the grey scale 
has the same range, and the contours are boxcar smoothed at $2 \hmpc$. The $\xi(\pi,r_p)$ contours indicate 
the constant probability of finding pairs at a given $\pi$ and $r_p$. The heavy solid line marks $\xi(\pi,r_p) = 1$, 
and contours are spaced with increment (inner) and decrement (outer) by a factor of 2. The effect of 
redshift-space distortion is clearly seen in all panels, manifested by their departure from isotropy 
(concentric circles). At small $r_p$, the contours are elongated along the line-of-sight ($\pi$) due to 
virial motions of galaxies in clusters. At large $r_p$, the contours are compressed in the $\pi$ direction 
due to the coherent large-scale streaming as galaxy infall into potential well. We recover the results 
of \cite{Zeh05}: red sequence galaxies show the strongest finger-of-God effect and the larger correlation 
amplitude; and all three subpopulations (of all samples and distribution) show clear signs of large-scale 
compression.

We see a clear finger-of-God effect for green valley galaxies, but not for blue sequence galaxies. In Figure 
\ref{fig:f3}, if one compares the first contour inwards from $\xi = 1$ (heavy line) for red galaxies to the 
$\xi = 1$ contour of green galaxies, green and red galaxies have identical kinematics, differing only by a 
scaling in the amplitude. The blue sequence appears to have a dynamical structure dominated by large-scale 
streaming distinct from both red and green galaxies, similar to the optical blue cloud sample of \cite{Zeh05}. 
We will consider the implications of this in the discussion below. The same conclusion can be drawn from the 
volume-limited sample shown on in Figure \ref{fig:f4}. The dust-corrected samples (bottom row) is 
substantially noisier. We also note that the contours of the blue sequence of 
the volume-limited full sample (top right panel) are roughly circular for $\sim 4 < \pi, r_p < 10 \hmpc$ where 
the finger-of-God elongation is balanced by compression due in infall. 

\begin{figure*} 
\includegraphics[width=0.45\textwidth,height=0.45\textwidth]{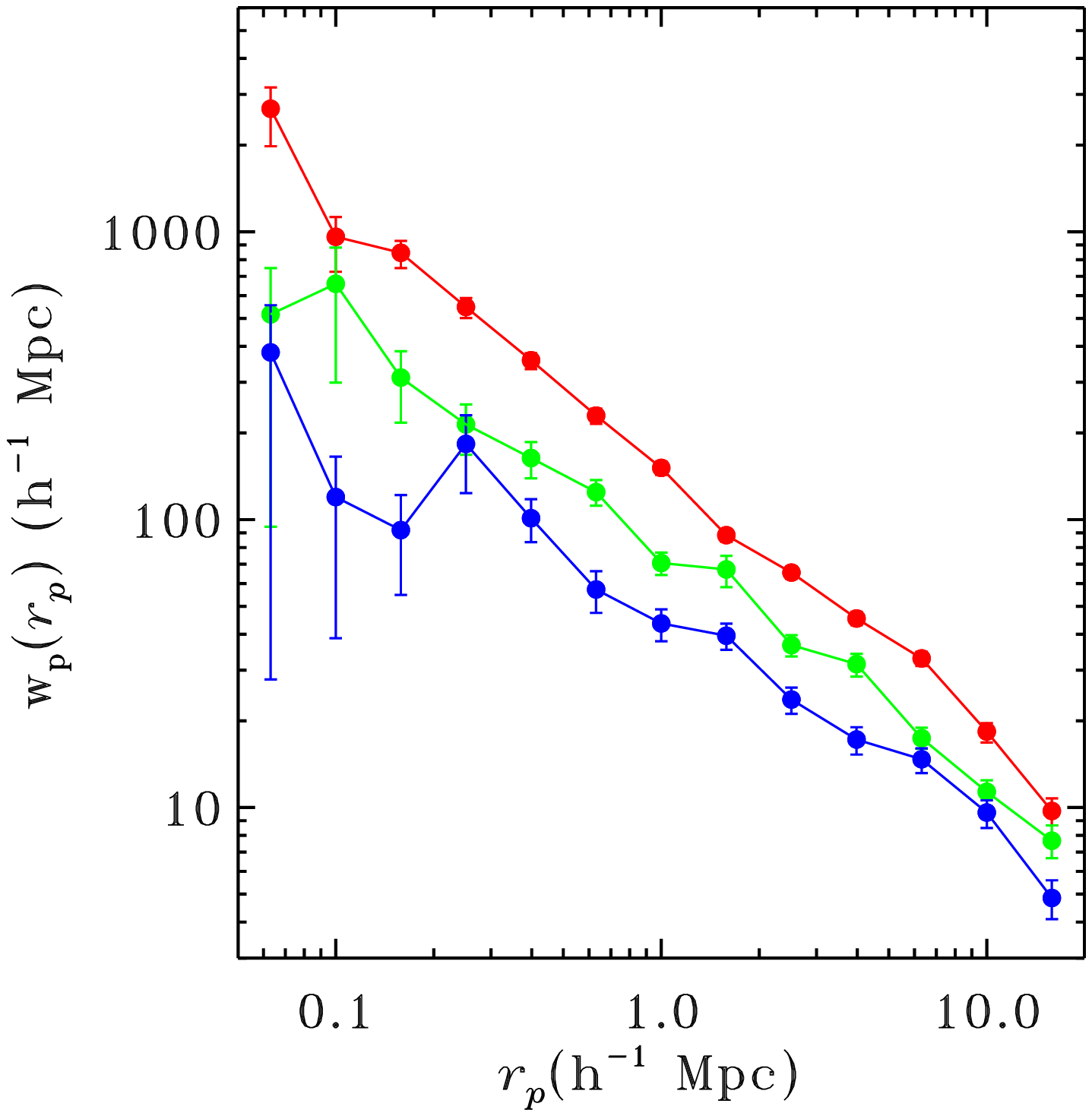}
\includegraphics[width=0.45\textwidth,height=0.45\textwidth]{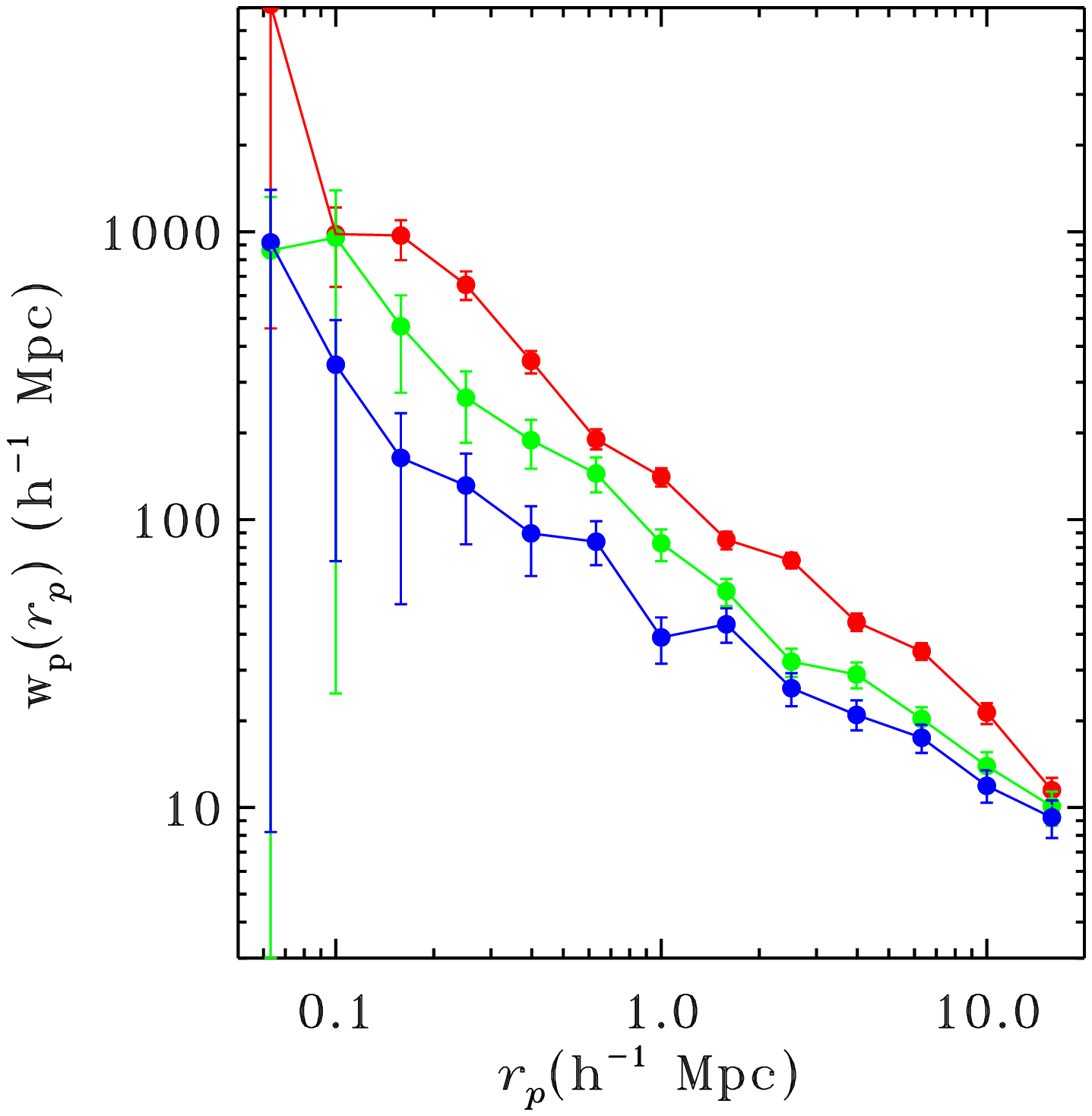}
\caption{The projected correlation function, $w_p(r_p)$, of red sequence, green valley,
 and blue sequence galaxies for the flux-limited sample.
The left (right) panel is from the full (dust-corrected) distribution. 
From red to blue, there is an increase in convexity, with the progressive shallowing of the slope on large scales, 
indicating an excess of two-halo signals at scale ($r_p \gtrsim 1 \hmpc$). 
\label{fig:f6}}
\end{figure*}
\begin{figure*} 
\includegraphics[width=0.45\textwidth,height=0.45\textwidth]{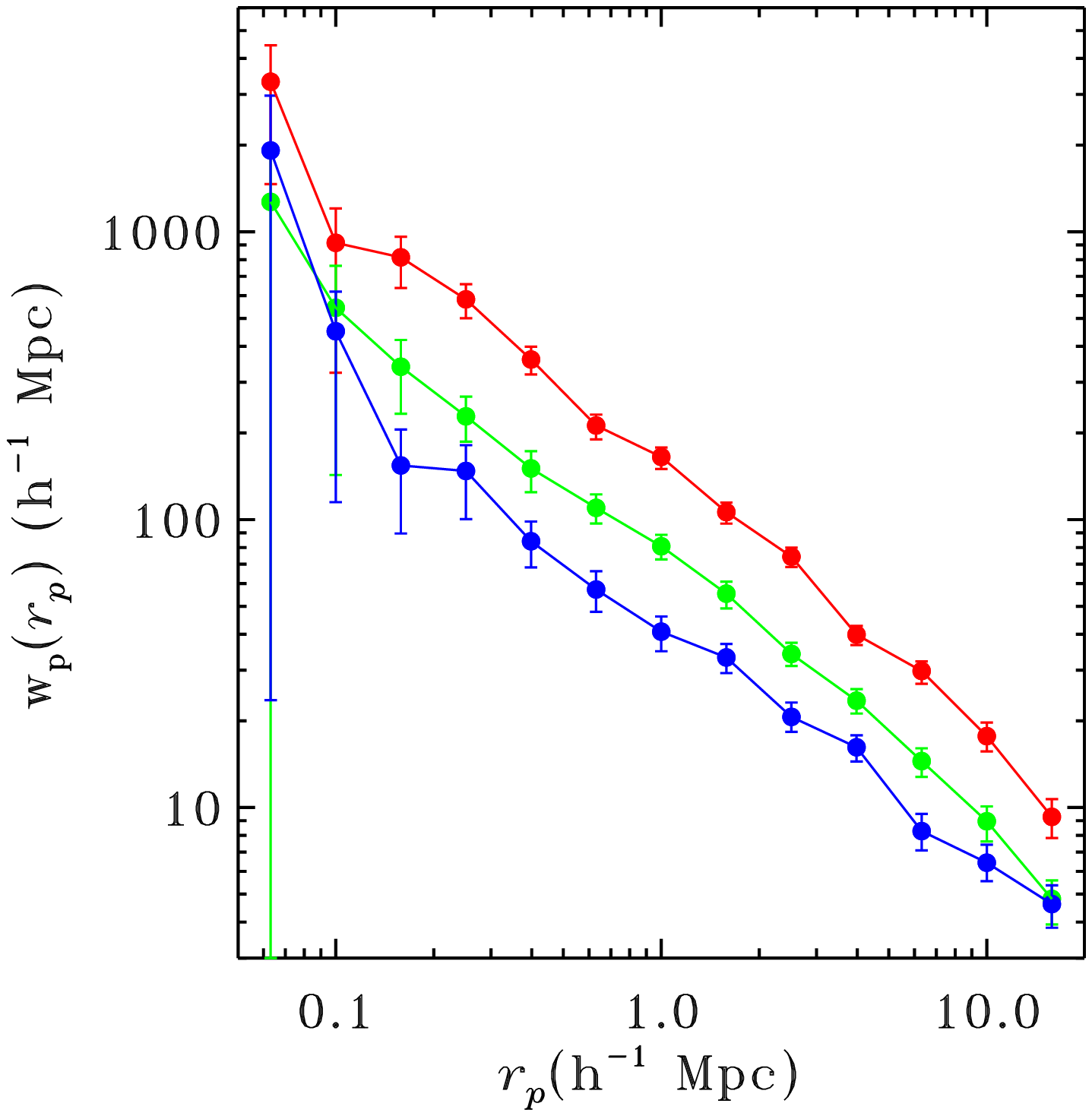}
\includegraphics[width=0.45\textwidth,height=0.45\textwidth]{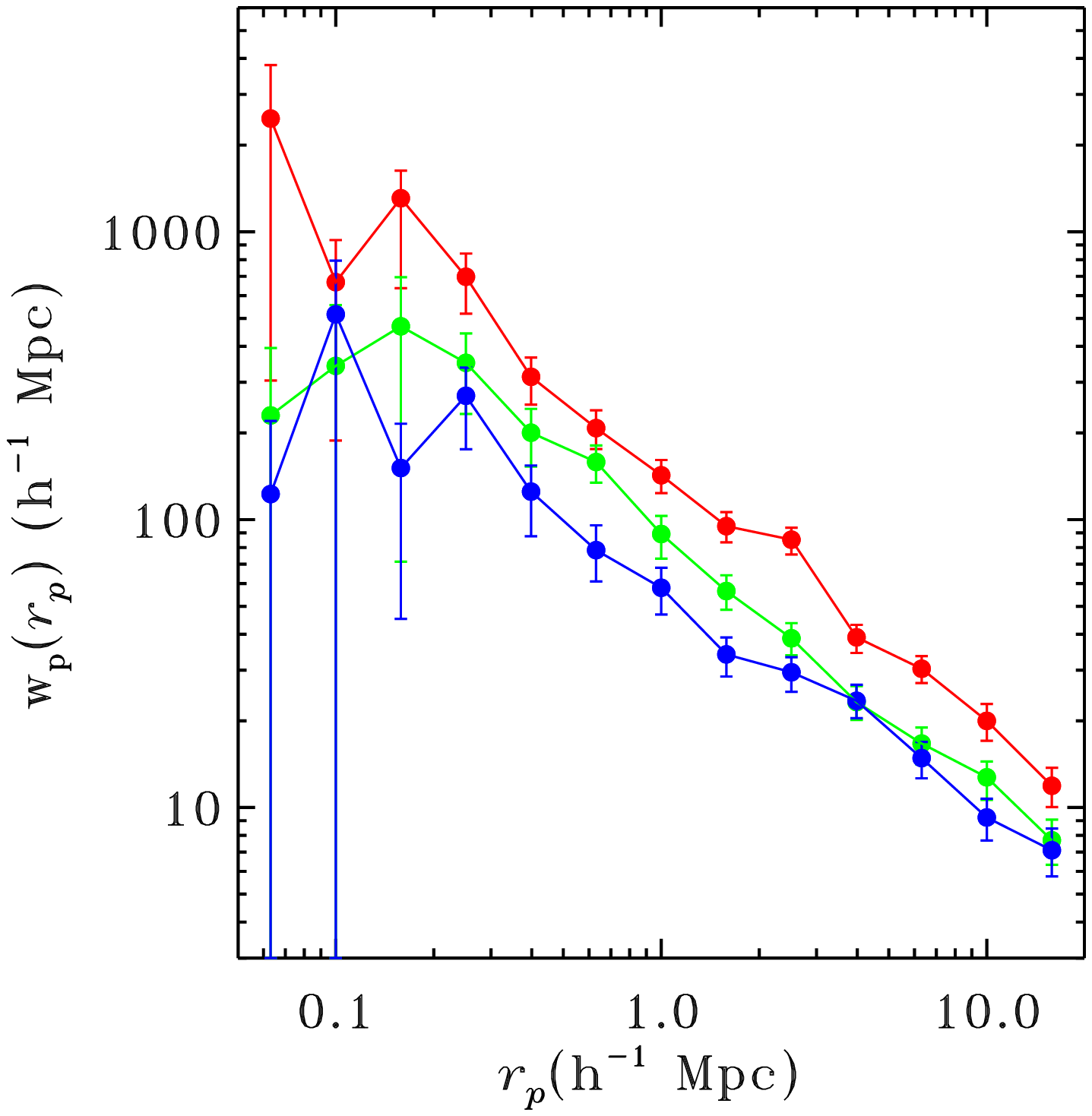}
\caption{The projected correlation function, $w_p(r_p)$, of red sequence, green valley, and blue sequence 
galaxies for the volume-limited sample.
Similar to Figure \ref{fig:f6}, from red to blue, there is an increase in convexity, with the progressive 
shallowing if the slope on large scale, indicating an excess of two-halo signals.
 at scale ($r_p \gtrsim 1 \hmpc$). 
For the full distribution (left panel), the green $w_p(r_p)$ is parallel to the red with a lower amplitude for scales $0.2 < r_p < 10 \hmpc$. 
For the dust-corrected distribution (right panel), the blue $w_p(r_p)$ approaches the green on large scales.
\label{fig:f7}}
\end{figure*}
Figure \ref{fig:f6} and \ref{fig:f7} shows the projected correlation function $w_p(r_p)$ for the flux-limited 
and volume-limited samples. The colored (red, green, blue) points are the respective measured correlation 
function. The panel on the left (right) is for the full (dust-corrected) distribution. As expected, in all 
four panels, the red sample has the largest amplitude, the blue the lowest, and green in between. 
On large scales ($r_p \gtrsim 1 \hmpc$), the ratio of projected auto-correlation function of the respective samples
are constant as a function of scales, as expected from linear theory. All correlation functions appear to have a form that is well fit by a 
broken power-law with the dust-corrected distribution having a more 
pronounce convexity. It is noteworthy that the flux-limited analysis (Figure \ref{fig:f6}) is more noisy than 
the volume-limited analysis (Figure \ref{fig:f7}), despite having a factor of two greater numbers. We believe 
that this is due to the inclusion of uncorrelated (in redshift space) galaxies, which dilute the clustering 
signal, especially for blue galaxies on small scales. From here onwards, we will restrict our analysis to the 
volume-limited sample.

\begin{figure*} 
\includegraphics[width=0.45\textwidth,height=0.45\textwidth]{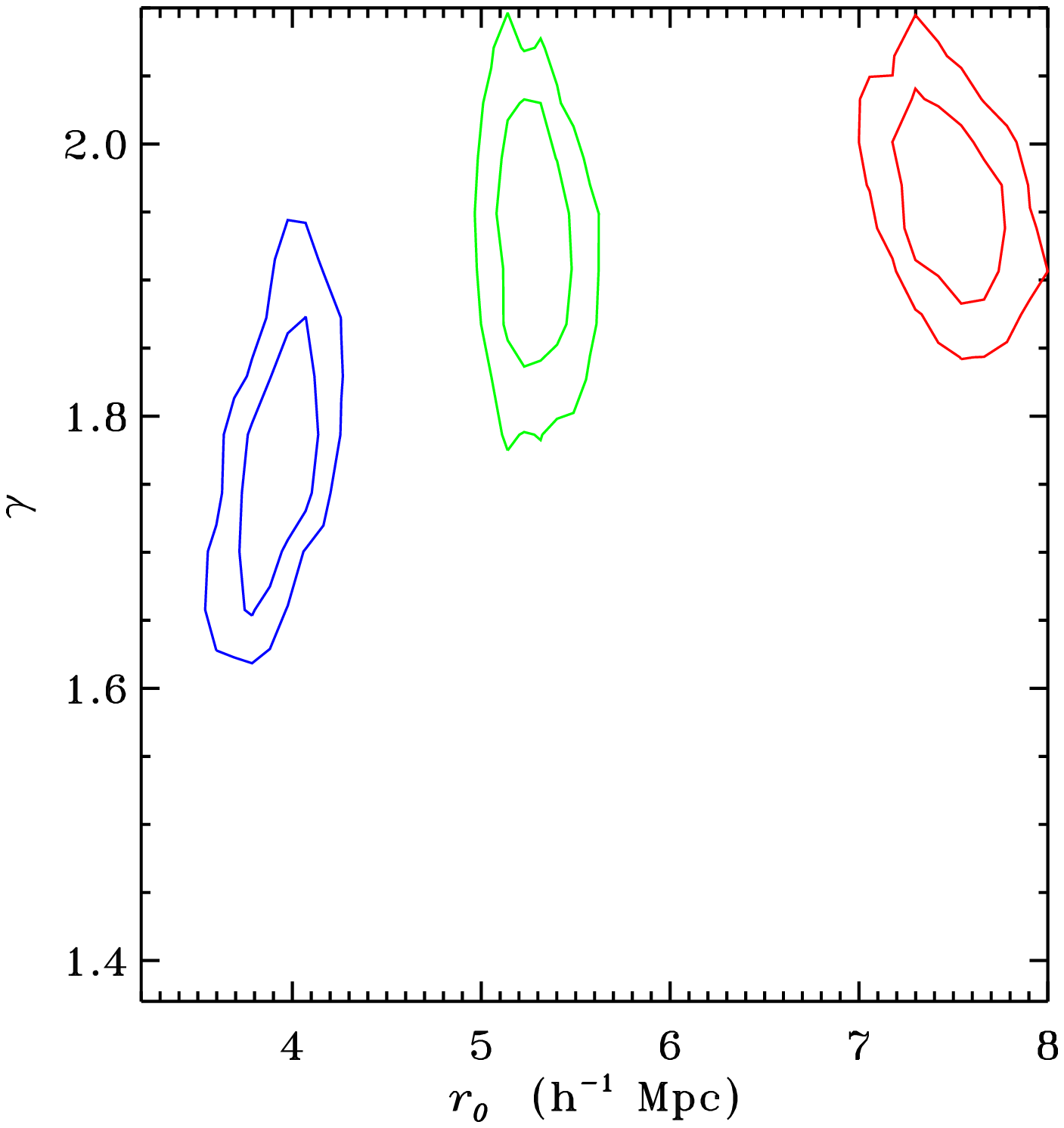}
\includegraphics[width=0.45\textwidth,height=0.45\textwidth]{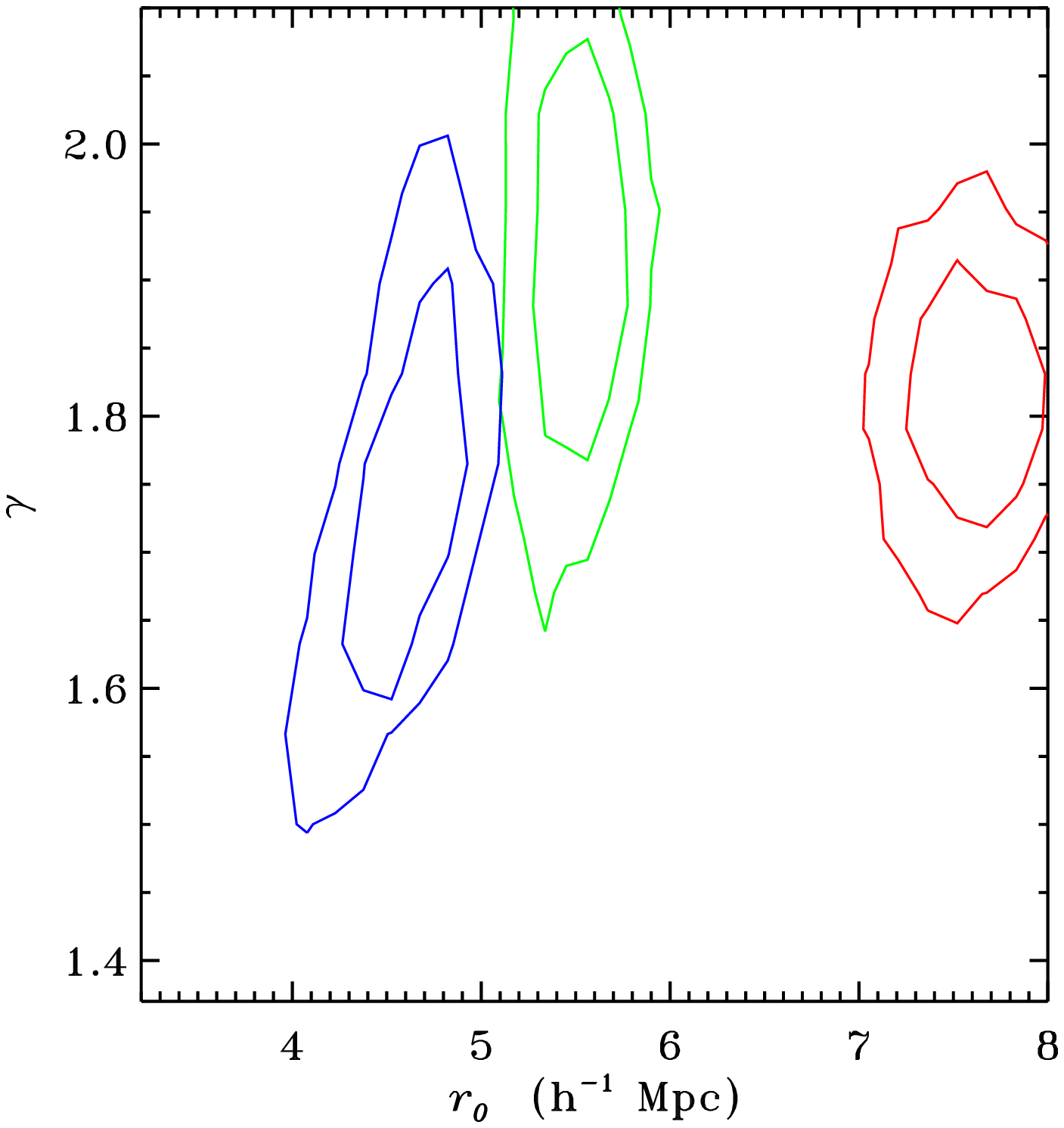}
\caption{Confidence contours of the power-law fits to $w_p(r_p)$ on scales $1 \hmpc < r_p < 10 \hmpc$ for the full (left panel) 
and dust-corrected (right) distributions from the volume-limited sample. The three colors are for red, green and blue subsamples, 
with the inner (outer) contours encircling the 68 (95) \% confidence region in the $r_0$--$\gamma$ space. 
\label{fig:f8}}
\end{figure*}
When we fit the projected correlation function of the full distribution with the standard two parameter power-law for 
scales from $1 < r_p < 10 \hmpc$, red and green galaxies appears to have similar slope with $\gamma \sim 1.93$, 
while blue galaxies have a substantially shallower slope, with $\gamma \sim 1.75$. The best fit correlation 
lengths $r_0$ are $7.5 \hmpc$ (red), $5.3 \hmpc$ (green), and $3.9 \hmpc$ (blue). 
The result of the covariance analysis using 999 bootstrap samples is shown as confidence interval 
contours -- 68\% (inner) and 95\% (outer) in Figure \ref{fig:f8} (left panel). These results are in excellent 
agreement with \cite{Hei08}. The correlation function from the dust-corrected distribution is substantially 
noisier, in part due to the smaller number of galaxies. Applying the same power-law fit, we obtain a larger 
uncertainty in $\gamma$, but comparable uncertainty in $r_0$. This is shown on the right panel of 
Figure \ref{fig:f8} and tabulated in Table 1. 

\begin{table*}
\caption{Power-law fit to the projected auto-correlation function of volume-limited samples ($1 \hmpc < r_p < 10 \hmpc$)}
\begin{center}
\begin{tabular}{cccccccc} \hline\hline
{\sc Full-Sample}&$N$&
\multicolumn{1}{c}{$z$}&{\sc median} $M_r$&$r_0$&$\gamma$&{\sc Relative Bias}&$\chi^2/dof$\\
\hline
Red..........&$1971$&$(0.03,0.12)$&$-20.9$&$7.5\pm0.27$&$1.94\pm0.07$&$1.53\pm0.08$&$1.36$\\
Green........&$1971$&$(0.03,0.12)$&$-20.9$&$5.3\pm0.19$&$1.93\pm0.08$&$1.08\pm0.06$&$0.40$\\
Blue.........&$1971$&$(0.03,0.12)$&$-20.9$&$3.9\pm0.25$&$1.73\pm0.10$&$0.81\pm0.06$&$1.54$\\
\hline
{\sc Face-On}&$$&
\multicolumn{1}{c}{$$}&{\sc }&{\sc }&{\sc }&{\sc }\\
\hline
Red..........&$1226$&$(0.03,0.12)$&$-20.9$&$7.6\pm0.38$&$1.81\pm0.11$&$1.47\pm0.10$&$2.47$\\
Green........&$1226$&$(0.03,0.12)$&$-20.9$&$5.5\pm0.24$&$1.92\pm0.15$&$1.13\pm0.09$&$0.48$\\
Blue.........&$1226$&$(0.03,0.12)$&$-20.9$&$4.6\pm0.31$&$1.74\pm0.17$&$0.92\pm0.09$&$1.17$\\
\hline
\end{tabular}
\end{center}
\end{table*}

On the left panel, the projected correlation function of green galaxies is evidently intermediate between 
the red and the blue on a range of scales ($0.2 < r_p < 6 \hmpc$), and runs faithfully parallel with the red 
correlation function, but with a lower amplitude. The picture for the dust-corrected distribution
is slightly different. While the green still runs in between the red and blue for $r_p < 1 \hmpc$ and mostly
parallel to the red with a lower amplitude for broad range of scales, for larger scales ($r_p \gtrsim 2 \hmpc$), 
it coincides with the blue. This is a consequence of the more prominent two-halo excess on large scales seen 
in the blue correlation function. 

\begin{figure}
\includegraphics[width=0.45\textwidth,height=0.45\textwidth]{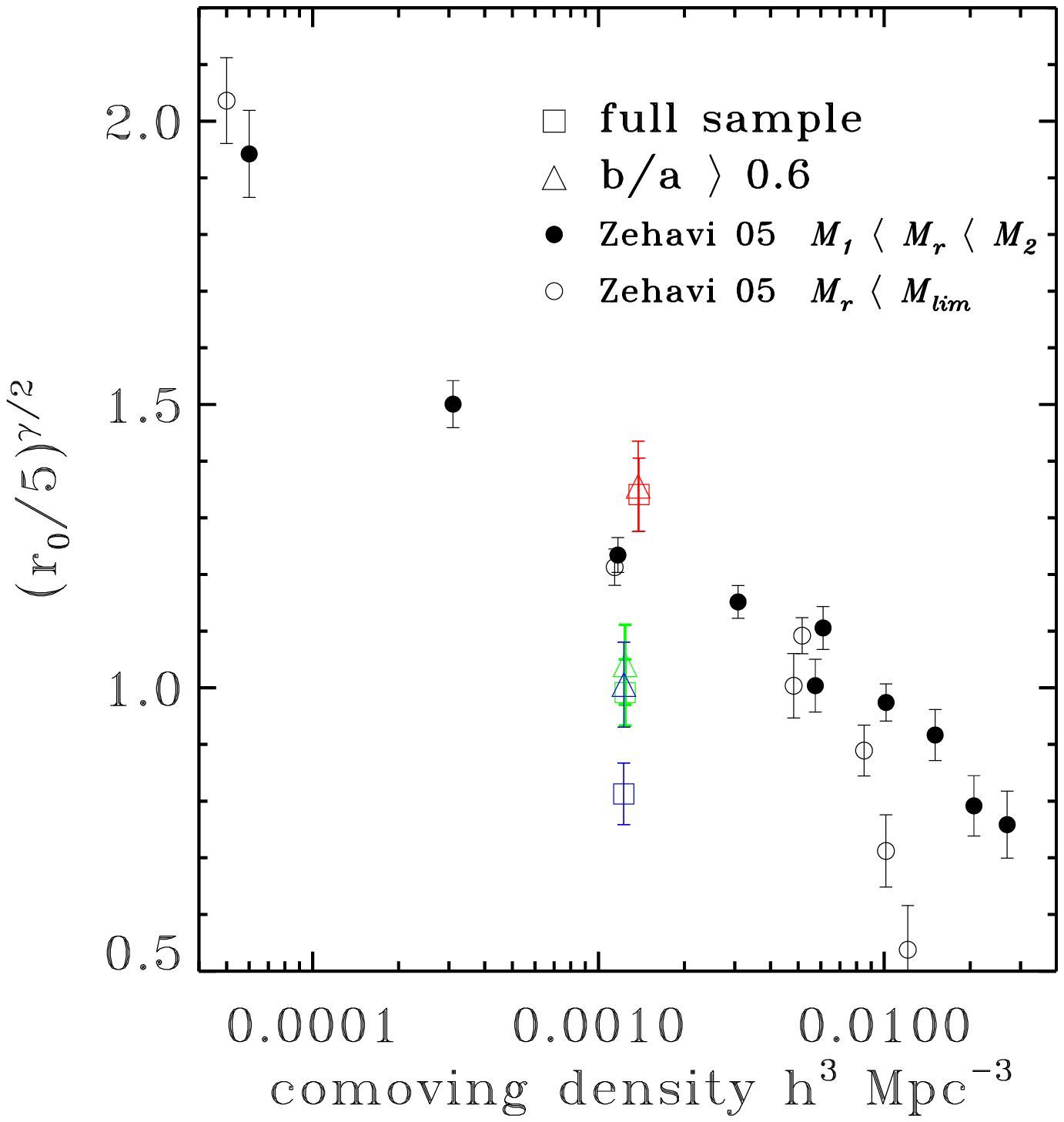}
\caption{Bias relative to the typical $L^*$ galaxy with a $\xi^{fid}(r) = (r/r_0)^{-1.8}$ at $r_p = 5 \hmpc$ versus co-moving number density.
We also plotted points from SDSS (Zehavi \etal 2005). The solid circles are from their luminosity range analysis, i.e. galaxy with absolute magnitude between
$M_1$ and $M_2$, while the open circles are from their luminosity threshold analysis, i.e. $M_r < M_{lim}$. 
While far from unique, the higher red bias suggests that on average red galaxies are more likely to be satellite galaxies then a typical galaxy with those number densities.
If we assume the lower blue and green galaxies are all central galaxies in the halo they reside in, the lower observed number densities suggest
that only a fraction of these halos, as described by their bias, host a blue or green galaxy, as their expected number density is much higher.
\label{fig:rdens}}
\end{figure}
One way to compare the clustering between subpopulations of galaxies is through their relative bias \citep[e.g.][]{Nor02}, defined 
as 
\beq
\frac{b_i}{b_*}(r) = \sqrt{\frac{\xi_i(r)}{\xi_*(r)}} \simeq
\frac{r_{0,i}^{\gamma_i/2}}{r_0^{\gamma/2}}r^{(\gamma - \gamma_i)/2}
\eeq
where $\xi_i$ is the correlation function of interest, $\xi_*$ is the fiducial correlation function that all correlation function 
is compare with. The approximate equality holds when the correlation function takes a power-law form.
\cite{Zeh05} found that a typical $L^*$ galaxies in the local
universe with $-20 < M_r < -21$ from the SDSS main spectroscopic sample have a fiducial power-law correlation function with 
$\xi_*(r) = (r/5 \hmpc)^{-1.8}$ on scale $1 < r_p < 10 \hmpc$. By fitting the correlation function using a 
constant $\gamma = 1.8$ on scales $1 < r_p < 10 \hmpc$, we obtained the bias relative at $r_p = 5\hmpc$ to a fiducial $L^*$ galaxies for 
each of our subpopulations: $b_{\tt red}/b_{*} = 1.53 \pm 0.08$,  $b_{\tt green}/b_{*} = 1.08 \pm 0.06$ and  
$b_{\tt blue}/b_{*} = 0.81 \pm 0.06$ for the full-sample, and $b_{\tt red}/b_{*} = 1.47 \pm 0.10$,  
$b_{\tt green}/b_{*} = 1.13 \pm 0.09$ and  $b_{\tt blue}/b_{*} = 0.92 \pm 0.09$ for the dust-corrected samples (Table 1).
In Figure \ref{fig:rdens}, we plot $b_i/b_*(r = 5 \hmpc) = (r_0/5\hmpc)^\gamma$ as a function of the co-moving number density and compare our
values with those obtained by \cite{Zeh05}. Co-moving number density of our sample is estimated using the $1/V_{max}$ method and further corrected to match
those obtained by \cite{Wyd07}. While all the samples by construction have almost identical co-moving density, there is a range of relative bias, with the red having
the highest bias, and blue the lowest. The spread among the bias are much smaller among the dust-corrected samples.
Red galaxies have clustering strength above the nominal strength (inferred from a typical SDSS galaxy) for a given number density. One plausible scenario
suggest that red galaxies have a larger than average satellite fraction. On the other hand, blue and green galaxies have lower observed number density compare to their clustering.
If we assume all blue and green galaxies are central galaxies in the halo they reside in, the lower observed number densities ($\sim0.001 \ihmpcC$)
suggest that only a small fraction of these halos, as described by their lower bias, host a blue or green galaxy, as their expected number density 
(inferred from SDSS to be $\sim0.01 \ihmpcC$) is much higher. These fractions are increased if the average phases of green and blue is shorter then the lifetime 
of the halos \citep{Hai01,Mar01}.  In the case of the green galaxies, the transitional nature of these galaxies may be closely related to the AGN they host, or to 
minor mergers and starbursts, with triggering cycles corresponding to the duty cycles for these respective phenomena.
We defer a more detail analysis of halo occupation and the life cycle of transition galaxies using 
star-formation rate tracers of different lags to a future paper (Heinis et al. in prep.).  
Using a theoretical bias function of dark matter halos \citep{Sel04} and normalizing to $M_{*} = 10^{12}\hmsolar$\footnote{The halo
mass that a typical $L^*$ galaxy resides in.}, red galaxies cluster similar to halos with mass $\sim 10^{12.7}\hmsolar$, 
green galaxies $10^{12.2}\hmsolar$, and blue galaxies $10^{11.6}\hmsolar$.

\subsection{Cross-Correlation Functions}\label{sec:cross}
\begin{figure*}  
\includegraphics[width=0.33\textwidth,height=0.33\textwidth]{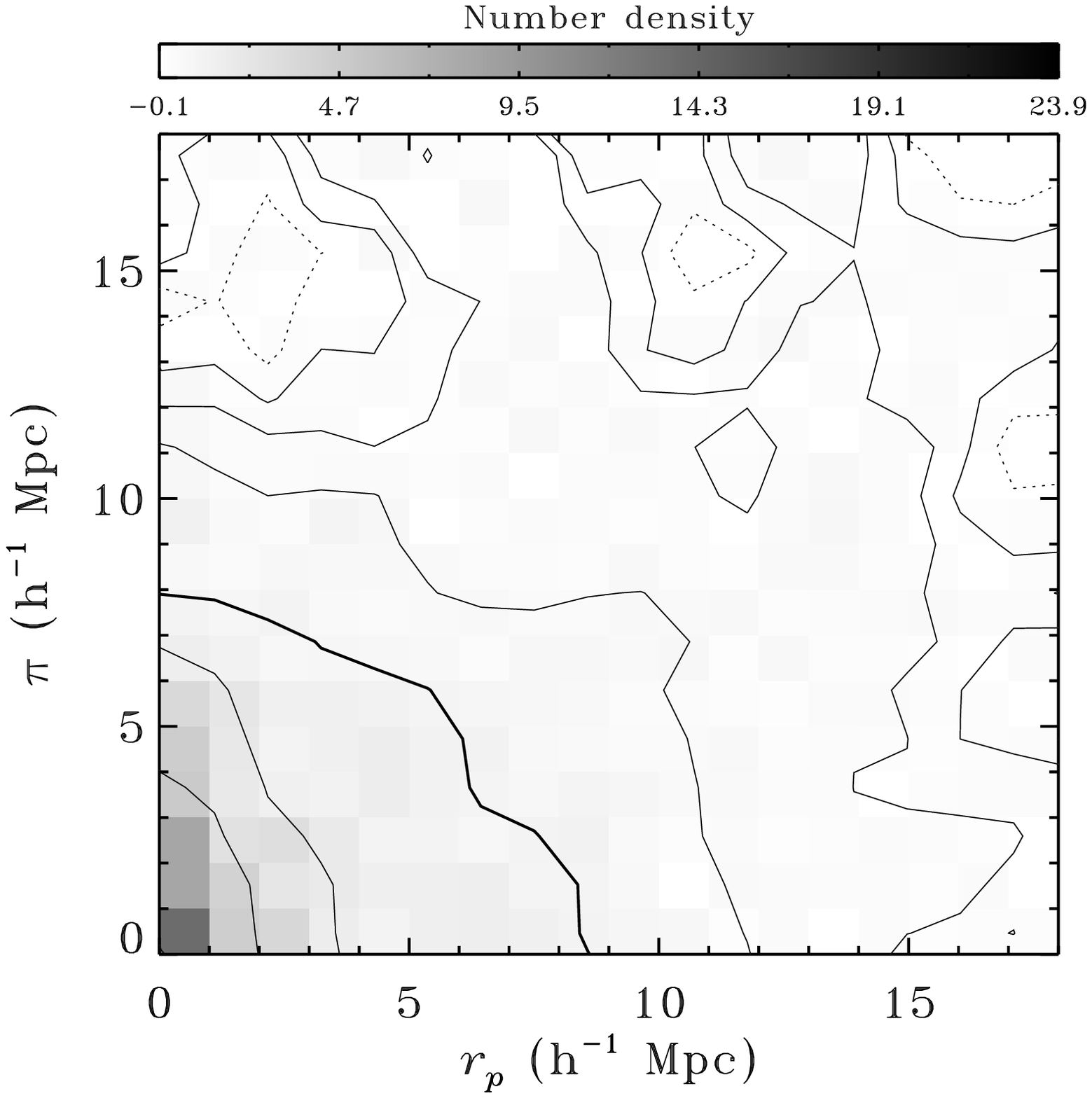}
\includegraphics[width=0.33\textwidth,height=0.33\textwidth]{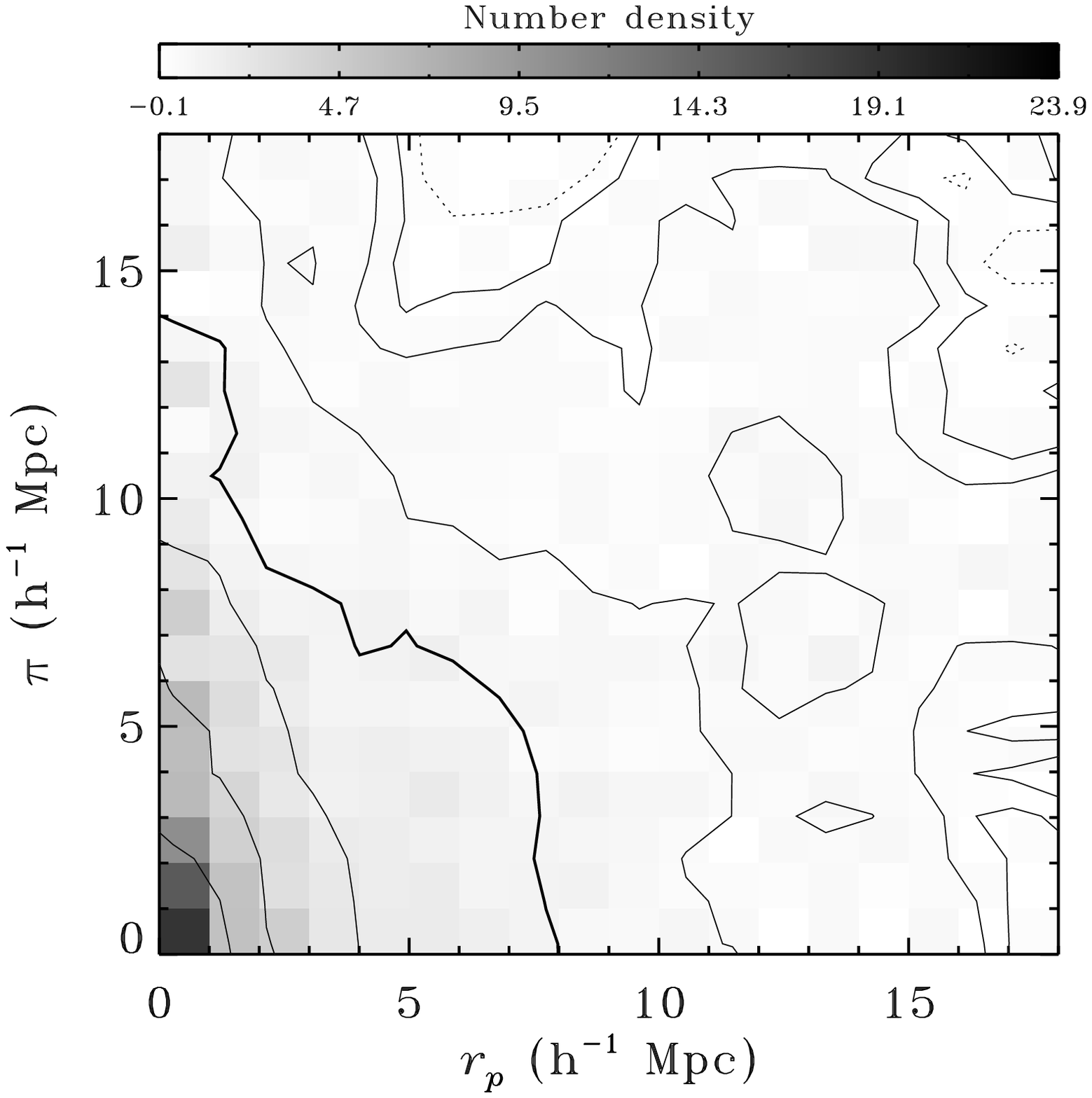}
\includegraphics[width=0.33\textwidth,height=0.33\textwidth]{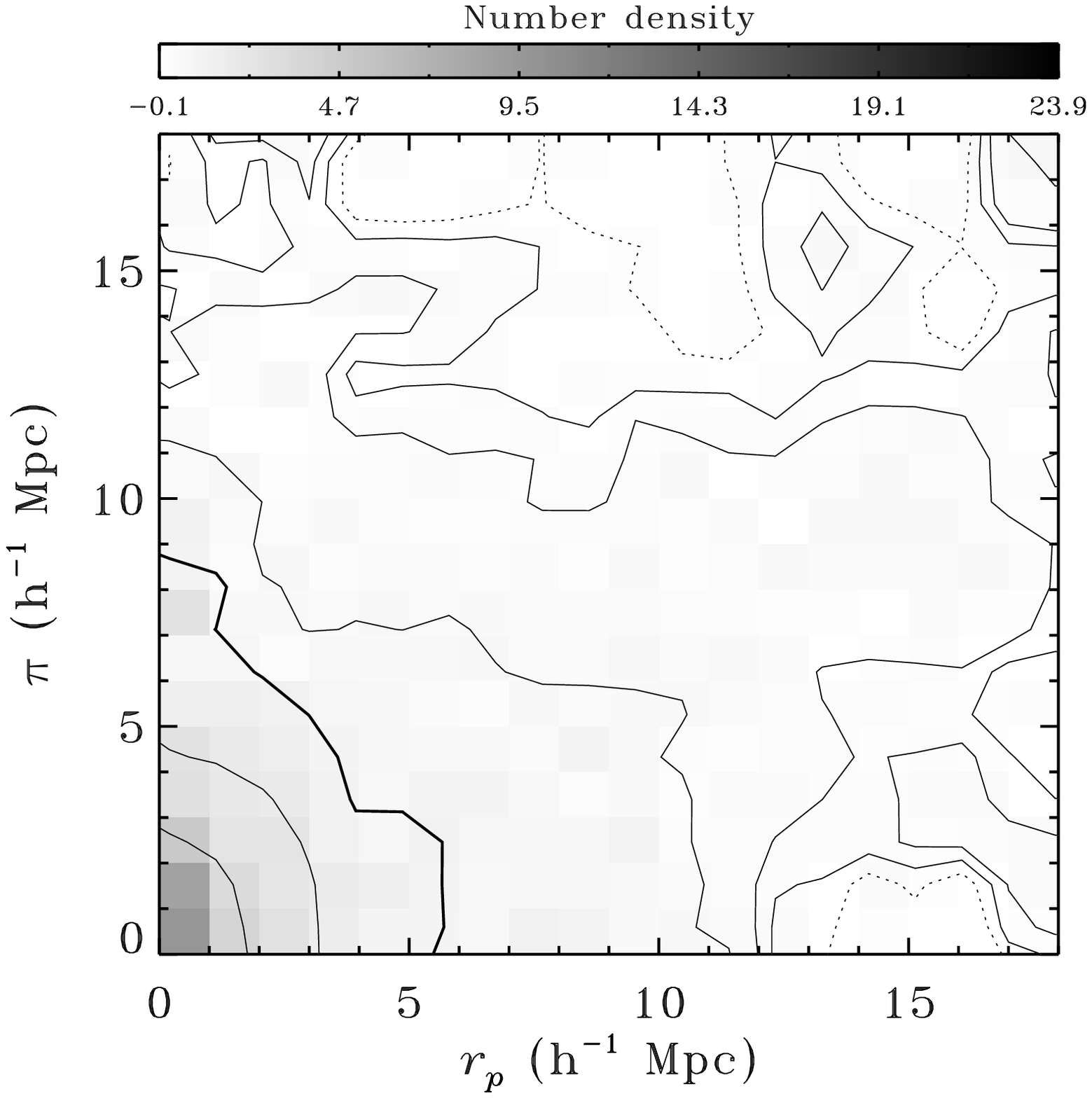}

\includegraphics[width=0.33\textwidth,height=0.33\textwidth]{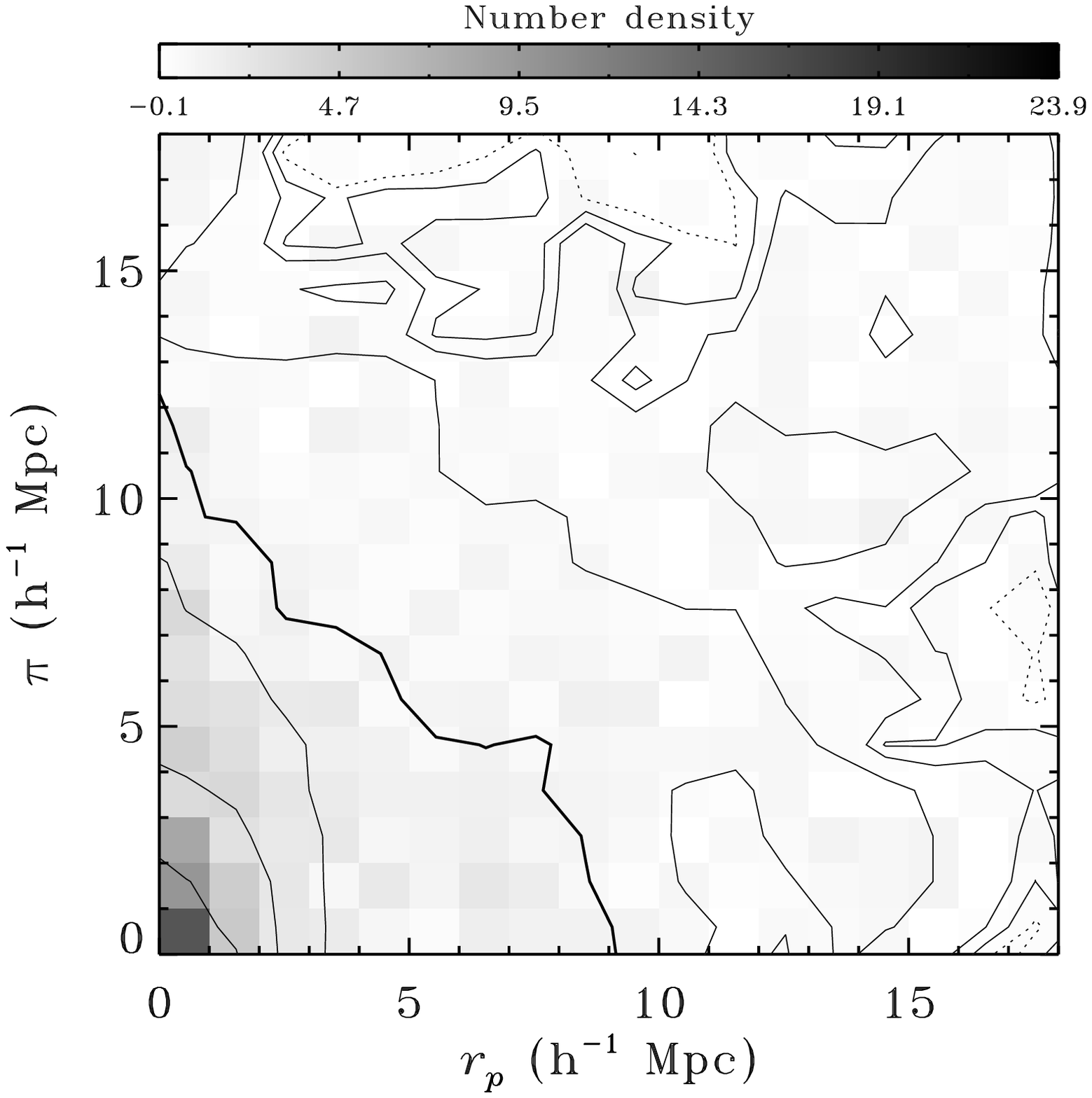}
\includegraphics[width=0.33\textwidth,height=0.33\textwidth]{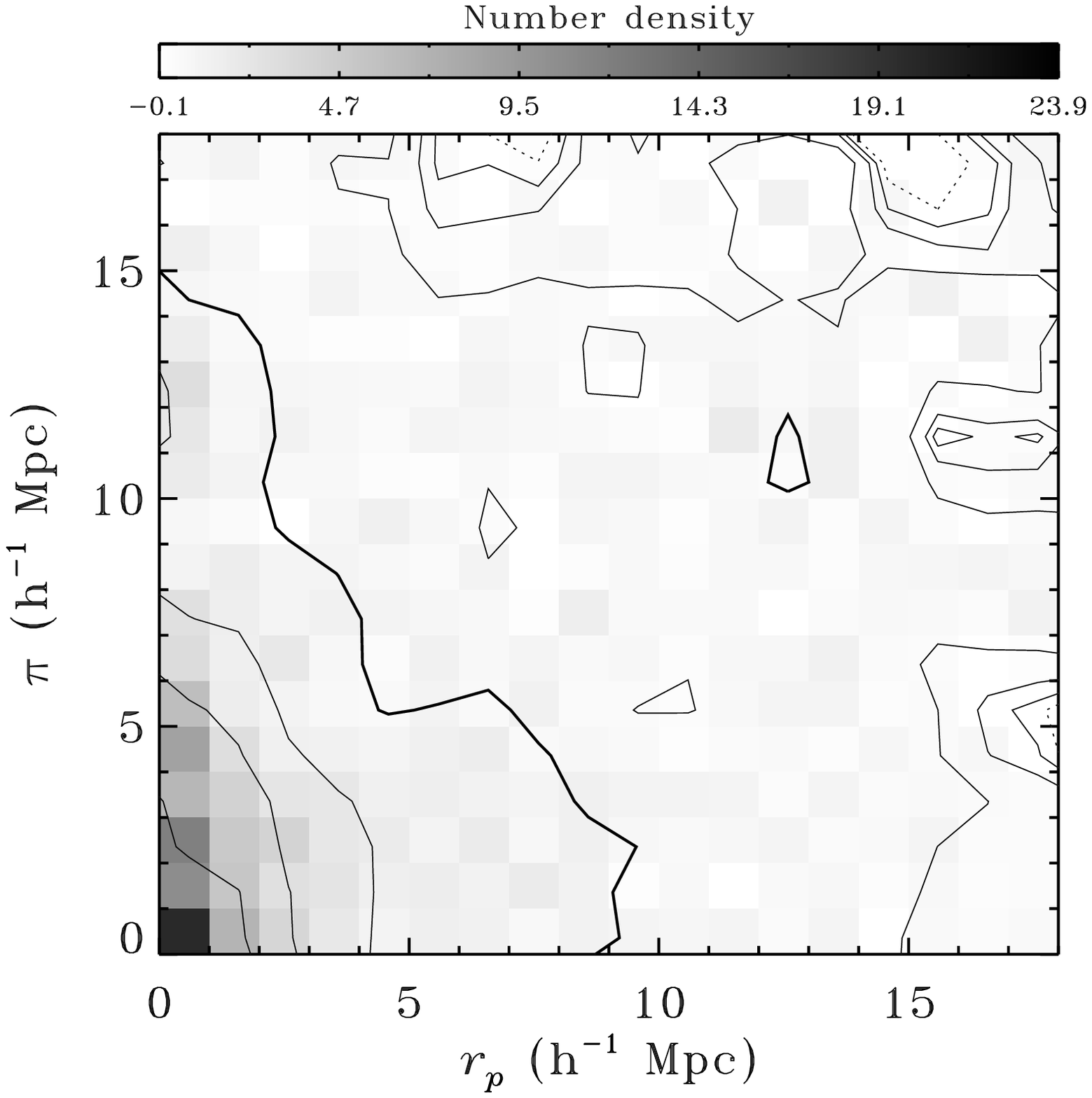}
\includegraphics[width=0.33\textwidth,height=0.33\textwidth]{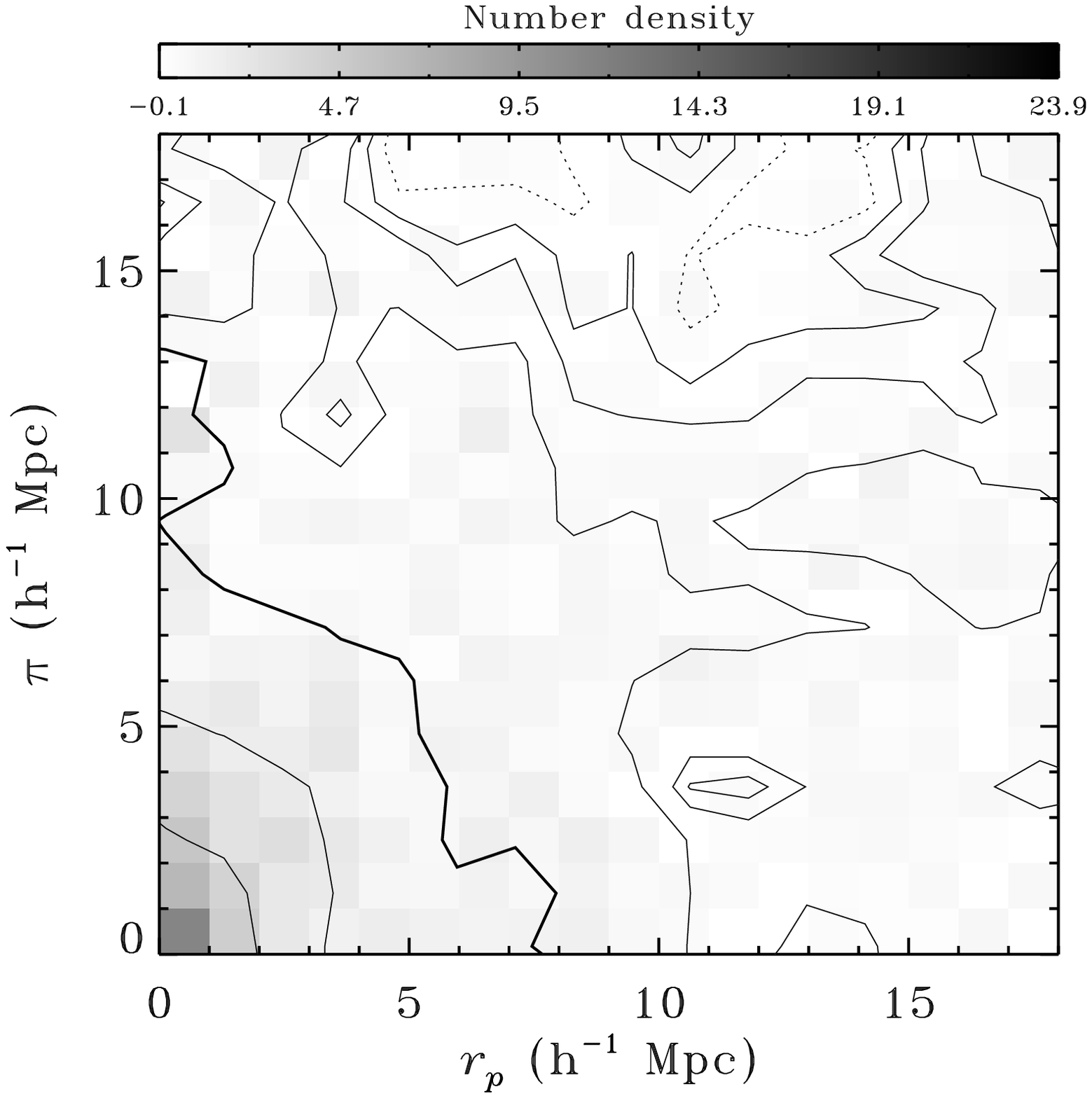}
\caption{Contours and normalized counts of the two-dimensional cross-correlation function $\xi(\pi,r_p)$ for
red-blue (left), green-red (middle) and green-blue (left) for the volume-limited sample. The panels on the 
top row are for the full distribution, bottom panels are restricted to 
face-on galaxies ($b/a > 0.6$). The contours obtained after 2 x 2 $\hmpc$ boxcar smoothing.
The levels are 0.0 (dotted lines), 0.25, 0.5, 1.0 (heavy lines), 2.0, 4.0, 8.0 and 16.0.
The large scale infall effect is seen in all three panels. On small scales, the finger-of-God effect is strongest 
for the red-green, intermediate for red-blue and weakest for green-blue, as expected. 
Note that red-blue CCF $\xi(\pi,r_p)$ has a weaker finger-of-God and stronger infall compression compare to the 
green valley auto-correlation function (Figure \ref{fig:f4} middle panels).
\label{fig:f5}}
\end{figure*}
\begin{figure}
\includegraphics[width=0.45\textwidth,height=0.45\textwidth]{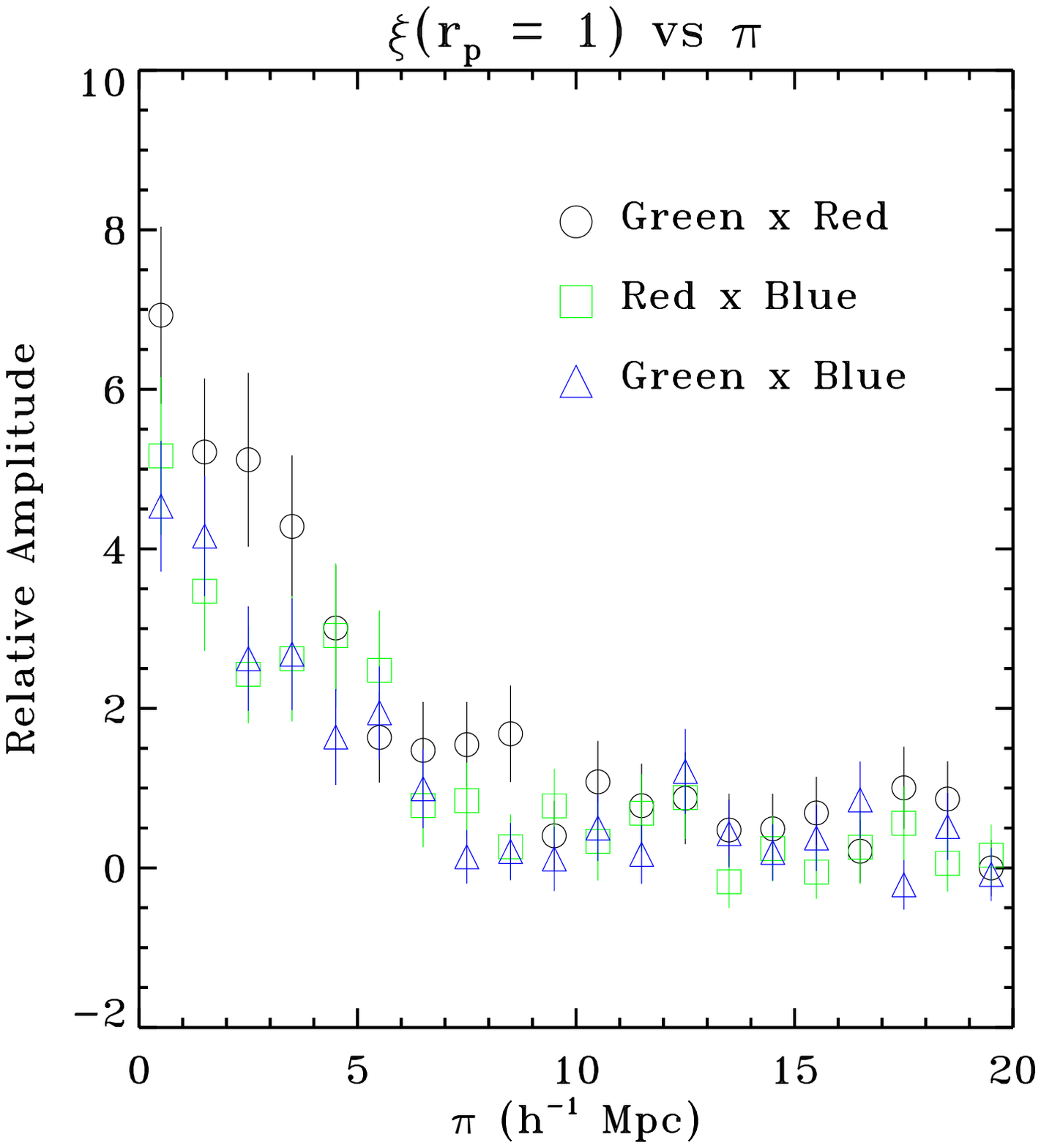}
\caption{This figure shows the relative amplitude of $\xi$ at projected distance $r_p = 1 \hmpc$ as a function of the line-of-sight distance $\pi$.
The circles are for green-red cross correlation, squares for red-blue, and triangles for green-blue. Green-red have largest amplitude for a wide range
of $\pi$, while red-blue and green-blue have smaller amplitudes. These results suggest that the finger-of-god component of green-red cross correlation
function is stronger then those from red-blue or green-blue.
\label{fig:fog}}
\end{figure}

Figure \ref{fig:f5} shows the two-dimensional $\xi(r_p,\pi)$ cross-correlation function (CCF) for the 
three cross-pairs of galaxies in 
our volume-limited sample, with the panels on the top for the full distribution, and the bottom panels 
for the dust-corrected distribution.  
The large scale infall effect is seen in all three panels suggesting that the galaxies, on average, trace 
a similar matter distribution as 
expected from linear theory. On small scales, the finger-of-God effect is strongest for the red-green CCF, 
intermediate for red-blue and green-blue. In Figure \ref{fig:fog} we plot the relative amplitude of $\xi$ at 
the projected distance $r_p = 1 \hmpc$ as a function of line-of-sight distance $\pi$. The circles are for green-red cross correlation, 
squares for red-blue, and triangles for green-blue. Green-red have largest amplitude for a wide range
of $\pi$, while red-blue and green-blue have smaller amplitudes. The stronger finger-of-god component suggest that the dynamics of
red and green are more strongly coupled compared with red-blue or green-blue pairs. 
Note that red-blue CCF have a very different $\xi(\pi,r_p)$ compared with the 
green auto-correlation function (ACF), 
the former has a weaker finger-of-God and stronger infall compression. 

\begin{figure*} 
\includegraphics[width=0.45\textwidth,height=0.45\textwidth]{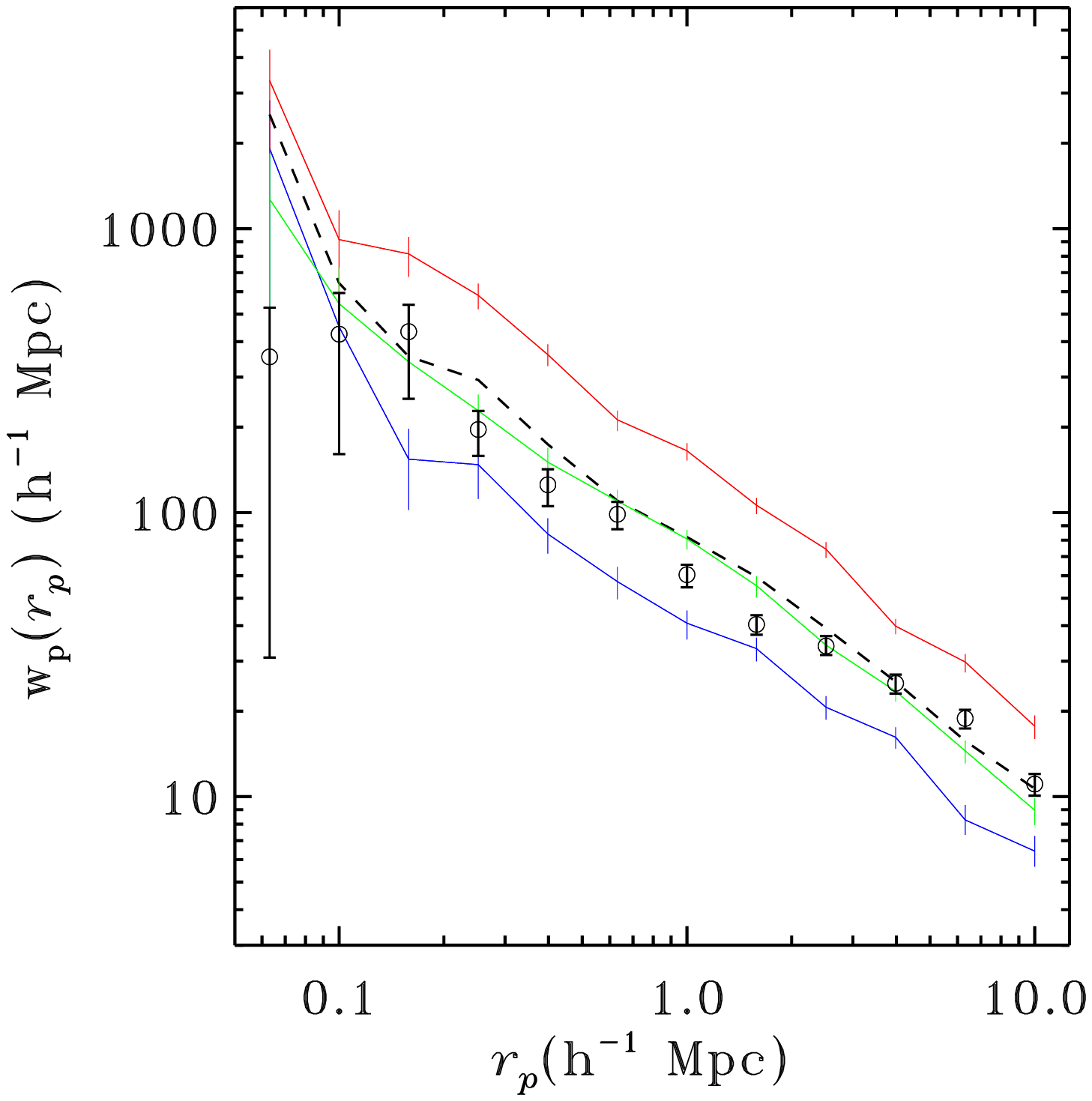}
\includegraphics[width=0.45\textwidth,height=0.45\textwidth]{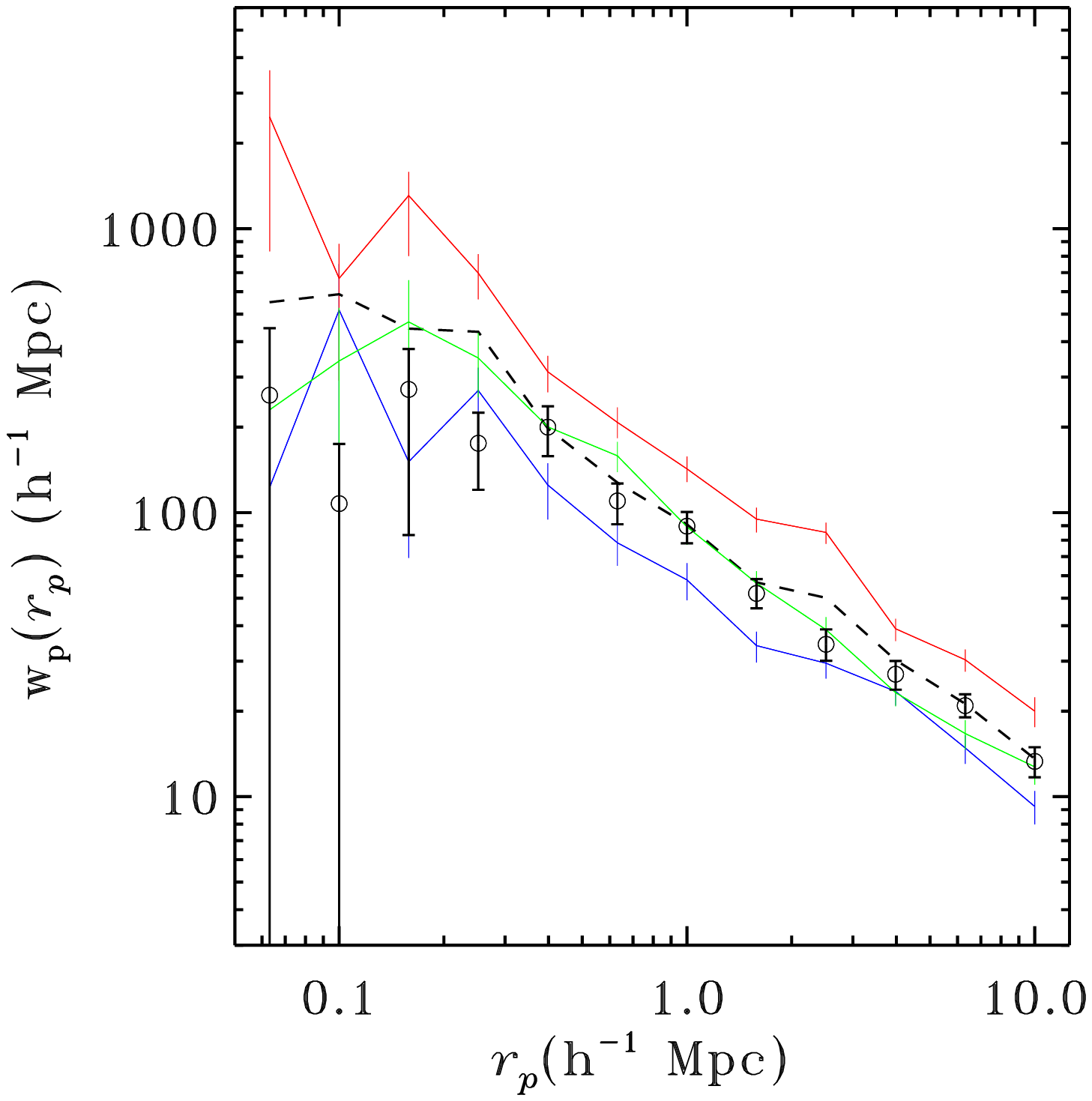}
\caption{The projected cross-correlation function between red and blue sequence galaxies $w_p^{r,b}(r_p)$ are shown as open circles. 
The solid lines are the auto-correlation function of the red and blue galaxies. The dashed line is the red-blue geometric mean $w_p^{\overline{r,b}}(r_p)$. 
On scales larger than $\sim 3 \hmpc$, the cross-correlation approaches the geometric mean. For scales $0.2 \hmpc < r_p < 2 \hmpc$,
the cross-correlation function is systematically below the geometric mean. This result suggests that on small scales, blue and red galaxies are spatially
segregated beyond what was expected from their auto-correlation function. This is consistent with the morphology density relation \citep{Dre80}.
\label{fig:f9}}
\end{figure*}
As discussed in section \ref{sec:cross_meth}, one way to understand the relationship between two populations 
is to compare the projected cross-correlation function between the two with their geometric mean. If the two 
populations are mixed evenly, the cross-correlation functions should trace the geometric mean. If they are
spatial segregated (partially) beyond what was expected from their respective auto-correlation function, 
the cross-correlation function should be systematically below the geometric mean. 
The projected cross-correlation functions for the red and blue galaxies $w_p^{r,b}(r_p)$ are plotted 
as open circles in Figure \ref{fig:f9}. Also plotted for comparison are the red and blue auto-correlation 
functions (solid lines), and the red-blue geometric mean $w_p^{\overline{r,b}}(r_p)$ (dashed lines). 
On scales larger than $\sim 3 \hmpc$, the $w_p^{r,g}(r_p)$ approaches the geometric mean. 
For scales $0.2 < r_p < 2 \hmpc$, $w_p^{r,g}(r_p)$ is systematically below the geometric mean for 
the full distribution (left panel). For the dust-corrected distribution (right panel), $w_p^{r,b}(r_p)$ 
starts to inch below the geometric mean only from $r_p \lesssim 0.7 \hmpc$ onwards. Our results are 
consistent with the partial morphology segregation within galaxy clusters \citep{Dre80}. At the one-halo regime ($r_p \lesssim 1-3 \hmpc$), 
the relevant scales for galaxy clusters, red galaxies tend to occupy the cores of the clusters, 
while blue galaxies tend to lie towards the periphery. The lower level of spatial mixing on these scales suppresses 
amplitude of the cross-correlation function. Note that the green auto-correlation function
(green solid line) does trace the red-blue geometric mean on large scales, suggesting that blue and red 
galaxies do mix evenly on these scales.  

\begin{figure*} 
\includegraphics[width=0.45\textwidth,height=0.45\textwidth]{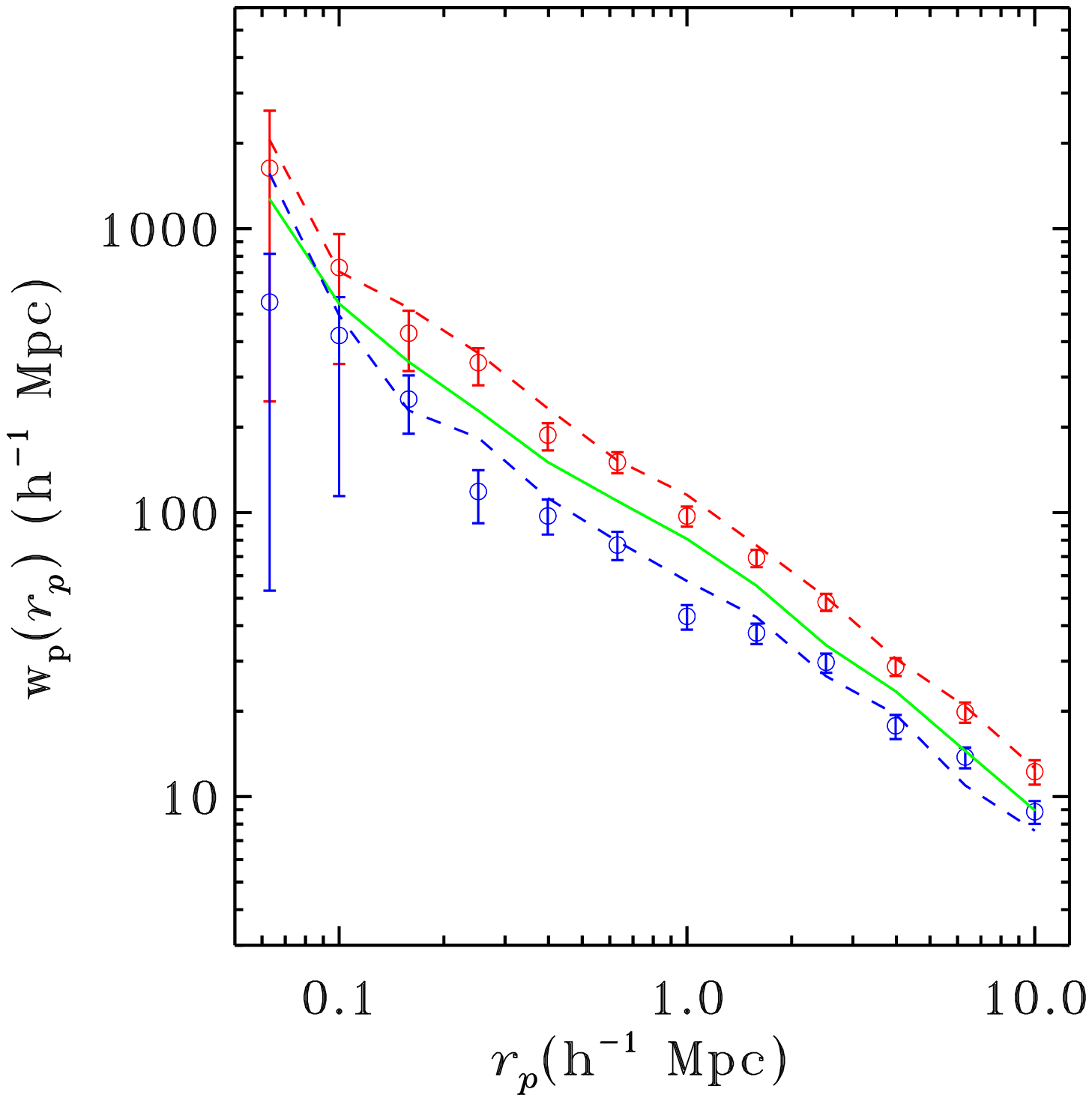}
\includegraphics[width=0.45\textwidth,height=0.45\textwidth]{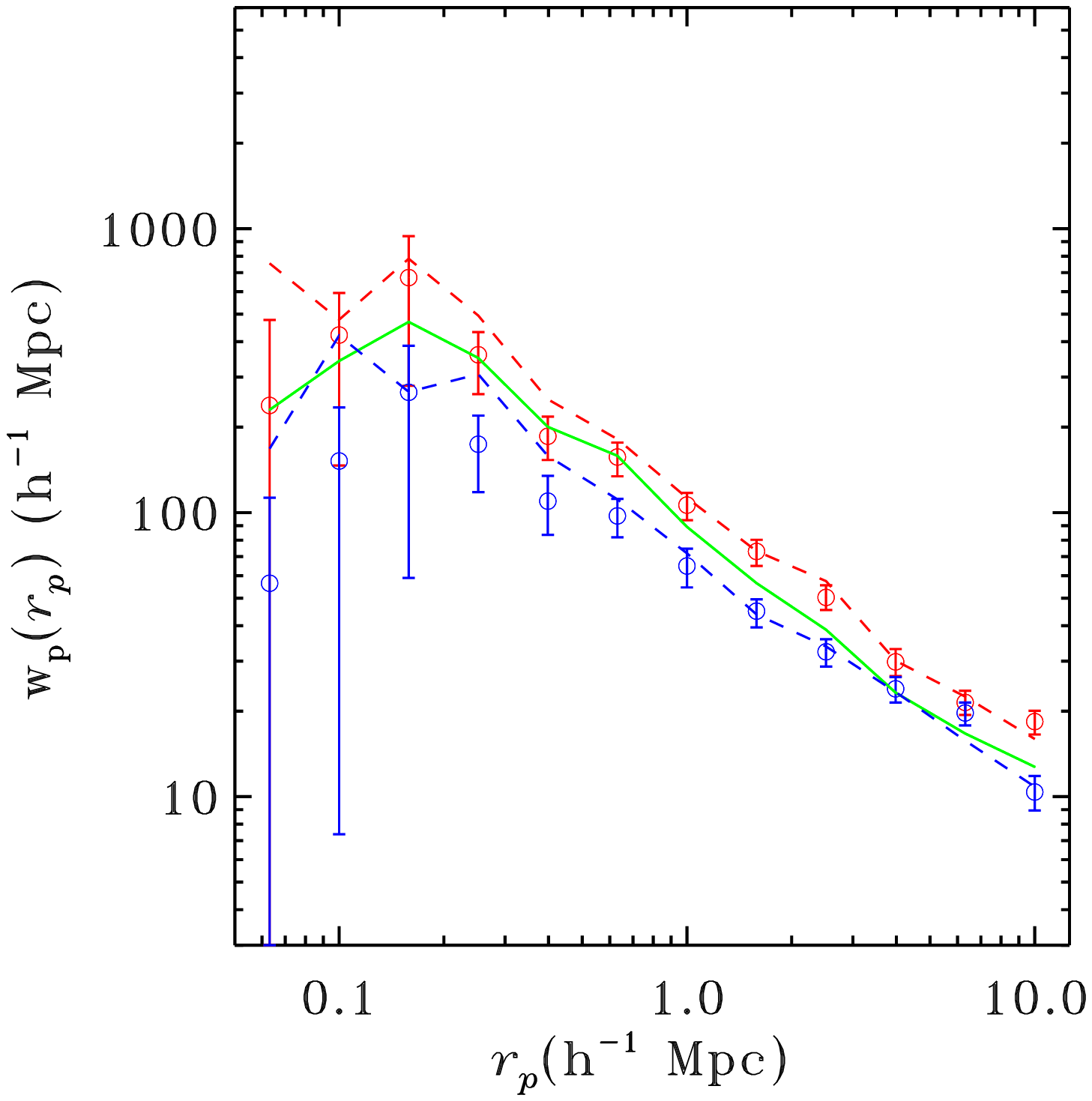}
\caption{The projected cross-correlation function of green sample with blue $w_p^{g,b}(r_p)$ (blue open circles) and 
red $w_p^{g,r})(r_P)$ (red open circles) subsamples.
The solid line shows the auto-correlation function of the green sample. The dashed lines are the geometric mean of the green-red (in red) and 
green-blue (in blue) auto-correlation. On the left (right) we show the analysis from the full (dust-corrected) distribution from the volume-limited sample.
(See Figure \ref{fig:f11} for more details.)
\label{fig:f10}}
\end{figure*}
\begin{figure*} 
\includegraphics[width=0.45\textwidth,height=0.45\textwidth]{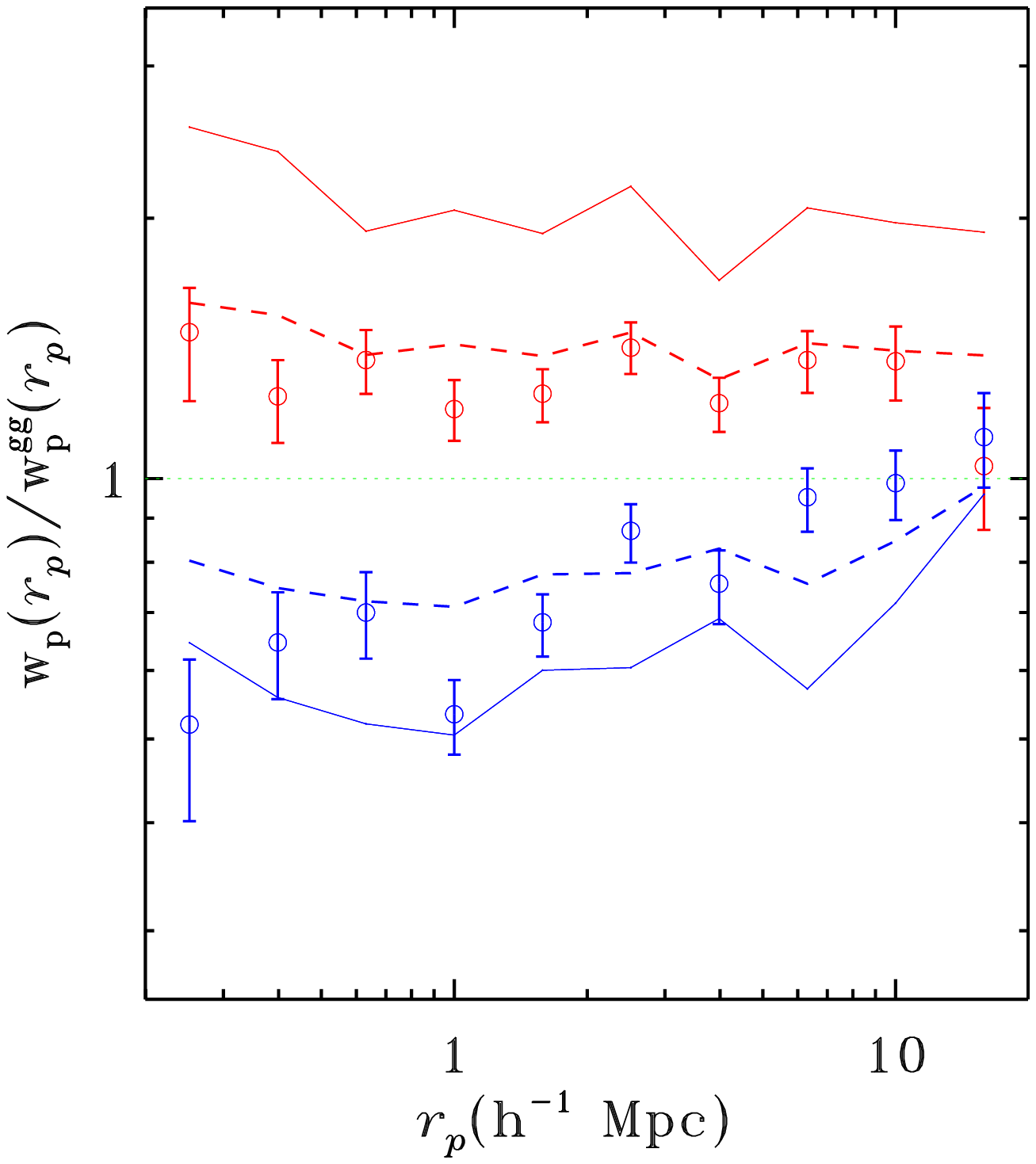}
\includegraphics[width=0.45\textwidth,height=0.45\textwidth]{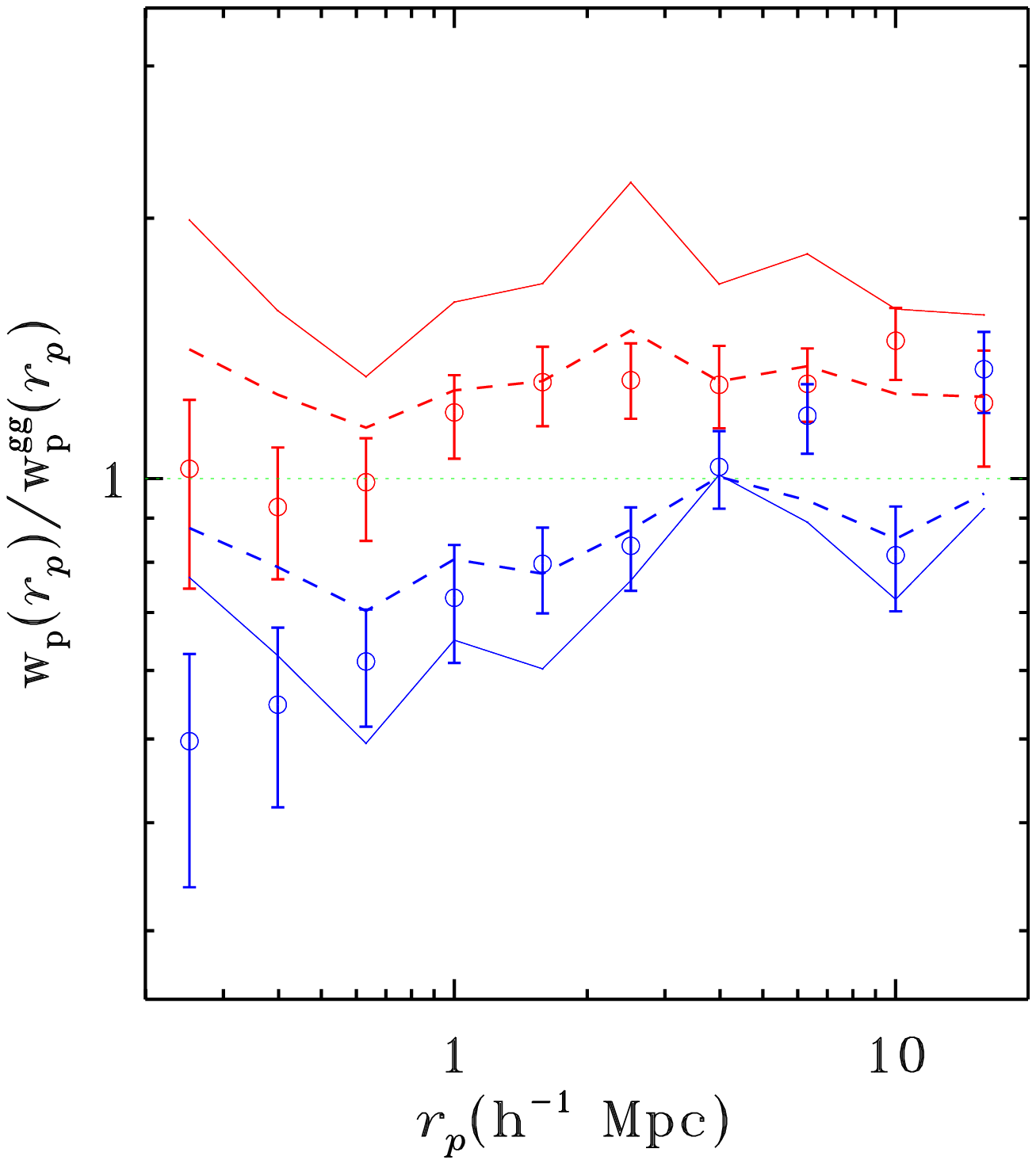}
\caption{The ratio of projected correlation function ($w_p(r_p)$) with the auto-correlation function ($w_p^{gg}(r_p)$) of the green subsample. 
The two ratios from the cross-correlation functions, $w_p^{g,r}(r_p)$ and $w_p^{g,b}(r_p)$, are 
indicated by colored open circles. Dashed lines are for geometric means, as in figure \ref{fig:f10}.  
The additional solid lines are ratios from the auto-correlation function of red and blue respectively. 
The fact that both $w_p^{g,r}(r_p)$ and $w_p^{g,b}(r_p)$ are below their respective geometric mean on small scales 
suggest that both red and blue galaxies are found in different environments compared to the green on these scales, albeit with
large uncertainties.  
\label{fig:f11}}
\end{figure*} 
Figure \ref{fig:f10} shows the projected cross-correlation function between the green and red $w_p^{g,r}(r_p)$, 
and green and blue $w_p^{g,b}(r_p)$. The solid green line is the auto-correlation function of the green 
sample ($w_p^{gg}(r_p)$), while the dashed lines are the geometric means: $w_p^{\overline{g,r}}(r_p)$ and 
$w_p^{\overline{g,b}}(r_p)$. For the full distribution (left panel), $w_p^{g,r}(r_p)$ is systematically below 
$w_p^{\overline{g,r}}$ for a range of scales within errors.  $w_p^{g,b}(r_p)$ are consistent, or slightly 
above $w_p^{\overline{g,b}}(r_p)$ for $r_p \gtrsim 1 \hmpc$, and systematically below for 
$0.3 \lesssim r_p \lesssim 1 \hmpc$. These inferences are shown more clearly in Figure \ref{fig:f11} 
(left for full, right for dust-corrected), where we plot the normalized cross clustering strength -- 
the ratio $w_p^{g,x}/w_p^{gg}$, where $x$ is either red, or blue -- relative to the auto-correlation of the 
green galaxies. The additional solid lines are the normalized ACF of the red and blue. This suggest that 
red and green are consistent with being drawn from the same statistical sample on average for a large range of 
scales with a slight anti-bias on small scales. For the green and blue, there is a stronger anti-bias for $r \lesssim 1 \hmpc$
as they avoid each other on these scales. Our results suggest that on scales typical of dark matter halos, 
galaxies drawn from the green and blue populations are less associated than would be predicted by their respective auto-correlation function.

\section{Discussion}\label{sec:discuss} 
\subsection{Comparison with DEEP2 $z \sim 1$} \label{sec:deep}
The GALEX $NUV$ selected sample used in our analysis is directly comparable to the high redshift 
($z \approx 1$) galaxy sample of the DEEP2 survey \citep{Coi08} since their optical selection mimics 
the rest-frame $NUV$. In that paper, clustering analysis was done on green valley galaxies for the 
first time. In their analysis, the green valley $\xi(r_p,\pi)$ appears to display kinematic structure 
intermediate between red and blue galaxies, with an intermediate finger-of-God effect and an intermediate 
overall clustering amplitude. When redshift space distortion is removed, the projected correlation 
function $w_p(r_p)$ shows a scale dependence convergence. At $r_p > 1 \hmpc$ the functions converge to 
those of red populations, while for $r_p < 1 \hmpc$ it tends toward the clustering of the blue population.

In contrast, we find that our green valley $\xi(r_p,\pi)$ has a strong finger-of-God effect consistent 
with that measured for red sequence galaxies, differing only in their amplitude. This can be seen most 
clearly in the $w_p(r_p)$ analysis for the volume-limited sample (Figure \ref{fig:f7}), where the green function 
has form similar to that of the red function for scales $0.2 < r_p < 15 \hmpc$, but is displaced to lower amplitude. 
It is noteworthy that the slope of the blue correlation function begins shallowing at $r \sim 1 \hmpc$, displaying the kind of 
one-halo and two-halo segregation expected from a correlation function dominated by central galaxies. In contrast 
to \cite{Coi08}, the green auto-correlation function converges to that of the blue at $r_p > 1 \hmpc$. 
We emphasize that the green galaxy sample of \cite{Coi08} is defined 
differently from ours.
We divide the CMD into three disjoint parts to separate our galaxies into the subpopulations while 
in \citeauthor{Coi08}, the green overlaps both the red and the blue. 

\subsection{Green Valley}
Many recent studies reveal that blue sequence mass has remained roughly constant since $z\sim 1$ 
\citep{Bla06,Fab07} because the average ongoing star formation over $0 < z \lesssim 1$ is 
balanced by mass flux off the blue sequence, presumably towards the build-up of the red sequence 
since $z \approx 1$ \citep{Bel04,Mar07}. 
Hence, green valley galaxies occupy a position where one expect to find many transitional galaxies. 
\citeauthor{Mar07} also note that the AGN fraction peaks at the green valley. For our green valley 
definition using the oblique color cuts from 
Figure \ref{fig:f1}, the AGN fraction is $\sim 50 \%$\footnote{This is a lower limit since the fraction 
of low luminosity AGN and composite objects is unknown.}.

In their study of AGN using SDSS, \cite{Con06} found that the redshift space two-point correlation function 
of Seyferts is less clustered than that of LINERS (low-ionization nuclear emission-line regions). 
However, \cite{Mil03} and \cite{Li06b} found that AGN as a whole 
cluster similarly as typical $L^*$ galaxies, if one takes into account luminosity bias. 
\cite{Wan08} argues that almost all galaxies in the local universe with stellar mass $\gtrsim 10 \hmsolar$ have active nuclei,
often LINERS with lines too weak to be detected spectroscopically in SDSS. 
Our green galaxies, with luminosities peaked at $M_r \sim -21$, have stellar masses well above $10 \hmsolar$, and
could be dominated by LINERS (either detected or undetected). This would in part explain the clustering effect we 
see in Figures \ref{fig:f3} and \ref{fig:f4} (middle panels) where those bulge-dominated LINERS display kinematics similar to red 
sequence (primarily non-active) galaxies. 
The reason that AGN from $r$-band selected survey \citep[e.g.][]{Mil03} cluster on average much like typical $L^*$ 
galaxies may merely be coincidental.

The lack of such behavior in the $z \sim 1$ green valley 
\citep{Coi08} may be attributable to evolution in the AGN population.  There may be fewer LINERS, or the red 
sequence galaxies may have been experiencing more gas infall, feeding their AGN.
We note that with a relative bias $\sim 1.1$, green galaxies cluster similarly to a typical galaxy
with $L^* - 0.5 \approx -21$, the median luminosity of our sample. 

As was discussed in section \ref{sec:dust}, a substantial fraction of the galaxies in the green valley 
are dusty star forming galaxies. Because dust content can modify the color-magnitude diagram we use to 
separate the galaxies into red, 
green and blue populations, this might potentially change the behavior of the correlation function. We 
argue here that to the extent that dust modifies the number counts of green galaxies, it is to promote 
the migration from the blue sequence to the green valley \citep{Cho07}. Our results from Figure 
\ref{fig:f4} (for the volume-limited sample) suggest that their influence in modest at best, and merely 
acts as additional poisson noise to the two-dimensional $\xi(\pi,r_p)$ without altering the kinematics 
in the sample. The projected correlation functions of Figure \ref{fig:f7} show similar characteristics.

\subsection{Blue ($UV$) vs. Blue ($\rm Optical$)}
In their studies using $g-r$ color, \cite{Zeh05} found that the correlation function of blue galaxies 
exhibits a lower amplitude and shallower correlation functions. By fitting the $w_p(r_p)$ with halo 
occupation distribution (HOD) models, they found that the majority ($\sim 70-90$\%) of blue galaxies 
are central galaxies in dark matter halos, usually halos with mass $\lesssim 10^{13} M_{\sun}$, in 
contrast to only the most luminous red galaxies being central objects of massive halos, while the 
majority of (less luminous) red galaxies are satellites. This fits well with the notion that blue galaxies are field 
galaxies -- the central objects of low mass halos; red galaxies, with the exception of the central 
galaxy in clusters and groups, are mostly satellites of massive halos.

One can infer a similar conclusion from the auto-correlation $w_p(r_p)$ plot for the blue galaxies 
(Figure \ref{fig:f7}). If we decompose the correlation function into two parts, one due to the 
one halo term and the other due to the two halo term, blue galaxies show strong two-halo excess on 
scales $\gtrsim 1 \hmpc$, implying that on those such scales, the majority of galaxy pairs are from 
different halos.
This was seen in the optical analysis of \cite{Zeh05} but prominently in our blue sequence sample. For 
the dust-corrected volume-limited sample (right panel of Figure \ref{fig:f7}), the amplitude of the 
auto-correlation function of the blue sample beyond $\sim 3 \hmpc$ actually rises to match the clustering 
strength of the green sample.

\cite{Li06} found that the dependence of $w_p(r_p)$ on optical $g - r$ color extends beyond $ 5 \hmpc$, 
suggesting that the conventional wisdom that clustering should converge at large scales may not occur 
until at a scales larger than $5 \hmpc$. Here, we argue that this is an effect due to the mixture of population 
between blue galaxies, and green and red; that the optical $g-r$ color does not have sufficient power to 
separate the green from the blue. Blue galaxies, by themselves, have a very pronounced two halo excess and 
are dominated entirely by central galaxies, flattening the correlation function substantially at large scales, 
compared to the red and green population. To the extent that one can eliminate or correct for the internal 
reddening due to dust, $NUV - r$ color is very efficient in isolating a sequence of purely star-forming galaxies.

\section{Summary and Conclusions}\label{sec:summary}
We have constructed a \galex and SDSS matched catalog, where we have used the GR3 catalog from \galex and 
the SDSS DR5 main spectroscopic galaxy sample. We construct the galaxy distribution of $NUV-r$ vs $M_r$ 
color-magnitude diagram, and divide the distribution into populations of red sequence, green valley and 
blue sequence. Since our main goal is to study the color dependence of clustering, we took substantial care in 
matching the luminosity distribution of each population.
For each population, we measure the two-dimensional correlation function $\xi(\pi, r_p)$, 
and the one-dimensional projected correlation function $w_p$. We also perform cross-correlation analyses 
between each of the sub-populations.  

Our principle finding is that the red sequence and green valley appear to show similar clustering properties, 
as expressed in the finger-of-God effect in the auto-correlation function. The projected correlation function 
is consistent with red and green galaxies residing as satellites of massive halos, while the blue sequence 
shows what appears to be a clear two-halo signature, hence primarily serving as central galaxies of less massive halos. 
The cross-correlation function also shows that green and blue galaxies, on small scales, are not a 
mere statistical mix, but are spatially segregated from each other.

The findings would appear to place the green valley population with the red sequence. The green valley would 
largely consist of massive galaxies that reside in massive halos, and which cluster like the red sequence. 
We note that \cite{Mar07}, \cite{Wyd07}, and \cite{Sal07} show that a large fraction of type II AGN are found 
in the green valley. \citeauthor{Sal07} show that in the plot of specific star formation rate vs stellar mass, 
the AGN tend to be found in massive ($>3\times 10^{10}M_\odot$) galaxies. The AGN occupy a region in these plots 
that strongly resembles that of the reddest class of galaxies, the ``no-H$\alpha$'' red sequence galaxies. 
Significantly, the AGN are clearly offset from the locus of the blue sequence, in the plot of specific star-formation 
rate (SFR) versus mass. The significance is that while a minority of AGN are found with properties that coincide 
with those of the more massive blue sequence galaxies, green valley galaxies -- the subsample with the 
largest AGN fraction --- exhibit properties similar to those of the red sequence, but showing mildly elevated star formation.
 
In this study, we have shown that the green valley population clusters in ways that are characteristic of, but 
also less strongly than, the red sequence. One may suggest that these studies paint a picture in which both the 
properties of the green valley and the ``demographics'' are different from those of the blue sequence, at the present 
epoch. These findings do not necessarily contradict the studies that find an increase in the total mass of red 
sequence galaxies since $z\sim 1$. They do suggest, however, that if blue sequence galaxies evolve by some process 
to the green valley, and ultimately to the red sequence, such evolution must also accompanied by
a transition from the field environment to a group/cluster environment.
Such a change could conceivably occur if the blue population resides along 
filaments that infall into clusters, over time. Our cross-correlation results show that green
galaxies avoid both red and blue galaxies on small scales is consistent with the change in environment hypothesis.
We note that models like ram pressure stripping \citep{Gun72}, starvation of (cold) gas, and the virial shock heating 
model of \citep{Dek06} naturally incorporate environmental factors in their mechanism for
color transformation in galaxies. 

It is also possible that the downsizing \citep{Cow96} effect is so strong that most star forming galaxies are evolving 
rapidly with redshift \citep[e.g.][]{Tin68}. One must recall that star forming activity at $z\sim 1$ resides in considerably more massive 
galaxies, and that a color-based population separation, as we have done, will refer to much higher masses; the rest-frame 
colors may be similar, but the fundamental nature of the galaxies, not. 

One may speculate that the the green valley is occupied by nominally red galaxies that experience the infall of a gas 
rich system that either induces star formation and/or fuels the AGN, rendering it visible via its emission lines. 
However, the feedback of an AGN might inhibit star formation and move a blue sequence galaxy to the green valley.  
Any number of environmental effects (e.g. harassment, starvation) might speed the consumption of gas in a disk, again 
moving a galaxy to the green valley. A small sample of optically quiescent members of the green valley that nominally have 
a UV excess show clear star formation signatures (spirals) when imaged in the UV using HST \citep{Ric09}.  
There is still the issue of the origin of the low mass red sequence, and the evolution of blue sequence, 
into the green valley and ultimately the red sequence, might have an important role in the growth of the lower mass portion of the
red sequence. This was partially addressed by semi-analytical work of \cite{Ben03} and \cite{Bow06}.

In contrast to the construction of color-magnitude diagrams for stellar populations, the environment, for galaxies, 
is a critical physical variable in their evolution. This is true both the in the sense of their dark matter environment 
as well as the presence of detectable companion stellar systems. In considering the major processes driving galaxy evolution, 
it would appear that evolution of both of these observables must be considered, as a function of look-back time.

\section*{Acknowledgments}
Y.S.L. would like to thank C. Hirata, S. Salim, C. Park, J. Kormendy and Z. Zheng for helpful discussions. 
This work has made extensive use of IDLUTILS\footnote{http://spectro.princeton.edu/idlutils}
and Goddard IDL libraries.  RMR acknowledges support from
grant GO-11182 from the Space Telescope Science Institute.

\galex ({\it Galaxy Evolution Explorer}) is a NASA Small
Explorer, launched in April 2003. We gratefully acknowledge
NASA's support for construction, operation, and science
analysis for the \galex mission, developed in cooperation
with the Centre National dEtudes Spatiales of
France and the Korean Ministry of Science and Technology.

{\it Facilities:} \galex, SDSS

\label{lastpage}
\end{document}